\immediate\write10{}
\immediate\write10{This file makes use of the commutative-diagram 
package "diagrams"} 
\immediate\write10{See beginning of this file for source and 
other instructions.}
\immediate\write10{}

\documentclass[12pt,titlepage]{article}
\usepackage{makeidx}
\input diagrams

\font\gothic=eufm10 at 12pt
\font\blackboard=msbm10 at 12pt
\font\blackeight=msbm8

\def\bbf#1{\hbox{\blackboard #1}}
\def\bbfeight#1{\hbox{\blackeight #1}} 
\def\goth#1{\hbox{\gothic #1}}

\def\lC{\bbf C}
\def\lCe{\bbfeight C}
\def\lF{\bbf F}
\def\lFe{\bbfeight F}
\def\lG{\bbf G}
\def\lH{\bbf H}
\def\lN{\bbf N}
\def\lR{\bbf R}
\def\lRe{\bbfeight R}
\def\lZ{\bbf Z}

\def\cA{{\cal A}}
\def\cB{{\cal B}}
\def\cC{{\cal C}}

\def\cF{{\cal F}}
\def\cG{{\cal G}}

\def\cL{{\cal L}}
\def\cM{{\cal M}}
\def\cO{{\cal O}}
\def\cP{{\cal P}}
\def\cU{{\cal U}}
\def\cV{{\cal V}}
\def\cX{{\cal X}}
\def\cY{{\cal Y}}
\def\cZ{{\cal Z}}

\def\gg{{\goth g}}
\def\gr{{\goth r}}
\def\gu{{\goth u}}

\def\Ad{\hbox{Ad}}
\def\Aut{\hbox{Aut}}
\def\Cl{{\cC\ell}}
\def\End{\hbox{End}}
\def\GA{\hbox{GA}}
\def\GL{\hbox{GL}}
\def\Hom{\hbox{Hom}}
\def\O{\hbox{O}}
\def\Pin{\hbox{Pin}}
\def\Tr{\hbox{Tr}}
\def\SO{\hbox{SO}}
\def\SU{\hbox{SU}}
\def\Spin{\hbox{Spin}}
\def\U{\hbox{U}}
\def\Id{\hbox{Id}}
\def\diag{\hbox{diag}\,}
\def\ppv#1#2{\frac{\partial}{\partial #1^#2}}

\newtheorem{theorem}{Theorem}

\newtheorem{eg}{Example}
\newenvironment{proof}{{\em Proof:\/}\ }{\ Q.E.D}
\newenvironment{example}{\begin{quote}\begin{eg}}{\end{eg}\end{quote}}

\makeindex
\title{Preparation for  Gauge Theory}
\author{George Svetlichny\thanks{Departamento de Matem\'atica, 
Pontif\'{\i}cia Universidade Cat\'olica, Rua Marqu\^es de S\~ao Vicente 225,
22453-900 G\'avea, Rio de Janeiro, RJ, Brazil\newline 
\centerline{e-mail: svetlich@mat.puc-rio.br}}}
\date{February 19, 1999}
\begin{document}
\maketitle
\pagestyle{empty}
\newpage\ \newpage
\pagestyle{plain}
\setcounter{page}{1}
\pagenumbering{roman}
\tableofcontents
\newpage
\pagenumbering{arabic}
\setcounter{page}{1}
\section{Introduction}
These are the class notes for a course preparatory to classical gauge
theory given at the Mathematics Department of the Pontif\'{\i}cia
Universidade Cat\'olica of Rio de Janeiro during the
(southern-hemisphere) spring term of 1997. They purport to provide the
necessary mathematical background at a beginning graduate level for
someone interested in gauge theory and who has an elementary
understanding of differentiable manifolds, Lie groups, Lie algebras, and
multilinear algebra. These notes are intended to take the reader to the
point at which he or she can understand what gauge theories {\em are\/},
but, unfortunately, stop short of doing anything with them. They do not
treat the
more advanced, and more
exciting topics, such as moduli spaces, topological invariants, and
quantum aspects.

Field theory was a scientific revolution initiated in the nineteenth
century with the development of Maxwell's electrodynamics. It gained a
strongly geometric character with Einstein's general theory
of relativity. A century later, the revolution still
continues with new insights and ideas appearing without abating. Much of
the motivation and inspiration comes from the still largely incomplete
theory of quantum fields and its generalizations, such as string and 
M-theory. These quantum  theories exercise a strong
influence on classical field theory and many aspects of research into
classical field theory are incomprehensible without understanding the
effort to gain insight into quantum theories. Since these notes are
directed toward classical gauge theory, certain construct may seem
arbitrary or unmotivated as their full appreciation can only come
through quantum theory. These would have to be taken mostly on faith
that they are appropriate and interesting, as we have no means to
explore the quantum aspects in these notes.

A second revolution in field theory occurred with the advent of gauge
theory to describe fundamental particle interactions. There were many
fundamental shifts in viewpoint, among them the realization that
physical fields, originally represented by functions on space-time, had
to be treated by topologically more sophisticated objects, technically
know as sections of fiber bundles. This change was already implicitly
presaged by general relativity, which for a long time led the process of
introducing geometric reasoning into physics. The full changeover
however had to await the flowering of gauge theory.

This ``gauge revolution" had remarkable and totally unexpected
mathematical consequences. The very same equations that were
instrumental in constructing the most successful theory of matter know
to date, provided subtle tools for exploring the structure of low
dimensional manifolds and objects therein, by means of entirely new
invariants. A mathematical revolution followed at the heels of the
physical one. These notes are dedicated to shorten the path of the
interested reader to the point where he or she can begin to appreciate
the fruits of these remarkable developments.

I am indebted to the participants of the course for many helpful remarks
and suggestions.

\section{Preliminaries, Notation, Conventions}\label{sec:prelim}

We assume the reader is familiar with the basic notions of general
topology, group theory, multilinear algebra, differentiable manifolds,
Lie groups and Lie algebras, and a smattering of categorical ideas.
Although fiber-bundle theory is treated in these notes, we assume the
reader has some elementary understanding of the standard bundles found
in manifold theory, such as the tangent bundle, the cotangent bundle,
the bundle of exterior forms, and tensor bundles.

We shall have occasion in these notes to deal with similar constructs in
several categories, the main ones being sets, topological spaces, and
\(\cC^\infty\) differentiable manifolds. By words such as ``map",
``morphism", ``isomorphism", etc.\  we shall mean the 
notion appropriate to the
category. Thus when talking about manifolds, by ``map" we shall mean a
\(\cC^\infty\) map, when talking about topological spaces, a continuous
map, and when talking about sets, just a map. Similarly for other
notions. Some expositions will be done in one category and the
translation to other categories, when it is straightforward, will be
left to the reader.

We shall use various notational conventions which are summarized in the
appendices. Mention to the appendices are made upon first use of the
conventions in the text. Appendix \ref{sec:bacon} resumes the main
conventions of these notes as a whole.

Most of the material in these notes is elementary and the proofs are
easy. We leave out some  proofs that are either very technical or do not
contribute to understanding the essentials of the subject.

\section{Groups}
\subsection{Group Action}
In this section we are in the category of sets. With the understanding
of Section \ref{sec:prelim}, most of the definitions and results can be
carried over without change when the objects are topological spaces or
manifolds.

Let \(G\) be a group and \(X\) a set.
By a {\em left action\/}
\index{action}%
\index{action!left}%
 of \(G\)
on \(X\) we mean a map \(\alpha: G\times X \to X\) which satisfies
\begin{enumerate}
\item \(\alpha(e,x) =x\)
\item \label{item:laccomp}\(\alpha(g, \alpha( h,x) ) = \alpha(gh,x)\).
\end{enumerate}%
For convenience we shall normally write \(g\cdot x\)
\index{\(G\cdot X\)@\(g\cdot x\)}%
 instead of
\(\alpha(g,x)\). 
\index{\(\alpha(g,x)\)}%
In this notation the two axioms become \(e\cdot x = x\)
and \(g\cdot(h\cdot x) = (gh)\cdot x\).  One easily sees that each map
 \(\alpha(g,\cdot)\) is  invertible with inverse
 \(\alpha(g^{-1},\cdot)\), and that by axiom (\ref{item:laccomp}) the map
 \(g\mapsto \alpha(g,\cdot)\) is a group homomorphism \(G \to \Aut(X)\).
Reciprocally, any group homomorphism of \(G\) into \(\Aut(X)\) defines a
left action.
Similarly, by a {\em right action\/}
\index{action!right}%
 of \(G\)
on \(X\) we mean a map \(\beta: X\times G \to X\) which satisfies
\begin{enumerate}
\item \(\beta(x,e) =x\)
\item \(\beta( \beta(x,g),h ) = \beta(x,gh)\).
\end{enumerate}%
For convenience we shall normally write \(x\cdot g\)
\index{\(X\cdot G\)@\(x\cdot g\)}%
instead of
\(\beta(x,g)\).
\index{\(\beta(x,g)\)}%
Properties analogous to those of left action hold also for right action,
however now
\(g\mapsto \beta(\cdot,g)\) is a group {\em anti-homomorphism\/} \(G \to
\Aut(X)\).

We have a natural left and a natural right action of a group G on
itself, given by multiplication: \(\alpha(g, h) = gh\) and \(\beta(h,g)
= hg\). 

If we have a left action on \(X\) by a group \(G\) and simultaneously a
right action by a group \(H\), we say the two  actions {\em commute\/}
\index{action!commutative}%
 if
\((g\cdot x)\cdot h = g \cdot (x \cdot h)\)  for all \(x\in X\), \(g\in
G\), and \(h\in H\).

Let \(G\) be a group and \(H\) a subgroup. We have a left action of
\(G\) on the right coset spaces \(kH\) of \(G\) defined by \(g\cdot (kH)
=gkH\), and we likewise have a right action of \(G\) on the left coset
spaces
by \((Hk)\cdot g = Hkg\).  When the subgroup \(H\) is normal,
then the left and
right coset spaces coincide and so one has both a left and a right action
on them and the two actions commute. The actions of \(G\) on its 
(left and right) coset
spaces defined above are called  {\em canonical\/}
\index{action!canonical}%
actions.

Let \((G,X,\alpha)\) and \((G,Y,\gamma)\) be
two left actions of the same group \(G\). A
morphism
\index{action!morphism}%
from the first action to the second is a map
 \(f:X \to Y\) such that 
\begin{equation}\label{eq:moract}
f(g\cdot x) = g\cdot f(x)
\end{equation}%
This is equivalent to the commutativity of the following diagram
\begin{diagram}
G\times X & \rTo^\alpha & X \\
\dTo^{\Id\times f} & & \dTo^f \\
G\times Y & \rTo^\gamma & Y \\
\end{diagram}%

We say such a morphism is an isomorphism if  \(f\) is invertible. 
Isomorphic group actions will
also be called {\em equivalent\/}.
\index{action!equivalent}%
Analogous definitions hold for right actions. 

If \(f\) is invertible then, given \(y\in Y\), one can 
write (\ref{eq:moract}) as 
\begin{equation}\label{eq:tranact}
g\cdot y = f(g\cdot f^{-1}(y))
\end{equation}%
This equation can be used, whenever one has an action in \(X\) but not 
in \(Y\),  to {\em define\/} one in  \(Y\) that is
equivalent to the one in \(X\). It is an easy exercise to show that
(\ref{eq:tranact}) does indeed define an action. 

Note that the definition of left and right action of a group 
does not use the
existence of the inverse in a group, and an identical definition can be
given for a semigroup \(S\). We thus have the more general notion of
{\em semigroup action\/},
\index{action!semigroup}%
which at times is useful to consider.

If \((G,X,\alpha)\) is a left action, then \(G_0=\{g\in G\,|\,\forall x \in
X, g\cdot x = x\}\) is a normal subgroup of \(G\), being the kernel of
the homomorphism \(G \to  \Aut (X)\). One then has  the action of
the quotient group \(G/{G_0}\) on \(X\) given by \(gG_0\cdot x = g\cdot
x\). It is easy to verify that this is well defined and that the axioms
are satisfied. An action for which \(G_0 = \{e\}\) is called {\em
effective\/}.
\index{action!effective}%
We shall normally deal only with effective actions though at times
non-effective actions will arise.

Given a left action of \(G\) on \(X\) and a point \(x\in X\) we call the
set \(\cO_x=\{g\cdot x\,| g\in G\}\)  the {\em orbit}
\index{orbit}%
\index{action!orbit}%
\index{\(O_x\)@\(\cO_x\)}%
of \(x\). We have
the following
\begin{theorem}
The set of orbits \(\{\cO_x\,|\,x\in X\}\) partitions the set \(X\).
\end{theorem}%
\begin{proof} As \(x\in \cO_x\) the set of orbits cover \(X\). Suppose
\(z\in \cO_x\cap\cO_y\) and \(w \in \cO_x\). We have \(z=g\cdot x\),
\(z = k\cdot y\)   and
\(w = h\cdot x\) for some elements \(g,h,k \in G\).  From this we find
\(w= hg^{-1}k\cdot y\in\cO_y\) and so \(\cO_x \subset \cO_y\)
but by symmetry we
also have \(\cO_y\subset \cO_x\) and so the two orbits coincide. Thus
two orbits either are disjoint or coincide and we have a partition.
\end{proof}%

We say an action is {\em transitive\/}
\index{action!transitive}%
if there is only one orbit, that
is, given any \(x_1,x_2 \in X\) there is a \(g\in G\) such that \(x_2  =
g\cdot x_1\).

If \(Y\subset X\) is a union of orbits and \(y\in Y\), then \(g\cdot y
\in Y\) for all \(g\in G\).  We can thus restrict the action of \(G\) to
\(Y\).   Clearly, the minimal subsets for which we can do this are 
single orbits.

For \(x\in X\)  define \(K_x=\{g\in G\,|\, g\cdot x = x\}\). Obviously
\(K_x\)
\index{\(K_x\)}%
is a subgroup of \(G\). It is called the {\em stability
subgroup\/}
\index{stability subgroup}%
of \(x\).

We say an action is {\em free\/}
\index{action!free}%
if \(K_x = \{e\}\) for all \(x\).  In
particular this means that given two points on an orbit, there is a
unique element of the group that connects the two.

\begin{theorem} Let \(y \in \cO_x\) then \(K_y = gK_xg^{-1}\) where
\(g\) is any element of \(G\) such that \(y=g\cdot x\).
\end{theorem}%
\begin{proof}
We have \(h\cdot y = y\) if and only if \(h g\cdot x = g \cdot x\) if
and only if \(g^{-1}hg \cdot x = x\).
\end{proof}%

\begin{theorem}The action restricted to
any orbit \(\cO_x\) is equivalent
to the canonical action of \(G\) on the right coset spaces of the
stability subgroup \(K_x\) of \(x\).
\end{theorem}%

\begin{proof} Let \(Z=\{gK_x\,|\,g\in G\}\) and define \(f:\cO_x \to Z\)
by associating to \(y=a\cdot x\in \cO_x\) the coset
\(aK_x\). One has \(a\cdot x = b\cdot x\) if and only if \(a^{-1}b \cdot
x = x\) that is, if and only if \(a^{-1}b \in K_x\), and this if and
only if \(aK_x = bK_x\). Thus the
correspondence is well defined and injective. It is obviously
surjective, and so bijective.   One has
\(f(g\cdot y) = f((ga)\cdot x) = gaK_x = g\cdot f(y)\) and so
\(f\)
defines an equivalence of actions.
\end{proof}%
 
\begin{example}
The special orthogonal group \(\SO(3)\) acts naturally on \(\lR^3\). The
orbit of any non-zero vector is the sphere centered at the origin that
 contains it. The
stability subgroup of \((0,0,1)\) is \(\SO(2)\), the group of rotations
of the \(x\)-\(y\) plane. One thus has for the coset space 
\[
\SO(3)/\SO(2)\simeq S^2
\]%
One should view this as an isomorphism of manifolds.
\end{example}

\begin{theorem}
Two canonical left actions on right coset spaces \(G/H\) and \(G/K\) are
equivalent if and only if the subgroups \(H\) and \(K\) are conjugate.
\end{theorem}%

\begin{proof}
Suppose \(f: G/H \to G/K\) establishes the equivalence. Let \(f(H) = wK\). 
For \(g\in H\) one has \(g\cdot f(H) = f(gH)=f(H)\)
and so \(gwK = wK\). Thus \(w^{-1}gw\in K\) and we conclude that
\(w^{-1}Hw \subset K\). On the other hand, given \(k\in K\) one has
\(f(wkw^{-1}H)=wkw^{-1}f(H)=wkw^{-1}wK=wK=f(H)\). As \(f\)
is an isomorphism one has \(wkw^{-1}H = H\) meaning that \(wKw^{-1}
\subset H\) which with the previous inclusion implies that
\(H=wKw^{-1}\).
\end{proof}%

We see thus that there are universal models for the action within any
single orbit, the canonical action on coset sets, and that the
equivalence classes of actions on  single orbits is in bijective
correspondence with
the set of conjugacy classes of subgroups of \(G\).  The actions within
an orbit can be studied
entirely within the structure of the group \(G\).
The
study of group actions thus divides neatly into two problems: the {\em
internal\/} problem of understanding the action within single orbits
(equivalent to studying the canonical action in coset spaces) and the
{\em external\/} problem of understanding how the orbits are put together
to form the set \(X\).

Since the orbits partition \(X\), they define an equivalence relation
and the quotient space by this relation is called
the {\em space of orbits\/}. Understanding the space of orbits is often
of utmost importance.

\subsection{Lie Groups and Lie Group Actions}\label{sec:liac}
In this section we review, without proofs, some basic facts about Lie
groups.
Readers desiring greater detail should consult appropriate textbooks on
the subject.

Let \(G\) be a Lie group.
When convenient,
we shall in this section denote the right action of \(G\) on \(G\) 
by \(\rho\). Thus
\(\rho_g
h = hg\).  Similarly for the left action, which we shall denote by
\(\lambda\), and write \(\lambda_gh= gh\).
These maps induce isomorphisms \(d\rho_g : T_hG \to T_{hg}G\) and
\(d\lambda_g: T_hG
\to T_{gh}G\). To facilitate notation we shall write \(d\rho_g(v) = v\cdot
g\)
and \(d\lambda_g(v) = g\cdot v\).
\index{\(V\cdot G\)@\(v\cdot g\)}%
\index{\(G\cdot V\)@\(g\cdot v\)}%

Of particular interest is the tangent space at the identity \(T_eG\)
which is known as the {\em Lie algebra\/} \(\gg\)
\index{\(G\)@\(\gg\)}%
of
\(G\). The tangent space \(T_gG\) at any other point can be canonically
identified with \(\gg\) in two ways, either by the map \(d\rho_{g^{-1}}\)
or by \(d\lambda_{g^{-1}}\). In these notes we shall conventionally use the
identification by \(d\rho\).
Under such an identification, a vector \(v\in T_gG\) will have the 
form \(v=L\cdot
g\)
\index{\(L\cdot g\)}%
 for some \(L\in\gg\).  Given a map \(\phi:G \to M\) where \(M\) is
any manifold, one has \(d\phi_g: T_gG \to T_{\phi(g)}M\). With the
identification of \(T_gG\) with \(\gg\) by right action one has \(d\phi_gv
= \tilde d\phi_g (v\cdot g^{-1})\) for a map \(\tilde d\phi_g = 
d\phi_g\circ d\rho_{g^{-1}}:\gg \to
T_{\phi(g)}M\). Thus 
\begin{equation}
v=L\cdot g\quad \Rightarrow\quad d\phi_gv = \tilde d\phi_g L
\end{equation}%
\index{\(D\phi\)@\(\tilde d\phi\)}%
We shall make frequent use of this construct.

The adjoint map 
\[
\Ad_g (L) = g\cdot L\cdot g^{-1}
\]
\index{\(Ad_g\)@\(\Ad_g\)}%
maps \(\gg\) into itself.

Given a vector \(L\in \gg\) one can extend it to a vector field
\(\cX^\lambda_L\)
on \(G\) by  \(\cX^\lambda_L(g) = g\cdot L\) using left action, and
likewise to a
vector field \(\cX^\rho_L(g) = L\cdot g\) by right action. These vector
fields are {\em invariant\/} in the sense that
\(g\cdot\cX^\lambda_L(h) =
\cX^\lambda_L(gh)\) and \(\cX^\rho_L(h)\cdot g = \cX^\rho_L(hg)\).  Given
these vector fields
one can introduce a Lie bracket in \(\gg\) by
\[
[L,K] = [\cX^\lambda_L,\cX^\lambda_K](e) = [\cX^\rho_L,\cX^\rho_K](e)
\]%
which happens to be the same using either
left-invariant or right invariant extension.
We recall the properties of the Lie bracket
\begin{eqnarray}\label{eq:lieassym}
& [L,K] = -[K,L] & \\ \label{eq:liejacob}
&[L,[K,M]] +[K,[M,L]]+[M,[L,K]] = 0 &
\end{eqnarray}%
Property (\ref{eq:liejacob}) is called the {\em Jacobi Identity\/}.

Furthermore, each vector field \(\cX^\lambda_L\) and \(\cX^\rho_L\) 
defines a flow,
\(\exp({t\cX^\lambda_L})\) and \(\exp({t\cX^\rho_L})\) respectively.
The integral curve passing through the identity \(e\) is the same for
the two vector fields and we define
\[
\exp(tL)  = \exp({t\cX^\lambda_L})e =\exp({t\cX^\rho_L})e
\]
Thus \(\exp(tL) \in G\), and \(L \mapsto \exp(L)\) defines a map
\(\exp:\gg \to G\) called the {\em exponential map\/}.
\index{map!exponential}%
 We also write
\(e^{L}\)
\index{\(E^{L}\)@\(e^{L}\)}%
 for \(\exp(L)\). One has
\(\exp({t\cX^\lambda_L})g = e^{tL}g\) and \(\exp({t\cX^\rho_L})g=ge^{tL}\)
so that both
flows are easily expressed through the exponential map. One has the
useful formula
\[
g \exp(L) g^{-1} = \exp(g\cdot L\cdot g^{-1})=\exp(\Ad_g L)
\]

Let now \(\alpha\) be a  left action  of \(G\) on a
manifold \(M\). The differential
\(d\alpha_{(g,x)}:T_{(g,x)}(G\times M)\to T_{g\cdot x}M\) 
can be written as
\((v,\xi) \mapsto d_1\alpha_{(g,x)}v + d_2\alpha_{(g,x)}\xi\)
\index{\(D_1\alpha\)@\(d_1\alpha\)}%
\index{\(D_2\alpha\)@\(d_2\alpha\)}%
where
\(d_1\alpha_{(g,x)}\) is the partial differential with respect to the
first variable, that is, the differential at \(g\) of the map \(g\mapsto
g\cdot
x\) for \(x\) fixed, and \(d_2\alpha_{(g,x)}\) is the differential at
\(x\) of the map \(x\mapsto g\cdot x\) for \(g\) fixed.  One has
\(T_{(g,x)}(G\times M) \simeq T_gG\times T_xM \simeq \gg \times T_xM\)
where we used the identification of \(T_gG\) with \(\gg\) by right
action. Thus we can write 
\begin{equation}\label{eq:dtilde}
d_1\alpha_{(g,x)}v = \tilde
d_1\alpha_{(g,x)}(v\cdot
g^{-1})=  \tilde d_1\alpha_{(g,x)}L
\end{equation}%
\index{\(D_1\alpha\)@\(\tilde d_1\alpha\)}%
 where \(L\in\gg\) corresponds to
\(v\in T_gG\) under the mentioned identification.

When \(G\subset \GL(X)\) for a finite-dimensional vector 
space \(X\) with the base field \(\lF\) being
 \(\lR\) or \(\lC\), the Lie algebra \(\gg\)
is the 
linear space of endomorphisms \(L\in\End(X)\) such that  
\(\exp (tL)\in G\) for all \(t\), where now 
\(\exp\) is the usual exponential  defined either by functional
calculus of by the exponential series
\[
\exp(tL) = \sum_{n=0}^\infty \frac{t^n}{n!}L^n
\]
The Lie bracket in this case is the commutator \([L,K] = LK-KL\). 
Each tangent space \(T_gG\) is a linear space 
consisting of endomorphisms of the form \(gL\), or equivalently of the form
\(Lg\), for \(L\in \gg\), and where the product is ordinary composition.
The differentials of the right and left action of \(G\) on itself are
likewise given by composition. Thus if \(v\in T_gG\),
then
\(g\cdot v = gv\) and \(v\cdot g= vg\). In particular \(\Ad_gL =
gLg^{-1}\). When \(G=\GL(X)\), then \(\gg = \End(X)\).

When \(X=\lF^n\),  then \(G\) is a matrix group, and all the spaces
mentioned above are spaces of matrices. Composition is ordinary matrix
multiplication. In particular, 
when \(G=\GL(n,\lF)\), then \(\gg = \cM(n,\lF)\), 
\index{\(M(n,\lF)\)@\(\cM(n,\lF)\)}%
the space of all \(n
\times n\) matrices over \(\lF\). 

\subsection{Group and Lie Algebra Representations}\label{sec:reps}

A particular type of group action is given by {\em group
representations\/}.
\index{group!representation}%
This is a left action of a group \(G\) on a {\em
vector space\/} \(X\) over a field \(\lF\) in which each map
\(x\mapsto g\cdot x\) is \(\lF\)-linear. One usually writes \(g\cdot
x =
R(g)x\) where \(R(g) \in \GL(X)\). 
One easily sees that the set of linear transformations
\(R(g)\)
\index{\(R(g)\)}%
satisfy \(R(e) =I\) and \( R(gh) = R(g)R(h)\), that is, one has a 
group homomorphism \(G\to\GL(X)\). Reciprocally,
given such a set of linear maps one has a group representation.
Whenever \(X\) is finite dimensional of dimension \(n\),  one can by a
choice of a fixed basis, represent each \(R(g)\) as an \(n\times n\)
matrix \(M(g) \in \GL(n,\lF)\). These matrices obviously satisfy \(M(e) =
I\)
and \(M(gh)=M(g)M(h) \). We say in this case that we have a {\em matrix
representation\/}
\index{representation!matrix}%
 of \(G\). A matrix representation is thus nothing more
than a group homomorphism \(G \to \GL(n,\lF)\).

Given two representations \(R_1\) and \(R_2\) of \(G\) on vector spaces
\(X_1\) and \(X_2\) over the same field, a linear map \(T:X_1 \to X_2\)
is called an {\em intertwiner\/}
\index{intertwiner}%
\index{representation!intertwiner}%
 for the two representations if
\begin{equation} \label{eq:intertw}
R_2(g)T = TR_1(g)
\end{equation}%

A similar notion holds for two matrix representation, in which case,
\(T\) is an \(n_2\times n_1\) matrix and \(M_i(g) \in \GL(n_i,\lF)\).

The intertwiner expresses the notion of a morphism of 
actions (\ref{eq:moract}), in the context of linear spaces.

We say two representations are {\em equivalent\/}
\index{representation!equivalent}%
 if they have an invertible 
intertwiner. 

We say a representation \(R\) is {\em reducible\/}
\index{representation!reducible}%
 if there is a non-trivial subspace \(Y\subset X\) 
invariant under \(R\), that is,
\(R(g)Y\subset Y\) for all \(g\in G\). For a matrix representation we
can take \(X=\lF^n\) and the same definition applies. A representation
that is not reducible is said to be {\em irreducible\/}.
\index{representation!irreducible}%

Given a representation \(R\) of a group \(G\) in a vector space \(X\)
one has a natural representation \(R^*\), called the 
{\em dual representation\/}
\index{representation!dual}%
 in the dual space \(X'\)
defined, for a linear functional \(\phi\), 
by \((R^*(g)\phi)(x)=\phi(R(g^{-1})x)\),
\index{\(R^*(g)\)}%
that is, \(R^*(g)=R(g^{-1})'\).

When \(G\subset \GL(X)\), there is a natural representation of \(G\) on
\(X\) given by \((g, x) \mapsto gx\).
When \(G\) is a matrix group, \(G \subset \GL(n,\lF)\) then there are
{\em two\/}  natural matrix representations on \(\lF^n\) 
defined by the actions 
\begin{eqnarray}\label{eq:canrep}(m,x) & \mapsto & mx \\ \label{eq:acanrep}
(m,x) & \mapsto & (m^t)^{-1} x
\end{eqnarray}%
Note that
representation (\ref{eq:acanrep}) is 
equivalent to the dual of (\ref{eq:canrep}).

Let \(\cL\) be a Lie algebra over a base field \(\lF\). 
A {\em representation\/}
\index{Lie algebra representation}%
 of \(\cL\) in an
associative algebra \(\cA\) over \(\lF\) is a map 
\(r:\cL \to \cA\) such that
\(r([K,L]) =r(K)r(L)-r(L)r(K).\) Note that in the
associative algebra \(\cA\), one can define a Lie bracket \([\cdot,\cdot]\)
as the commutator \([a,b]=ab-ba\). With this definition one has 
\(r([K,L])=[r(K),r(L)]\).
Particularly important cases are when
\(\cA\) is \(\End(X)\) for a vector space \(X\) over \(\lF\)
and when \(\cA\) is
\(\cM(n,\lF)\) the algebra of \(n\times n\)-matrices over
\(\lF\).

Let now \(G\) be a Lie group and suppose \(X\) finite dimensional with
the base field being either \(\lR\) or \(\lC\). Let
\(L\in\gg\). One can now define a representation \(\gr\)
\index{\(R\)@\(\gr\)}%
of \(\gg\)
in \(\End(X)\) by
\begin{equation}\label{eq:derrep}
\gr(L) = \left.\frac{d}{dt}R(e^{tL})\right|_{t=0}
\end{equation}%
One easily finds that \(\gr\) is indeed a representation and that
\begin{eqnarray*}
R(g)\gr(L)R(g^{-1}) &=& \gr(\Ad_g L) \\
R(\exp(L)) &=& \exp(\gr(L))
\end{eqnarray*}

\subsection{Affine Actions}
Let \(X\) be a vector space over a field \(\lF\). An {\em affine map\/}
\index{map!affine}%
 \(A:X\to X\) is one of the form \(Ax=Bx+a\) where \(B\) is linear and
\(a\in X\). Denote an affine map by the pair \((B,a)\). 
The set of affine maps is an algebra over \(\lF\) with the linear
structure given by 
\(\alpha(B,a)+\beta(D,e)= (\alpha B+\beta D, \alpha a+\beta e)\) for 
\(\alpha, \beta\in \lF\), and the
product by composition: \((B,a)(D,e)=(BD,
a+Be)\). An affine map \((B,a)\) is invertible if and only if \(B\) is
invertible, in which case the inverse is \((B^{-1}, -B^{-1}a)\). 
The set of all invertible affine maps is called the {\em
general affine group\/}
\index{group!general affine}%
 of \(X\) and will be denoted by \(\GA(X)\). 
\index{\(GA(X)\)@\(\GA(X)\)}%
For \(X=\lF^n\) we write \(\GA(n,\lF)\)
\index{\(GA(n,\lF)\)@ \(\GA(n,\lF)\)}%
and for \(X=\lR^n\) we write \(\GA(n)\).
\index{\(GA(n)\)@\(\GA(n)\)}%
An {\em affine representation\/}
\index{representation!affine}%
 of a group \(G\) is a homomorphism of 
\(G\) into
\(\GA(X)\) for some vector space \(X\). The pairs \((B(g), a(g))\) then
satisfy \(B(gh)=B(g)B(h)\) and \(a(gh) = a(g)+B(g)a(h)\). Note that
\(B\) is then a representation of \(G\) as defined in Section
\ref{sec:reps}. 

\section{Bundles}
\subsection{Fiber Bundles}
A {\em fiber bundle\/}
\index{bundle!fiber}%
 is a mathematical object with the following ingredients:
\begin{enumerate}
\item Three topological spaces: the {\em total space\/} \(E\),
\index{space!total}%
the {\em base space\/} \(X\),
\index{space!base}%
and the {\em fiber\/} \(F\).
\index{fiber}%
\item A map \(\pi:E \to X\),
\index{\(\pi\)}%
the {\em projection\/}.
\index{projection}%
\item A covering \(\cU\)
\index{\(U\)@\(\cU\)}%
of \(X\) by a family \((U_\alpha)_{\alpha\in A}\),
of open sets.
\item For each \(U_\alpha\) in \(\cU \) a homeomorphism, the {\em local
trivialization\/}, 
\index{trivialization}%
\[
h_\alpha: \pi^{-1}(U_\alpha ) \to U_\alpha  \times F
\]
\index{\(H_\alpha\)@\(h_\alpha\)}%
such that for
 \(U_\beta\) in
\(\cU \) with \(U_\alpha\cap U_\beta  \neq \emptyset\)  the map
\[
h_\alpha  \circ h_\beta^{-1}: (U_\alpha \cap U_\beta ) \times F \to
(U_\alpha \cap U_\beta ) \times F
\]
is given by 
\begin{equation}\label{eq:trans}
h_\alpha  \circ h_\beta^{-1}(x,f) = (x,h_{\alpha\beta}(x)(f))
\end{equation}%
\index{\(H_{\alpha\beta}\)@\(h_{\alpha\beta}\)}%
where
\[
h_{\alpha\beta}: U_\alpha \cap U_\beta  \to \Aut(F)
\] 
and are called the
{\em transition maps\/}.
\index{map!transition}%
\end{enumerate}%

Note that \(h_{\alpha\beta}\) depends on the {\em ordered\/} pair
\((\alpha,\beta)\).
It's clear that these maps
satisfy the following relations for all \(U_\alpha \) in \(\cU\), for all
 pairs \(U_\alpha ,U_\beta \) in
\(\cU\) such that \(U_\alpha \cap U_\beta  \neq \emptyset\), and 
for all
triples \(U_\alpha ,U_\beta ,U_\gamma\) in \(\cU\) such
that \(U_\alpha \cap U_\beta \cap U_\gamma\neq\emptyset\):
\begin{eqnarray}   \label{eq:ho}
h_{\alpha\alpha}(x) &=& \Id_F  \\ \label{eq:houv}
h_{\beta\alpha}(x) &=&  h_{\alpha\beta}(x)^{-1}\\ \label{eq:hocycle}
h_{\alpha\beta}(x)\circ h_{\beta\gamma}(x)\circ h_{\gamma\alpha}(x)
 &=& \Id_F
\end{eqnarray}%
These three relations are not independent as either (\ref{eq:ho}) or
(\ref{eq:houv}) follows from the other two, but it is conceptually
useful to write down all three.

For simplicity we shall sometimes use abbreviated expressions such as
``fiber bundle \(E\)" or ``fiber bundle \(E\) over \(X\) with fiber
\(F\)" without specifying all of the data required by the definition.
It is to be understood however that such data is always present.

The transition maps can be construed as {\em gluing instructions\/}
\index{gluing instructions}%
 by
which the total space \(E\) is constructed by gluing together
the cartesian products
\(U_\alpha  \times F\).

In fact, let  
\[
\tilde E = \coprod_{\alpha \in A }U_\alpha \times
F=\bigcup_{\alpha \in A}U_\alpha \times F\times\{\alpha \}
\]
 be the
disjoint union and let \(\sim\) be the equivalence relation
generated by the equivalences 
\begin{equation}\label{eq:glue}
(x,f,\alpha) \sim (x,
h_{\alpha\beta}(x)(f),\beta)
\end{equation}%
 for all \(x \in
U_\alpha \cap U_\beta \) and all \(U_\alpha ,U_\beta  \) in \(\cU \)
with non-empty intersection.
Let
\(E=\tilde
E / \sim\) be the quotient space.
We now have:
\begin{theorem}\label{th:glue}
Let \(X\) and \(F\) be topological spaces, \(\cU \) a family
\((U_\alpha)_{\alpha \in A}\) of  open sets covering
\(X\), and \(h_{\alpha\beta}\) maps satisfying
(\ref{eq:ho}--\ref{eq:hocycle})
above, then the quotient space \(E\) constructed in the preceding paragraph
is a
fiber bundle satisfying the given data.
\end{theorem}%
The proof is utterly straightforward and is left to the reader.

Two very familiar examples of such gluing leads to the Moebius strip 
and to the tangent bundle of a manifold. 

\begin{example}\label{ex:moebius}
Consider the circle \(S^1\) as the set of unimodular complex numbers,
and let \(\phi(x)=e^{i\pi x}\).
Let \(U_1=\phi(-\frac{3}{4},\frac{3}{4})\) 
and \(U_2=\phi(\frac{1}{4},\frac{7}{4})\) be an
open cover. The intersection \(W=U_1\cap U_2\) consists of two components
\(W_1=\phi(-\frac{3}{4},-\frac{1}{4})\), and 
\(W_2=\phi(\frac{1}{4},\frac{3}{4})\). Let the fiber
\(F\) be the interval \([-1,1]\) and define the transition map
\(h_{21}:W\to \End(F)\) to be \(h_{21}(x)f=f\) for \(x \in W_1\) and
\(h_{21}(x)f=-f\) for \(x \in W_2\). 
\end{example}%
It is easy to see that the resulting space
\(E\) is the Moebius strip. This should be considered as the
prototypical example of gluing together 
a fiber bundle, the general one is seen to be
the result of gluing together many cartesian products ``twisted" by the
transition maps. If all the transition maps \(h_{\alpha\beta}(x)\) are
the identity then the construction of the previous theorem obviously
leads to \(E=X\times F\). 
\begin{example}\label{ex:tm}
Let \(M\) be a manifold of dimension \(n\),
and consider an atlas \(\cU\) given by a family
of open sets
\((U_\alpha)_{\alpha\in A}\), along with coordinate functions
\(x_\alpha^1,\dots,x_\alpha^n\) in each \(U_\alpha\). For \(x\in
U_\alpha\) a tangent vector is given by
\[
v=\sum_jv_\alpha^j\frac{\partial}{\partial x_\alpha^j}
\]
If \(x\in U_\alpha\cap U_\beta\) then also 
\[
v=\sum_jv_\beta^j\ppv{x_\beta}{j}
\]
where one has
\[
v_\beta^j = \sum_k \frac{\partial x_\beta^j}{\partial x_\alpha^k}
v_\alpha^k
\]
We can use this last formula to define the transition functions for a
bundle with fiber \(\lR^n\) and base space \(M\). For \(x\in
U_\alpha\cap U_\beta\), and \(f=(f^1,\dots,f^n)\in\lR^n\) define
\[
(h_{\beta\alpha}(x)f)^j=
\sum_k \frac{\partial x_\beta^j}{\partial x_\alpha^k}(x)f^k
\]
\end{example}%
Property (\ref{eq:houv}) follows from the 
inverse function  theorem and (\ref{eq:hocycle}) from the chain rule. 
The resulting bundle is the {\em tangent bundle\/}
\index{bundle!tangent}%
\(TM\) of the manifold, which, as a set, consists of all the tangent vectors
at all points of \(M\).
The local trivialization map \(h_\alpha:\pi^{-1}(U_\alpha)\to
U_\alpha\times\lR^n\) is, for a vector \(v\) at a point \(x\in U_\alpha\),
given by \(h_\alpha(v) = (x,(v_\alpha^1, \dots, v_\alpha^n))\). The
tangent bundle is a manifold of dimension \(2n\) 
with local coordinates in \(\pi^{-
1}(U_\alpha)\) being \((x_\alpha^1,\dots,x_\alpha^n,v_\alpha^1, \dots,
v_\alpha^n)\).

We now extend the notion of local trivializations beyond the meaning
that
occurs in the definition.
By a
{\em local trivialization\/}
\index{trivialization}%
in the extended sense  we now mean a pair \((W,h_W)\) consisting of
an open subset \(W \subset
X\) and a homeomorphism \(h_W:\pi^{-1}(W) \to W\times F\)
\index{\(H_W\)@\(h_W\)}%
such that for
any \(U_\alpha \) in \(\cU\) such that \(U_\alpha \cap W \neq \emptyset\),
the map \(h_W\circ
h_\alpha^{-1}: (U_\alpha \cap W) \times F \to (U_\alpha\cap W) \times F\)
has the form
\begin{equation}\label{eq:loctriv}
h_W\circ h_\alpha^{-1}(x,f) = (x, h_{W\alpha}(x)(f))
\end{equation}%
where
\(h_{W\alpha}:U_\alpha \cap W \to
\Aut(F)\).
 Each of the local
trivializations \(h_\alpha\) for \(U_\alpha \) in \(\cU\)
specified in the definition
of course continues being a local trivialization in this extended sense.
Note also that if \((W_1, h_1)\) and \((W_2, h_2)\) are two local
trivializations in the extended sense, with \(W_1\cap W_2 \neq
\emptyset\), then the transition
 map \(h_2\circ h_1^{-1}: 
(W_1\cap W_2)\times F  \to (W_1\cap W_2) \times F\) also has the same
form, to wit, \(h_2\circ h_1^{-1}(x,f) = (x, h_{21}(x)(f))\) where 
\(h_{21}:W_1 \cap W_2 \to \Aut(F)\). This means that if we add to the
family of defining local trivializations 
\(((U_\alpha, h_\alpha))_{\alpha\in A}\) any
family of local trivializations in the extended sense, we again have
data defining a fiber bundle with a larger set of local trivializations.
By definitions that follow shortly, this new bundle will be
equivalent to the original one. In what follows we shall conventionally
drop the adjective ``local" from the expression ``local trivialization",
and speak simply of a ``trivialization".
\index{trivialization}%

Given a bundle \(\pi: E \to X\) we write \(F_x = \pi^{-1}(\{x\})\) and
call \(F_x\) the {\em fiber over \(x\)\/}. 
\index{fiber!over \(x\)}%
\index{\(F_x\)}%
 Each \(F_x\) is homeomorphic
to the fiber \(F\) though not necessarily in any canonical way.

We shall in general deal with bundles defined over some fixed base set
\(X\). These form a category in which 
 the morphisms, also called {\em bundle maps\/},
\index{map!bundle}%
 are maps \(\phi:E_1 \to
E_2\) such that the following diagram commutes:
\begin{diagram}
E_1 & & \rTo^\phi & & E_2 \\
 & \rdTo_{\pi_1} & & \ldTo_{\pi_2} &         \\
 & & X & & \\
\end{diagram}%
Note that this definition implies that \(\phi\) maps fiber to fiber, 
\(\phi(F_{1x})\subset F_{2x}\). 
An isomorphism in this category is a map \(\phi\) as above
which is an isomorphism between the total spaces \(E_i\). 
Isomorphic bundles are said to be {\em equivalent\/}.
\index{bundle!equivalent}%

We shall generally
consider fiber-bundle theory as concerning bundles only up to
isomorphism.
With this notion we
first establish isomorphism of bundles that merely change the system of
 trivializations. Suppose we have 
the complete data  \((E,F,X,\pi,\cU,(h_\alpha)_{\alpha\in
A})\) of a fiber bundle as given in the definition. Let
now \(\cV\) denote another family \((V_\lambda)_{\lambda\in \Lambda}\) of
open sets covering
 \(X\)  and
\((k_\lambda)_{\lambda\in\Lambda}\)  a family of corresponding 
trivialization homeomorphisms.
It is easily seen
that  \((E,F,X,\pi,\cV,(k_\lambda)_{\lambda\in\Lambda})\) provides a
complete set of
data for a fiber bundle as given by the definition. It is also easily
seen that the identity map \(\Id_E:E\to E\) is a bundle isomorphism. By
this observation we can now free ourselves of the original defining set
of  trivializations  and pass on to any other set defining an
equivalent bundle. A particular case of this is to pass on to a {\em
refinement\/}
\index{refinement}%
 of \(\cU\), that is a cover \(\cV\) such that each
\(U_\alpha\) is a union of a subfamily of \(V_\lambda\) in \(\cV\).
Consider now the cover by the family \(V_{(\alpha,\lambda)}=V_\lambda\)
indexed by the subset of pairs
\((\alpha,\lambda)\) in \(A\times \Lambda\) such that
 \(V_\lambda\subset U_\alpha\). Define \(h_{(\alpha,\lambda)}\) to be
\(h_\alpha\) restricted to \(V_\lambda\). The new resulting covering family
of
open sets and  trivializations defines an isomorphic bundle.
With this in mind, if we now have a finite set of bundles over
the same base space, we can, by choosing a common refinement of all
of the covering families of  the individual bundles, consider that all
bundles have the same covering family of open sets over which the 
trivializations are defined.

Returning now to the notion of a bundle morphism,
let \(\cU\) be a covering
with respect to which both bundles trivialize. Since \(\phi\) maps fiber
to fiber one has the following commutative diagram  for \(U_\alpha\)
in \(\cU\)
\begin{diagram} \pi_1^{-1}(U_\alpha) &  \rTo^{h_\alpha^{(1)}} & 
U_\alpha\times F_1 \\
\dTo^\phi & & \dTo_{\,{\rm Id}\times\phi_\alpha} \\
\pi_2^{-1}(U_\alpha) & \rTo^{h_\alpha^{(2)}} & U_\alpha\times F_2 \\
\end{diagram}%

Here the right vertical arrow should be read as
\[
\Id\times \phi_\alpha: (x,f) \mapsto (x,\phi_\alpha(x)(f))
\]
 where
the maps
\(\phi_\alpha: U_\alpha\to  {\Hom}(F_1,F_2)\)
\index{\(\theta_\alpha\)@\(\phi_\alpha\)}%
{\em represent\/}
\index{represent}%
 \(\phi\) in
the  trivializations.

We now have \(h_\alpha^{(2)}\circ \phi= (\Id\times\phi_\alpha)\circ
h_\alpha^{(1)}\) from which 
\[
\phi= (h_\alpha^{(2)})^{-1}\circ(\Id
\times\phi_\alpha)\circ
h_\alpha^{(1)}
\]
 valid in \(\pi_1^{-1}(U_\alpha)\). One has a similar
expression for
\(\phi\) in \(\pi_1^{-1}(U_\beta)\) for another open set \(U_\beta\in\cU\).
If now
\(U_\alpha\cap U_\beta \neq\emptyset\) then one has
\[
(h_\alpha^{(2)})^{-1}\circ(\Id\times\phi_\alpha)\circ
h_\alpha^{(1)} = (h_\beta^{(2)})^{-1}\circ(\Id\times\phi_\beta)\circ
h_\beta^{(1)}
\]
valid in \(\pi_1^{-1}(U_\alpha\cap U_\beta)\). From this we deduce
\[
(\Id\times\phi_\alpha)\circ
h_\alpha^{(1)}\circ (h_\beta^{(1)})^{-1}  = h_\alpha^{(2)}\circ
(h_\beta^{(2)})^{-1}\circ(\Id\times\phi_\beta)
\]
By (\ref{eq:trans}) this can now be easily 
expressed in terms of the transition maps
\begin{equation} \label{eq:fibmorx}
h^{(2)}_{\alpha\beta}(x)(\phi_\beta(x)(f)) = \phi_\alpha(x)(
h^{(1)}_{\alpha\beta}(x)(f))
\end{equation}%
for all \(x\in U_\alpha\cap U_\beta\) and \(f\in F_1\).
One should note the resemblance of formula (\ref{eq:fibmorx}) to that of the
morphism of
group actions (\ref{eq:moract}) and the intertwiner of group
representations (\ref{eq:intertw}).

Reciprocally, we have the following theorem whose proof is utterly
straightforward:
\begin{theorem}\label{th:fiblocmor}
Let \(E_1\) and \(E_2\) be two fiber bundles with fibers
\(F_1\) and \(F_2\) over the same base space \(X\). Given a family 
\((U_\alpha)_{\alpha\in A}\) of open sets covering \(X\) over which both
bundles trivialize, and given maps \(\phi_\alpha: U_\alpha \to
\Hom(F_1,F_2)\)
satisfying (\ref{eq:fibmorx}) then there is a unique bundle morphism
\(\phi:E_1\to E_2\) for which the \(\phi_\alpha\) are the local
representatives. 
\end{theorem}%

We shall now adopt a convention by which we 
abbreviate (\ref{eq:fibmorx}) to
\begin{equation}\label{eq:fibmor}
h^{(2)}_{\alpha\beta}\circ\phi_\beta = \phi_\alpha \circ
h^{(1)}_{\alpha\beta}
\end{equation}%
In this convention, maps such as the ones appearing in (\ref{eq:fibmor})
are considered as {\em parameterized maps\/},
\index{map!parameterized}%
 that is a family of maps
indexed by elements of some set (\(U_\alpha\cap U_\beta\) in this
particular case) and equation such as (\ref{eq:fibmor}) is supposed to
hold at each point of the indexing set. This translates to
(\ref{eq:fibmorx}). Details of this convention are to found in 
Appendix \ref{sec:parmaps}. This convention will be generally in 
force from now on.

In the particular case that \(\phi\) is an isomorphism then \(F_1\) and
\(F_2\) are isomorphic and we can identify the two and denote each by
\(F\). Each \(\phi_\alpha\) is invertible, belonging to \(\Aut(F)\). In
this
case we can now write
\begin{equation}\label{eq:fibeq}
h^{(2)}_{\alpha\beta}= \phi_\alpha\circ
h^{(1)}_{\alpha\beta}\circ\phi_\beta^{-1}
\end{equation}%

We say a fiber bundle is {\em trivial\/}
\index{bundle!trivial}%
 if it is equivalent to the {\em
product bundle\/}
\index{bundle!product}%
 \(X\times F\) with one single defining 
trivialization
\(h_X = \Id_{X\times F}\).
We see from (\ref{eq:fibeq}) that a bundle is
trivial if and only if its transition maps can be written as
\[
h_{\alpha\beta} =
\phi_\alpha\circ\phi_\beta^{-1}
\]
 for a family of maps  \(\phi_\alpha:
U_\alpha \to \Aut(F)\).
\begin{example}\label{ex:trivnottriv}
Consider the trivial bundle \(E=[0,1]\times\lR\) with base space
\([0,1]\),
fiber \(\lR\), and the single identity map \(\Id:E\to E\) 
as the defining family of  trivializations.
\end{example}%

We shall find it instructive to consider the maps \(h_i,\,i=1,2,3\) 
from \(E\) to \(E\) given by
\begin{eqnarray*}
h_1(x,f) &=& (x, (1+x^2)f) \\ 
h_2(x,f) &=& (x, 2x+(1+x^2)f) \\ 
h_3(x,f) &=& (x, (1+x)f + f^3)
\end{eqnarray*}%
each of these is a bundle isomorphism. This example call attention to
the fact that though a bundle may be isomorphic to a product bundle, one
should not think that certain relations that are natural to the
cartesian product carry over to a trivial bundle. Thus while it is
natural to think of the subset \([0,1]\times\{1\}\subset
[0,1]\times\lR\) as ``horizontal" being of the form \(\pi_{\lRe}^{-1}
(\{1\})\) where \(\pi_{\lRe}:[0,1]\times\lR\to \lR\) is the canonical
projection, in a trivial bundle there is in general 
no canonical projection on the
fiber.  Thus the image of the same set by the \(h_i\) are in no way
``horizontal". Metaphorically speaking, a trivial bundle is a cartesian
product which lost one of its canonical projections.
Though a trivial bundle has a  trivialization over the 
full base space, which could be called a ``global" trivialization,
there is in general no one canonical such.

\begin{example}
Let \(G\) be a Lie group. The tangent bundle \(TG\) is trivial, a
global trivialization \(TG \to G \times \gg\) is given by the map
that associates to \(v\in T_gG\) the pair \((g, v\cdot g^{-1})\).
Associating the pair \((g, g^{-1}\cdot v)\) defines a second
trivialization.
\end{example}

By a {\em local section\/} 
\index{section!local}%
 of a bundle we mean an open set \(U\subset X\)
and a map \(\sigma: U \to E\) such that \(\pi\circ\sigma (x) = x\) for
all \(x\in U\).  We shall denote by \(\Gamma(U)\)
\index{\(\chi(U)\)@\(\Gamma(U)\)}%
 the set of local
sections over U. A {\em global section\/}
\index{section!global}%
 is a section in which \(U=X\).
A bundle may not have any global sections. A simple obvious 
example of this is
the double cover \(S^1\to S^1\) which can be realized as the map
\(z\mapsto z^2\) of  unimodular complex numbers. 
This is a
bundle with fiber being a two point set.

Let \(\sigma:U \to E\) be a local section of a bundle and consider a
trivialization over an open set \(W\subset U\). The map \(h_W\circ
\sigma:W\to W\times F\) then has the form 
\begin{equation}\label{eq:locsec}
h_W\circ\sigma(x) = (x, s_W(x))
\end{equation}%
 The map \(s_W:W \to F\) is called the {\em representative of
\(\sigma\)\/} 
\index{section!representative}%
in the given  trivialization. Local sections are therefore
generalizations of (partial) maps of \(X\) to \(F\).
Local sections are also often know as {\em fields\/}.
\index{field}%
The value of a field at \(x\in U\) being \(\sigma(x)\in F_x\). Familiar 
examples are, for instance, local
sections of \(TM\) known as {\em vector fields\/} or of \(T^*M\),  known
as {\em covector fields\/} or  {\em \(1\)-forms\/}. 
Note that the value of a field in general lies in
a space, the fiber \(F_x\) over the  point \(x\), that varies 
with the point. 
In a trivialization 
though,
the representative of the section takes its value in a fixed space, the
fiber \(F\). Introducing local coordinates \(y^1,\dots,y^m\) in \(F\), 
the representative  can
then be described by a set of {\em components\/} 
\(s_W(x)= (s^1_W(x),\dots,s^m_W(x))\). In the sequel we shall usually
drop the adjective ``local" and speak simply of a section. Global
sections will be referred to as such. 

We shall in the sequel often have to deal with how mathematical
structures related to fiber bundles appear, or as is usually said,
are {\em represented\/} in  trivializations. For
a  trivialization \(h_W\) in an open set \(W\), such local
representatives
\index{representative}%
 should in principle carry some indication that
identifies the  trivialization (such as the index \(W\) on the map
\(s_W\) in (\ref{eq:locsec}) that represents a local section \(\sigma\)). 
The systematic use
of such indices would generally overburden the notation and we shall
often drop them whenever one is dealing only with one  
trivialization.
We shall of course be forced to introduce them any time we must compare
local representatives in two  trivializations over intersecting open
sets. Thus for local representatives for \(\sigma\) 
 one has
\begin{equation}\label{eq:tranf}
s_V(x) = h_{VW}(x)(s_W(x))
\end{equation}
 Such {\em transition formulas\/}
\index{transition formula}%
 will be common in what follows. It should also be pointed out that
although a  trivialization consists of a {\em pair\/}, an 
open set \(W\)
and a homeomorphism \(h_W:\pi^{-1}(W)\to U\times F\), it is often customary
to use only the open set \(W\) as a label. This is somewhat awkward
as it is absolutely legitimate to consider two different 
trivializations over the same open set \(W\), that is, to consider two
different homeomorphisms. This should be kept in mind in understanding 
transition formulas such as (\ref{eq:tranf}), and  
interpret it also 
as expressing a relation for two 
trivializing homeomorphism over the same set.

\subsection{\protect\(G\protect\)-bundles}
Let \(E\) be a fiber bundle with base space \(X\) and fiber \(F\). 
Let \(G\) be a group and suppose we have a
(usually taken to be effective) left action of
\(G\) on \(F\)  and let \(\rho:G \to \Aut(F)\) the the corresponding
group homomorphism. Suppose \(h_{\alpha\beta}(x) \in
\rho(G)\) for all \(x\in U_\alpha\cap U_\beta\)
and all \((\alpha,\beta)\) with \(U_\alpha\cap U_\beta\neq\emptyset\), 
then we say that \(G\) is the {\em structure group\/}
\index{structure group}%
of the bundle.  This expression is a bit abusive
as the structure group is not uniquely defined, and a bundle has in
general many structure groups. One can of course always consider
\(\Aut(F)\) as the structure group.  Just to what
extent a structure group of a bundle is to be considered as part of its
structure depends mostly on the application one has in mind. It is often
useful to consider bundles with a  given structure group \(G\) in mind, in
which case one speaks of a  {\em \(G\)-bundle\/}.
\index{gbundle@\(G\)-bundle}%
This is particularly true  in gauge theory.
 
To be more precise,  a \(G\)-bundle with a given left action of
\(G\) on \(F\) is one where the transition maps have the form 
\(h_{\alpha\beta}(x)(f) = g_{\alpha\beta}(x)\cdot f\) and where the maps
\(g_{\alpha\beta}:U\cap V \to
G\)
\index{\(G_{\alpha\beta}\)@\(g_{\alpha\beta}\)}%
are required to satisfy

\begin{eqnarray}   \label{eq:co}
g_{\alpha\alpha}(x) &=& e  \\ \label{eq:couv}
g_{\beta\alpha}(x) &=&  g_{\alpha\beta}(x)^{-1}\\ \label{eq:cocycle}
g_{\alpha\beta}(x)g_{\beta\gamma}(x)g_{\gamma\alpha}(x) &=& e.
\end{eqnarray}%
These conditions automatically follow from properties
(\ref{eq:ho}-\ref{eq:hocycle}) of transition maps if the action is
effective.
Condition (\ref{eq:cocycle}) above is know as the (\v{C}ech)
\index{cech@\v{C}ech}%
 {\em
\(1\)-cocycle}
\index{\(1\)-cocycle}%
condition. By abuse of language, we shall also refer to the
maps \(g_{\alpha\beta}\) as {\em transition maps\/},
\index{map!transition}%
 or when convenient
to specify the group, as {\em \(G\)-transition maps\/}.
\index{map!gtransition@\(G\)-transition}%

When  dealing with a \(G\)-bundle, we can relativize some of the notions
defined above to reflect the fixed structure group.
Thus by a  trivialization of a \(G\)-bundle
in an open set \(W\) we shall now
mean one in which the maps \(h_{W\alpha}\) in equation 
(\ref{eq:loctriv})  are of the
form \(h_{W\alpha}(x)(f) = g_{W\alpha}(x)\cdot f\) for a map
\(g_{W\alpha}:U_\alpha\cap W \to G\).
\index{\(G_{W\alpha}\)@\(g_{W\alpha}\)}%
Likewise an isomorphism of two
\(G\)-bundles with the same base-space and same
fiber will be given by maps \(\phi_\alpha\) appearing in
(\ref{eq:fibeq}) of the form \(\phi_\alpha(x)(f) = g_\alpha(x)\cdot f\)
where
\(g_\alpha:U_\alpha\to G\). 
\index{\(G_\alpha\)@\(g_\alpha\)}%
Equation (\ref{eq:fibeq}) is now to be
read in terms of group multiplication as
\begin{equation}\label{eq:gfibmor}
g^{(2)}_{\alpha\beta}= g_\alpha
g^{(1)}_{\alpha\beta}g_\beta^{-1}
\end{equation}%

 In \v{C}ech cohomology, a
set of maps \(g_\alpha:U_\alpha\to G\) is called a {\em \(0\)-cochain
 with values in
\(G\)\/}.
\index{\(0\)-cochain}%
Given such a cochain, the maps
\(g_\alpha g_\beta^{-1}\) are easily seen to satisfy conditions for
being
the \(G\)-transition maps of a \(G\)-bundle, in particular they define a
\(1\)-cocycle. This cocycle is called the {\em coboundary\/}
\index{coboundary}%
 of the
cochain. Thus a \(G\)-bundle is trivial if and only if its cocycle is a
coboundary. Similarly we can say that two bundles are equivalent if the
cocycles are intertwined by a cochain.

Under the viewpoint of considering the structure group as part and
parcel of fiber-bundle structure,
isomorphism of bundles may now depend on the
structure group chosen. The largest isomorphism classes of course
correspond to \(G=\Aut(F)\). In Example \ref{ex:trivnottriv} if we take the
structure group  \(G\) to be \(\GL(1)\) then \(h_1\) is a \(G\)-bundle
isomorphism while \(h_2\) and \(h_3\) are not. Expanding \(G\) to 
\(\GA(1)\),  makes \(h_2\) a \(G\)-bundle
isomorphism, while \(h_3\) continues not being, and expanding \(G\) to
\(\hbox{Diff}(\lR)\), the diffeomorphism group of \(\lR\), 
 now includes \(h_3\).

\subsection{Structured  Fibers}
Very often the fiber \(F\) of a bundle has additional structure that is
transferred to each individual fiber \(F_x\). This
happens when each of the transition maps \(h_{\alpha\beta}\) is an
isomorphism of the structure in \(F\). In this section  we shall be
concerned only with certain algebraic structures, such as
 vector space and algebra. Other structures will appear in subsequent
sections. Whenever \(F\) is a vector space and each transition map
\(h_{\alpha\beta}\) is a linear isomorphism then each fiber \(F_x\) is
also canonically a vector space. This is because the equivalence classes
defined by the gluing instructions (see Theorem \ref{th:glue} and 
(\ref{eq:glue})) are
compatible with the linear operations in \(F\),
that is, given that
\((x,f_1,\alpha) \sim (x, h_{\alpha\beta}(x)f_1,\beta)\) and
\((x,f_2,\alpha) \sim (x ,h_{\alpha\beta}(x)f_2,\beta)\) and that \(a_1\)
and \(a_2\) are elements of the base field, then
\[
(x,a_1f_1+a_2f_2,\alpha) \sim (x,
a_1h_{\alpha\beta}(x)f_1+a_2h_{\alpha\beta}(x)f_2,\beta)
\]
so that in \(F_x \) one can
define
\[
a_1[(x,f_1,\alpha)]+a_2[(x,f_2,\alpha)]=[(x,
a_1f_1+a_2f_2,\alpha)]
\]
which makes \(F_x \) canonically into a vector
space. A bundle with this structure is called a {\em vector
bundle\/}. 
\index{bundle!vector}%
Note that the {\em zero section\/}
\index{section!zero}%
 which takes each \(x\in
X\) to the zero element of \(F_x\) is a well-defined global section.
This global section, by abuse of notation, is usually denoted simply by
\(0\). A vector \(G\)-bundle arises whenever the transition maps are of
the form \(h_{\alpha\beta} = R(g_{\alpha\beta})\) for a representation
\(R\) of \(G\) and \(G\)-transition maps \(g_{\alpha\beta}\).

A slightly weaker notion is that of an {\em affine bundle\/}, 
\index{bundle!affine}%
which
arises whenever \(F\) is a vector space but the transition maps
\(h_{\alpha\beta}\) are affine isomorphisms. One now has
\(h_{\alpha\beta}f =
B_{\alpha\beta}f+ c_{\alpha\beta}\) where the \(B_{\alpha\beta}\) are
linear isomorphism of \(F\) and \(c_{\alpha\beta}\in F\). In this case
the \(B_{\alpha\beta}\) are linear transition maps and the
\(c_{\alpha\beta}\) satisfy
\(B_{\alpha\beta}c_{\beta\gamma}+c_{\alpha\beta}=c_{\alpha\gamma}\),
whenever the open sets corresponding to the indices have a 
non-empty intersection. Note that in this process, part of the structure
of \(F\) is lost when going to the fibers \(F_x\). Due to the presence
of the terms \(c_{\alpha\beta}\), the linear structure is weakened
as there is no way to canonically identify the zero
element of \(F_x\), but linear relations between differences of elements
of \(F_x\) continue to be well defined. Vector
bundles are obviously special case of affine bundles, those for which
\(c_{\alpha\beta}=0\) for all \((\alpha,\beta)\).
 An affine \(G\)-bundle would be one in which
\(h_{\alpha\beta}=H(g_{\alpha\beta})\) for an affine representation
 \(H\) of 
 \(G\) and \(G\)-transition maps \(g_{\alpha\beta}\). 

Finally one has an {\em algebra bundle\/},
\index{bundle!algebra}%
 or, a {\em bundle of
algebras\/},
whenever the fiber is an algebra \(\cA\) and each \(h_{\alpha\beta}\) is
an \(\cA\)-isomorphism.  In particular an algebra bundle is a vector
bundle. An algebra \(G\)-bundle arises whenever there is a group
homomorphism \(\rho:G\to \Aut(\cA)\) and one has 
\(h_{\alpha\beta} =\rho(g_{\alpha\beta})\) for 
\(G\)-transition maps \(g_{\alpha\beta}\).

Obviously the above type of constructions can be
extended to a wide variety of algebraic structure.

Let \(E\) be a vector bundle over \(X\) and \(U\in X\) an open set.
Let \(\cF(U)\)
\index{\(F(U)\)@\(\cF(U)\)}%
be the set of maps \(U\to \lF\) of \(U\) to the base field
\(\lF\)
of \(F\). \(\cF(U)\)  is naturally an \(\lF\)-algebra under pointwise
operations.

The
set of local sections \(\Gamma(U)\) is obviously a vector space under
pointwise linear operations. One also defines a pointwise product of
elements \(f\in\cF(U)\) and \(\sigma\in\Gamma(U)\) by
 \((f\sigma)(x)=f(x)\sigma(x)\). 
If \(E\)
is an algebra bundle, then
\(\Gamma(U)\) is in addition an algebra under pointwise multiplication.

\subsection{Principal Bundles}
Let \(G\) be a topological group. 
A very special kind of bundle with structured
fibers is a {\em principal
\(G\)-bundle\/}
\index{bundle!principal}%
 in which each fiber is a homogeneous space with free
transitive action of G.
Let there be given a topological space   \(X\), a family \(\cU\) of  
open sets
that cover
\(X\), and  a family of \(G\)-transition maps
subordinate to
\(\cU\).  The group \(G\) acts on itself by left
multiplication and so we can now use this action to construct, by gluing, 
a
bundle with fiber \(G\) and structure group \(G\) using the given
transition maps. This bundle is called a {\em principal \(G\)-bundle\/}
normally denoted by \(PG\).
\index{\(PG\)}%
  For a given base space the principal
\(G\)-bundles are not necessarily unique as they depend on the chosen
cocycle and there are in general many inequivalent ones.

\begin{example} \label{ex:pz2}  There are two
inequivalent principal \(\lZ_2\)-bundles over \(S^1\). One has
\(E=S^1\coprod S^1\) with \(\pi\) being the identity on each copy.
The other has \(E=S^1\) as the double cover over \(S^1\). 
Identifying \(S^1 \) with the unimodular complex numbers, 
\(\pi\) is the map \(z\mapsto z^2\).
\end{example}%

A group acts upon itself also by multiplication on the right and this
commutes with the left action. This right action thus is compatible with
the gluing procedure dictated by the transition maps, that is the relation
\((x,g,\beta) \sim (x,g_{\beta\alpha}(x)g,\alpha)\) is invariant
under right multiplication of \(g\) by \linebreak \(h\in G\).
Using this, there is a natural {\em right\/} action of \(G\) on \(PG\).
Under this action each fiber is a single orbit and the stability
subgroup of any point is the trivial subgroup \(\{e\}\).  Thus \(G\)
acts transitively and freely on each fiber. 

Let now \(F\) be any topological space and \(\rho\) a left action of
\(G\) on \(F\). Consider the space \(PG \times F\). This is easily seen
to be a bundle with projection \linebreak
\(\hat\pi:PG\times F \to X\) given by
\(\hat\pi(p,f) = \pi(p)\), with fiber \(G\times F\), and the same cocycle
as that of
\(PG\) using the left action \((g,(h,f))\mapsto (gh, f)\) of \(G\) on
\(G\times F\). Define now on \(PG \times F\) another left action
\(\gamma(g, (p,f)) =g\cdot(p,f) =(p\cdot g^{-1},g\cdot f)\). 
Denote by \(PG \times_\rho F\)
\index{\(PG \times_\rho F\)}%
the space of orbits by this action and by \(\pi_\rho: PG \times F \to
PG \times_\rho F\) the canonical map to the quotient. It is easy to see
that \(\hat\pi\) factors trough \(\pi_\rho\) and we have a map
\(\tilde\pi:PG \times_\rho F\to X\).

\begin{theorem}\label{th:assocb}
\(PG \times_\rho F\) is a \(G\)-bundle isomorphic to the bundle with
fiber \(F\),  projection \(\tilde\pi\), group action \(\rho\), and the
cocycle of \(PG\).
\end{theorem}%
\begin{proof}
Consider the action map \(\rho: G\times F \to F\) taking \((g,f)\) into
\(g\cdot f\). Consider also on \(G\times F\) the action 
\( \eta(h,(g,f)) =h\cdot(g,f) =
(gh^{-1}, h\cdot f)\) and let \(Q\) be the space of orbits under this
action endowed with the quotient topology. Now \(\rho (gh^{-1}, h\cdot f) =
gh^{-1}\cdot(h\cdot f) = g\cdot f\) so \(\rho\) is constant
on the orbits and  thus factors through \(Q\):
\begin{diagram}G\times F & \rTo^\rho & F \\
\dTo_{\pi_\eta} &\ruTo_{\sigma}  & \\
Q & & \\
\end{diagram}%
where \(\pi_\eta\) is the canonical map to the quotient.

Since \(\rho\)
maps \((e,f)\) to \(f\), it is surjective, and thus so is
\(\sigma\).  Suppose now that \(g_1\cdot f_1 = g_2 \cdot f_2\). One
has \( (g_2^{-1}g_1)\cdot(g_1,f_1) = (g_1(g_2^{-1}g_1)^{-1},
(g_2^{-1}g_1)\cdot f_1) = (g_1g_1^{-1}g_2, g_2^{-1}\cdot(g_2\cdot f))
=(g_2, f_2)\). So \((g_1,f_1)\) and
\((g_2,f_2)\) are on the same orbit. Since one has
\(\sigma(\cO_{(g,f)})=g\cdot f\) this
result means that  \(\sigma\) is also
injective and hence bijective.  The continuity of \(\sigma\) follows
from the universal property of the quotient topology. We now prove it is
open. A subset \(A\subset Q\) is open if and only if the union of orbits
\(W=\pi_\eta^{-1}(A)\)
is an open subset of  \(G\times F\). If now \((g,f)\in W\) then
\((e,g^{-1}\cdot f)\in W \cap (\{e\}\times F) = \{e\} \times W_e\) where
\(W_e\) is an open set in \(F\). It is now clear that 
\[
W=\bigcup_{f\in
W_e}\cO_{(e,f)}
\]
 and that \(\sigma (A) = W_e\) so
that indeed \(\sigma\) is open and thus a homeomorphism. Using this
fact, we transfer, by (\ref{eq:tranact}), the action \(\rho\) on \(F\) 
to an equivalent action 
\(\rho_Q\) on
\(Q\).

We shall now show that \(PG\times_\rho F\) is a \(G\)-bundle with fiber
\(Q\), with the group action  \(\rho_Q\), and the cocycle of \(PG\).
Since \(Q\) and \(F\) are homeomorphic and \(\rho\) and \(\rho_Q\)
equivalent, this will prove the theorem.

Let \(h_\alpha\) be a defining local trivializing homeomorphism of
\(PG\), then one has the map
\[
h_\alpha\times{\rm Id}:\hat\pi^{-1}(U_\alpha) 
\to U_\alpha\times G \times F\\
\]
which gives a trivialization of \(PG\times F\). This map is an 
isomorphism of
actions \(\gamma\) and \(\Id\times \eta\) so one can pass to quotients
and get a homeomorphism 
\[
\tilde h_\alpha: \pi_\rho^{-1}(U_\alpha) \to U_\alpha \times Q
\]
which we take as a defining local trivialization.
One now has the following diagram:
\begin{diagram}\hat\pi^{-1}(U_\alpha\cap U_\beta) & 
\rTo^{h_\alpha\times{\rm Id}} & U_\alpha\cap U_\beta\times G \times
F & \rTo^{{\rm Id}\times\pi_\eta} & U_\alpha\cap U_\beta\times Q\\
\dTo_{\,{\rm Id}} & & \dTo_{\,{\rm Id}\times g_{\beta\alpha}\times 
{\rm Id}} &  &\\
\hat\pi^{-1}(U_\alpha\cap U_\beta) & 
\rTo^{h_\beta\times{\rm Id}} & U_\alpha\cap U_\beta\times G \times
F & \rTo^{{\rm Id}\times\pi_\eta} & U_\alpha\cap U_\beta\times Q\\
\end{diagram}%

 Note that  \(\rho(g_{\beta\alpha}(x)g, f) =
(g_{\beta\alpha}(x)g)\cdot f =
g_{\beta\alpha}(x)\cdot \rho(g,f)\) which,  composing with
\(\sigma^{-1}\), implies
\(\pi_\eta(g_{\beta\alpha}(x)g, f)=
g_{\beta\alpha}(x)\cdot\pi_\eta(g,f)\).
Thus the above diagram can be
completed with
a rightmost downward vertical arrow \((x,q) \mapsto (x,
g_{\beta\alpha}(x)\cdot q)\), establishing the proper transition map.
This
completes the proof.
\end{proof}%

With this construction we can now form all the bundles based on a given
cocycle in a global fashion starting with the  principal
bundle \(PG\). Such bundles are said to be {\em associated\/} 
to the principal
bundle.
\index{bundle!associated}%
We can also in this way extend certain constructions on the
principal bundle to all its associated bundles in a systematic fashion.

\begin{example}
Let \(M\) be an \(n\)-dimensional manifold and \(\cU\) and atlas. Let
\(U,V \in \cU\) with \(U\cap V\neq \emptyset\) and let \(x^1,\dots,x^n\)
and
\(y^1,\dots,y^n\) be local coordinates in \(U\) and \(V\) respectively.
One then has for \(x\in U \cap V\) the Jacobian matrix
\begin{equation}\label{eq:jacocy}
J^i{}_j(x) = \frac{\partial y^i}{\partial x^j}(x) \in \GL(n)
\end{equation}
As was seen in Example \ref{ex:tm}, these matrices satisfy the 
conditions for being
the transition maps of a bundle. The principal \(\GL(n)\)-bundle
\(\cF(M)\)
\index{\(F(M)\)@\(\cF(M)\)}%
 formed
from this cocycle is called the {\em frame bundle\/}
\index{bundle!frame}%
 of \(M\).
\end{example}%

The reason for the name {\em frame bundle\/}
\index{bundle!frame}%
 is that one can identify
the elements of each fiber \(F_x\) of the bundle with ordered bases
\(e_1,\dots,e_n\) of \(T_xM\).  If we use the coefficients
\(\epsilon^i{}_j\) in the local expansion 
\[
e_i = \sum_k
\epsilon^k{}_i\ppv{x}{k}
\]
 as local coordinates of the ordered base
 at \(x\), then one has 
\[e_i =
\sum_k\sum_\ell J^k{}_\ell\epsilon^\ell{}_i  \ppv{y}{k}
\] 
Now \(\epsilon^i{}_j \in \GL(n)\) and
\(\epsilon^i{}_j  \mapsto \sum_\ell J^k{}_\ell\epsilon^\ell{}_i  \) is a
left action of \(J\in \GL(n)\) on \linebreak
\(\epsilon \in \GL(n)\) by ordinary
matrix multiplication.  From this one sees that indeed the set of frames
is a principal \(\GL(n)\)-bundle with cocycle \(J\). One has the right
action of \(\GL(n)\) on the bundle of frames by having \(A^i{}_j\in \GL(n)
\) transform the frame \(e_1,\dots,e_n\) to the frame \(f_1,\dots,f_n\)
by \(f_j = \sum_i e_iA^i{}_j\).  Note that we were able to define the
right action without recourse to local coordinates.

Consider now the actions
by the two natural representations 
\(\rho(A,z) = Az \) and \(\rho^*(A,z) = (A^t)^{-1}z\)
of \(\GL(n)\) on \(\lR^n\).  
One has the identifications:
\begin{eqnarray*}
\cF(M)\times_\rho {\lR}^n  &\simeq &  TM \\
\cF(M)\times_{\rho^*} {\lR}^n &\simeq &  T^*M
\end{eqnarray*}%

For a bundle to be associated to a principal bundle is not very special.
All that is necessary, by Theorem \ref{th:assocb}, is to interpret the
transition maps as due to the action of a topological group. We shall
only need this for vector bundles \(E\)  with
a finite dimensional fiber \(F\).  One has \(h_{\alpha\beta}\in \GL(F)\)
and one can construct the corresponding principal bundle \(P\GL(F)\).
The action \(\rho:\GL(F)\times F\to F\) is the natural one
\((T,f)\to Tf\). As a corollary of Theorem \ref{th:assocb} one has
\begin{theorem}\label{th:eispgl}
\[
P\GL(F)\times_\rho F\simeq E
\] 
\end{theorem}

\subsection{Bundle Operations}\label{sec:bundleops}
Consider a family of  fiber bundles \((E_\lambda)_{\lambda\in\Lambda}\)
over the same base space \(X\) and
with fibers \(F_\lambda\). It is often possible to extend an operation
that produces a new
space \(F\) from the \(F_\lambda\) to act fiber-wise on the spaces
 \(E_\lambda\)  to produce a new fiber bundle \(E\) with fiber
\(F\). This is the case whenever one can trivialize all the
\(E_\lambda\)  over the open sets of the same covering family and then
 combine the transition maps \(h^\lambda_{\alpha\beta}\)
of each bundle \(E_\lambda\) to  transition maps \(h_{\alpha\beta}\) for
the space \(F\). The first step is always possible if the family is
finite, and we shall only deal with this case in these notes. So assume
there is a fixed family of open sets \((U_\alpha)_{\alpha\in A}\)
covering \(X\) and let \(h^\lambda_\alpha\) and
\(h^\lambda_{\alpha\beta}\) be respectively the  trivialization
and the transition maps of the bundles \(E_\lambda\).

A simple example of such a construction is the cartesian product
\(F={\prod}_\lambda F_\lambda\). Define a new bundle \(E\)
with transition maps
\(h_{\alpha\beta}={\prod}_\lambda
h^\lambda_{\alpha\beta}\). This bundle is called the {\em product
bundle\/}
\index{bundle!product}%
 and one writes \(E={\prod}_\lambda E_\lambda\). Note that this 
notation is misleading as the total space \(E\) is {\em not\/} 
the cartesian 
product of the total spaces \(E_\lambda\). The product bundle 
is the product 
object in the category of bundles over a fixed base space. The canonical 
categorical projections \linebreak \(p_\mu : E \to E_\mu\) are 
bundle maps whose local 
representatives \(U_\alpha \times \prod_\lambda F_\lambda \to 
U_\alpha\times 
F_\mu\) are \((x,f)\mapsto (x,f_\mu)\). Theorem \ref{th:fiblocmor} then
provides a bundle map.
A special case of this construction 
is when each \(E_\lambda\) is a principal \(G_\lambda\)-bundle, 
\(E_\lambda=PG_\lambda\). It is easy to see that \(E={\prod}_\lambda 
E_\lambda\) is a principal \(\prod_\lambda G_\lambda\) bundle. 

For vector bundles with the fibers being vector spaces over the same base
field  one can form the direct sum, also known as the
Whitney sum, and the tensor product of bundles. The direct sum has fiber
\(F=\bigoplus_\lambda F_\lambda\) with transition maps \(h_{\alpha\beta}=
{\bigoplus}_\lambda
h^\lambda_{\alpha\beta}\) and the tensor product has fiber
\(F=\bigotimes_\lambda F_\lambda\) with transition maps \(h_{\alpha\beta}=
{\bigotimes}_\lambda
h^\lambda_{\alpha\beta}\). The corresponding bundles are denoted by
\(\bigoplus_\lambda E_\lambda\) and
\(\bigotimes_\lambda E_\lambda\) respectively. Notations such as
\(E_1\oplus E_2\), or \(E_1\otimes E_2\otimes E_3\) are also used and
are self-explanatory. As a topological fiber bundle,  
\(\bigoplus_\lambda E_\lambda\) 
coincides with \(\prod_\lambda E_\lambda\). It is the algebraic structure 
that justifies a different notation. For a fixed vector bundle \(E\) one
can also form the exterior powers \(\bigwedge^p(E)\) defined by the
transition maps \(\bigwedge^p(h_{\alpha\beta})\).

Another construct for vector bundles \(E_1\) and \(E_2\) is to consider
\(F=\Hom (F_1,F_2)\), the space of linear maps \(F_1\to F_2\). Define
for \(\phi\in\Hom (F_1,F_2)\) the transition formula
\(h_{\alpha\beta}\phi = 
h^2_{\alpha\beta}\circ\phi\circ h^1_{\alpha\beta}{}^{-
1}\). These are the transition maps for the vector bundle
\(\Hom(E_1,E_2)\) of linear homomorphisms.

Consider now
\(F'\), the dual of  \(F\).
For \(\phi\in F'\) define
\begin{equation}\label{eq:trandual}
h_{\alpha\beta}'\phi = \phi\circ h_{\alpha\beta}{}^{-1}
\end{equation}%
These are the
transition maps of the bundle \(E'\), the dual bundle. This is a
particular case of \(\Hom(E_1,E_2)\) where for \(E_2\) we take the
trivial bundle \(X\times \lF\) where \(\lF\) is the base field of the
fiber.

For the case of \(G\)-bundles a simplification occurs whenever all the
bundles use the same \(G\)-transition maps and only differ in the
action of \(G\) defined on each fiber \(F_\lambda\).
Provided these actions can
be combined to an action on \(F\), the resulting bundle continues to be
a \(G\)-bundle for the same group. Thus for vector bundles, one can form
direct sums and tensor products of group representations and so the
Whitney sum and the tensor product of vector \(G\)-bundles can again
be considered as being
\(G\)-bundles. The same is true of the bundle of linear homomorphisms,
the dual bundle, and the exterior powers.

Starting with the tangent bundle \(TM\), the dual bundle is the
cotangent bundle \(T^*M\). Tensor products of these two bundles leads to
the usual tensor bundles. The exterior powers \(\bigwedge^p(T^*M)\) are
the familiar bundles of exterior \(p\)-covectors whose local sections
are \(p\)-forms. All these bundles can be construed as \(\GL(n)\)-bundles. 
In Section \ref{sec:pseudoriemannian} we shall see how this group
may be reduced. 

If \(E_i\) for \(i=1,\dots,n\) are vector bundles with fibers \(F_i\),
and \(V\) is a vector bundle with fiber \(W\), all defined over the same
base field, then a  bundle map \(\alpha:E_1\times\cdots\times 
E_n \to V\) is
said to be {\em multilinear\/}
\index{map!bundle!multilinear}%
or {\em \(n\)-linear\/}
\index{map!bundle!nlinear@\(n\)-linear}%
if each fiberwise restriction \(F_{1x}\times\cdots F_{nx}\to W_x\)
of \(\alpha\) 
is an \(n\)-linear map. Such bundle maps obviously form a vector space.

If all the \(E_i\) are the same bundle \(E\) with fiber \(F\), then the
natural  \(S_n\) (permutation group) action  on the
\(n\)-fold cartesian product \(F\times\cdots\times F\) gives rise,
by Theorem \ref{th:fiblocmor}, 
 to an
action on \(E\times\cdots\times E\). We say an \(n\)-linear map 
\(\alpha:E\times\cdots\times E \to V\) is {\em  antisymmetric\/}
\index{map!bundle!antisymmetric}%
if \(\alpha(\pi\cdot p) = \sigma_\pi \alpha(p)\) for \(\pi \in S_n\)
where \(\sigma_\pi\) is \(\pm 1\) depending on whether \(\pi\) is even
or odd.

\subsection{Connections}\label{sec:connect} 

Consider a fiber bundle  \(\pi:E \to M\) with fiber \(F\) and
in which all spaces are manifolds.
At each point \(p\in E\) there is a canonical subspace \(V_pE \subset
T_pE\)
\index{\(V_p\)}%
of the tangent space at \(p\) to \(E\) called the {\em vertical
subspace\/}
\index{subspace!vertical}%
 consisting of all vectors \(v\) such that \(d\pi\,v = 0\).
These are vectors tangent to the fiber \(F_{\pi(p)}\), which is a
submanifold.
On the other
hand, there is in general no canonical way of choosing a complementary
``horizontal" subspace (see the discussion following
Example \ref{ex:trivnottriv}).  By a {\em connection\/}
\index{connection}%
 on \(E\) we mean a
smooth
choice (to be shortly explained) at each \(p\in E\)
of a subspace \(H_pE\subset T_pE\)
\index{\(H_p\)}%
complementary to \(V_pE\).  Given such a connection then at each
point \(p\in E\) we have projections \(\pi^v_p\)
\index{\(\pi^v_p\)} and \(\pi^h_p\)
\index{\(\pi^h_p\)}%
onto
the vertical and horizontal subspaces respectively. We shall assume
 that the distribution \(H_pE\) varies in
a \(\cC^\infty\) way with \(p\), which means that for any
\(\cC^\infty\) vector field \(\cX\) on \(E\), the horizontally projected
field
\(\pi^h\cX\) is likewise \(\cC^\infty\).

Let us examine this in a  trivialization. One has
\(T_{(x,f)}(U\times F) \simeq T_xU\times T_fF\). Let \((\xi, y)\in
T_xU\times T_fF\). Such a vector is vertical if and only if \(\xi=0\).
Thus \(\pi^v_{(x,f)}\) must have
the form 
\[
\pi^v_{(x,f)}(\xi,y) =(0,y+\Gamma(x,f)\xi)
\] 
where \(\Gamma(x,f)\)
\index{\(\chi(x,f)\)@\(\Gamma(x,f)\)}%
is
a {\em linear\/} map \(T_xU \to T_fF\). From this  it follows that 
\[
\pi^h_{(x,f)}(\xi,y) = (\xi,-\Gamma(x,f)\xi)
\]

 It is useful to calculate the change
in the connection map \(\Gamma(x,f)\) in a different  
trivialization.
In this case define the map
\[
\psi(x,f) =(x, h_{VU}(x)(f))
\]
 The differential of this map must
intertwine the local representatives  of \(\pi^v\),
the projection maps on the
vertical tangent subspaces, specifically, \linebreak
\(d\psi_{(x,f)}\circ \pi^v_{(x,f)} = \pi^v_{\psi(x,f)}\circ
d\psi_{(x,f)}\).
   One has
\[
dh_{VU}(x)(f)(\xi,y)= (\xi, d_1h_{VU}(x)(f)\xi +
d_2h_{VU}(x)(f)y)
\]
The intertwining relation with \(\pi^v\) gives us
\[
d_1h_{VU}(x)(f) + \Gamma_V(x, h_{VU}(x)(f))=
d_2h_{VU}(x)(f)\Gamma_U(x,f)
\]
 which can be solved for
\begin{eqnarray}\label{eq:contran} \nonumber
\Gamma_V(x,f) &=& 
 d_2h_{VU}(x)(h_{VU}^{-1}(x)( f))\Gamma_U(x,
h_{VU}^{-1}(x)(f)) \\ \label{eq:conntr}
& & -d_1h_{VU}(x)(h_{VU}^{-1}(x)(f))
\end{eqnarray}%

Reciprocally, one can use (\ref{eq:contran}) to define a
connection. 

\begin{theorem}
Let \(\pi:E\to X\) be a bundle with  fiber \(F\). Suppose
we have a set of  trivializations \(h_U:\pi^{-1}(U)\to U\times F\) 
such that the open sets \(U\) cover \(X\).  Suppose that for each
trivialization and each \((x,f)\in U\times F\) we have linear maps 
\(\Gamma_U(x,f):T_xU\to T_fF\) varying smoothly with respect to 
\((x,f)\) and such
that relation (\ref{eq:contran}) is satisfied for each pair of
trivializations such that \(U\cap V\neq\emptyset\). Then there is a
unique connection of the bundle for which the  representative with
respect to the given trivializations are the given \(\Gamma_U(x,f)\).
\end{theorem}%
The proof is entirely straightforward. 

In the case of a \(G\)-bundle, one has
\(h_{UV}(x)(f)=g_{UV}(x)\cdot f=\alpha(g_{UV}(x),f)\) where for
convenience we have explicitly introduced  the action 
\(\alpha\) of \(G\)
on \(F\). Equation (\ref{eq:conntr}) now becomes
\begin{eqnarray} \nonumber
\Gamma_V(x,f) &=&
  d_2\alpha(g_{VU}(x), g_{VU}^{-1}(x)\cdot f)\Gamma_U(x,
g_{VU}^{-1}(x)\cdot f) \\ \label{eq:gconntr}
& & -d_1\alpha(g_{VU}(x), g_{VU}^{-1}(x)\cdot f)
dg_{VU}(x)
\end{eqnarray}%
where \(dg_{VU}\) is the differential of the map \(g_{VU}:U\cap V \to
G\).

We now specialize to vector bundles. In this case one has a
canonical identification  \(V_pE \simeq F_{\pi(p)}\) as follows:
If \(q\in F_{\pi(p)}\) then it makes sense to consider the curve
\(\gamma(t) = p+tq \in F_{\pi(p)}\) as this  space is a vector space.
Identify \(\gamma'(0)\in V_pE\) with \(q\in F_{\pi(p)}\).
We shall say that a connection is {\em
linear\/}
\index{connection!linear}%
 if \(\pi^h_p\) varies affinely with \(p\) in a fixed fiber
\(F_x\), and the canonical zero section of \(E\) is
horizontal. 

In a locally trivialized vector bundle using the mentioned identification
\(T_{(x,f)}(U \times F) \simeq T_xU \times T_fF\simeq T_xU \times F\)  for
a connection to be linear one must have \(\Gamma(x,f)\xi =
\Gamma(x)(\xi)f\) where \(\Gamma(x)\)
\index{\(\chi(x)\)@\(\Gamma(x)\)}%
can now be interpreted as a
linear map \(T_xU \to \End(F)\), in other words an \(\End(F)\)-valued
\(1\)-form on \(U\). Such {\em vector-valued\/} forms are
explained in Appendix \ref{sec:parmaps}. 
\index{\(1\)-form!\(\End(F)\)-valued}%

We now have
\begin{eqnarray}
\pi^h_{(x,f)}(\xi,w) &=& (\xi, -\Gamma(x)(\xi)f) \\
\pi^v_{(x,f)}(\xi,w)  &=& (0,w+\Gamma(x)(\xi)f)
\end{eqnarray}%

Using the parameterized 
map convention of Appendix \ref{sec:parmaps},  (\ref{eq:conntr}) now reads
\begin{equation}\label{eq:vcontran}
\Gamma_V =
h_{VU}\Gamma_Uh_{VU}^{-1} 
-dh_{VU}h_{VU}^{-1}
\end{equation}
This
has a form analogous to a   {\em gauge
transformation\/}, which  shall be defined
in section \ref{sec:gaugetra}. 

Another useful expression is that of (\ref{eq:gconntr}) for 
a linear connection on a vector \(G\)-bundle.
The
action is now \(g\cdot f = R(g)f\) where \(R\) is a group representation.
Using the conventions introduced in section \ref{sec:liac}, 
equation (\ref{eq:dtilde}), we have
 \(d_1\alpha (g,f)L\cdot g^{-1} = \tilde d_1\alpha (g,f)L =
\gr(L)f\) where \(L\in \gg\). Also \(d_2\alpha(g,f)y = R(g)y\). From
these we find, again using the parameterized map convention of Appendix 
\ref{sec:parmaps}, that
\begin{equation}\label{eq:gammatrans}
\Gamma_V = R(g_{VU})\Gamma_UR(g_{VU})^{-1} - \gr
(dg_{VU}\cdot g_{VU}^{-1})
\end{equation}%
 Just as (\ref{eq:vcontran}) this
has the form of a {\em gauge
transformation\/}, defined in section \ref{sec:gaugetra}. 
\index{gauge transformation}%

Let now \(G\) be a Lie group and  \(PG\)  a principal \(G\)-bundle over an
\(n\)-dimensional
manifold \(M\). In this case  one has a canonical
identification \(V_pPG \simeq \gg\) as follows.   Let \(p\in PG\) and
\(L\in \gg\). Consider the curve \(\gamma(t) = p\cdot \exp(tL)\) in
\(PG\). Identify \(L_p=\gamma'(0)\in V_pPG\) with \(L\in \gg\).
With this identification, the vertical projection \(\pi^v_p\) 
is represented for every \(p\in PG\) by a map
\(\omega_p:T_pPG \to \gg\) which is thus a {\em
\(\gg\)-valued \(1\)-form\/}
\index{\(1\)-form!\(\gg\)-valued}%
 on \(PG\), which we shall call the {\em
connection \(1\)-form\/}.
\index{\(1\)-form!connection}%
Such a \(1\)-form obviously satisfies
\begin{equation} \label{eq:proj}
\omega_p(L_p) = L
\end{equation}%
A connection is said to be {\em invariant\/}
\index{connection!invariant}%
 if the
horizontal subspaces \(H_p\) satisfy
\(H_{p\cdot g} =
dR_g(H_p)=H_p\cdot g\), where \(R_g \) is the right action
map \(p\mapsto p\cdot g\) on \(PG\).
This means that if at \(p\) one has \(v=v_V\oplus
v_H\),
a decomposition of a tangent vector into the vertical and horizontal
components, then \(v\cdot g=v_V\cdot g \oplus v_H\cdot g\)
is the decomposition at \(p\cdot g\).

One has
\[
 L_p\cdot g = \left.\frac{d}{dt}p\cdot\exp(tL)g\right|_{t=0} =
\left.\frac{d}{dt}p\cdot g(g^{-1}\exp(tL)g)\right|_{t=0}  =
(\Ad_{g^{-1}}L)_{p\cdot g}
\]
 so the action of \(dR_g\) on the spaces
\(V_p\), when transferred to \(\gg\), translates to \(\Ad_{g^{-1}}\).
Now for an invariant connection, one
must have \(\pi^v_{p\cdot g}(v\cdot g) = \pi^v_p(v)\cdot g\), which by what
was said
above means
\begin{equation} \label{eq:ad}
\omega_{p\cdot g}(v\cdot g) =\Ad_{g^{-1}}\omega_p(v)
\end{equation}%

Let now \(\omega\) be an invariant connection \(1\)-form on \(PG\).
Let \(h_U\) be a trivialization map on an open set \(U\).
We shall
analyze the structure of
\(\omega^U = (h_U^{-1})^*(\omega)\) on
  \(U\times G\). Since the trivialization map
commutes with right action, the image form also corresponds to an
invariant connection. One has the 
identification
\[
T_{(x,h)}(U\times G) \simeq T_xU\times T_hG \simeq
T_xU \times
\gg 
\]
where of course the identification of \(T_hG\) with \(\gg\) is via
\(dR_{h^{-1}}\).  Assuming this identification for the time being,
consider
 \(v=(\xi,L)\in  T_xU \times
\gg\). One has 
\(v\cdot g=(\xi, \Ad_{g^{-1}}L)\), and 
by (\ref{eq:proj})
\(\omega^U_{(x,h)}(\xi,L)= \omega^U_{(x,h)}(\xi,0) + L\).
By (\ref{eq:ad}) one has
\[
\omega^U_{(x,h)}(\xi, h^{-1}\cdot L\cdot h) = h^{-1}\cdot
\omega^U_{(x,e)}(\xi,L)\cdot h
=h^{-1}\cdot (\omega^U_{(x,e)}(\xi,0) +  L)\cdot h
\]
 so that
\(\omega^U_{(x,h)}(\xi,0) = h^{-1}\cdot A_U (x)(\xi)\cdot h\) where
\(A_U \) is a \(\gg\)-valued \(1\)-form on \(U\). One finally has
has
\begin{equation} \label{eq:oa}
\omega^U_{(x,h)}(\xi,L) = h^{-1}\cdot (A_U (\xi) +
L)\cdot h
\end{equation}%
In the trivialized bundle therefore an invariant connection is
represented by the \(\gg\)-valued \(1\)-form \(A_U \)
\index{\(A\)}%
on
\(U\), which we shall call the {\em
local principal gauge potential\/}.
\index{potential!local principal gauge}%
Note that in doing so we have
passed from an object defined on the total space \(PG\) to one
defined on the base space \(M\). As in physics \(M\) is
generally space-time, this is in keeping with the notion that physical
field theory  deals with objects defined directly on space-time.
 In the physical literature, for a matrix group \(G\), 
what is know as the {\em gauge potential\/} 
\index{potential!gauge}%
arises  from  \(A_U \)
after a choice of a local chart, as will be explained 
in Section \ref{sec:lagrange}.

We will now find the relation between \(A _U\) and
\(A _V\) in \(U \cap V\), corresponding to two trivializations.
Let \(\psi_{VU} = h_V\circ h_U^{-1}\). One has
\begin{equation}\label{eq:trano}
\omega^V_{(x,g_{VU}h)}(\tilde d\psi_{VU}(\xi,L)) =
\omega^U_{(x,h)}(\xi,L)
\end{equation}%
where  by \(\tilde d\psi_{VU}\) we mean the
translation of  \(d\psi_{VU}\)
under the
identification of \(T_hG\) with \(\gg\).
Now \(\psi_{VU}(x,h) = (x, g_{VU}(x)h)\).
We can then write
\(d\psi_{VU}(\xi,v) = (\xi, g_{VU}\cdot  v +
dg_{VU}\xi \cdot h) \). One has
\(v=L\cdot h\) for some \(L\in\gg\), which  gives \(d\psi_{VU} (
\xi, L\cdot h) = (\xi, g_{VU}\cdot L\cdot h
+ dg_{VU}\xi \cdot h)\). The vertical component is at  the point
\((x,g_{VU}(x)h)\)
and we must bring it to \(\gg\) via right
action. One thus has: \(\tilde d\psi_{VU} (\xi, L) = (\xi,
g_{VU}\cdot L\cdot g_{VU}^{-1}
+ dg_{VU}\xi \cdot g_{VU}^{-1})\)
Using this and equations (\ref{eq:oa}) and (\ref{eq:trano}), one deduces
the
transformation law of the local principal gauge potentials:
\begin{equation}\label{eq:locg}
A _V =
g_{VU}\cdot A _U\cdot g_{VU}^{-1} -
dg_{VU}\cdot g_{VU}^{-1} = \Ad_{g_{VU}}A _U -
dg_{VU}\cdot g_{VU}^{-1}
\end{equation}%

We shall see in section \ref{sec:gaugetra} that this has the form of a {\em
gauge
transformation}.
\index{gauge transformation}%

Just as before, we can use (\ref{eq:locg}) to define an invariant
connection on a principal \(G\)-bundle.

\begin{theorem}\label{th:gluea}
Let \(\pi:PG\to X\) be a principal \(G\)-bundle. Suppose
we have a set of  trivializations \(h_U:\pi^{-1}(U)\to U\times G\) 
such that the open sets \(U\) cover \(X\).  Suppose that for each
trivialization we have a \(\gg\)-valued \(1\)-form 
\(A_U\) such
that relation (\ref{eq:locg}) is satisfied for each pair of
trivializations such that \(U\cap V\neq\emptyset\). Then there is a
unique invariant connection in \(PG\) for which the  representatives with
respect to the given trivializations are the given \(A_U\).
\end{theorem}%
Again, the proof is entirely straightforward. It is often through such
local representatives that an invariant connection on a principal
bundle is defined.

Given a Lie group \(G\) and  a principal \(G\)-bundle over a
manifold \(M\), let \(F\) be a manifold that carries a left
action \(\rho\) of \(G\).  Let \(PG\times_\rho F\) be the
associated bundle with map \(\pi_\rho: PG \times F \to PG \times_\rho
F\).
If now \(\omega\) is an invariant connection on
\(PG\), there is a canonical induced connection on \linebreak
\(PG\times F\) given
by \(H_{(p,w)}(PG\times F) = H_pPG \times \{0\}\). This distributions of
tangent subspaces is obviously invariant by the action \((p,f) \mapsto
(p\cdot g^{-1}, g\cdot f)\) used in the construction of the associated
bundle and so descends
to \(PG\times_\rho F\) by the differential of \(\pi_\rho\),
providing us with the {\em induced
connection\/}
\index{connection!induced}%
in \(PG \times_\rho F\).

Consider now the case in which \(F\) is a vector
space and \(\rho\) is given by a representation \(g\cdot f = R(g)f\) of
\(G\). Let \(\gr\) be the corresponding representation of \(\gg\) (see
Section \ref{sec:liac}).
In a  trivialization, \(\pi_\rho\) is represented by
\(\Id\times\rho: U \times G \times F \to U \times F\).
The horizontal subspace at \((x,f) \in U\times F\) is the image by the
differential of \(\Id\times\rho\) of the horizontal subspace at
\((x,e,f)\). Now \((\xi, L, y) \in T_xU\times \gg \times F\) is
horizontal at \((x,e,f)\) if and only if \(y=0\) and
\(A (x)(\xi)+L = 0\),
furthermore, the differential of \(\Id\times\rho\) at this point acts as
\((\xi,L,y) \mapsto (\xi,\gr(L)f + y)\). One now sees that the
horizontal subspace at \((x,f)\) consists of vectors of the form \((\xi,
-\gr(A (x)(\xi))f)\). As the second component is linear in \(f\)
we see that the induced connection is linear. One also has the following
expression for the projection onto the vertical subspace
\begin{equation}\label{eq:inprov}
\pi^v_{(x,f)}(\xi, y) = y+\gr(A (x)(\xi))f
\end{equation}%

There is something like a reciprocal to this construction. Let \(E\) be
a vector bundle  and \(\gamma\) a linear
connection. By Theorem \ref{th:eispgl}, \(E\) is naturally an 
\(\End(F)\)-bundle. Equations (\ref{eq:gammatrans}) and (\ref{eq:locg})
are now identical and by Theorem \ref{th:gluea} one has an invariant
connection on \(P\End(F)\) which by (\ref{eq:inprov}) induces the 
original connection \(\gamma\) on \(E\). This situation will always be
assumed for a vector bundle if the structure group in not explicitly
indicated. 

Consider now  bundle operations as discussed in  
Section \ref{sec:bundleops}. 
If each bundle \(E_\lambda\) has a connection, it is often possible to 
combine them in a canonical fashion to define a connection in \(E\). 
We shall 
only need this for the case of the product bundle \(\prod_\lambda 
E_\lambda\). One defines the {\em product connection\/}
\index{connection!product}%
 by stipulating that 
\(v\in H_pE \) if and only if \(dp_\lambda v\in 
H_{p_\lambda(p)}E_\lambda\).
It is easy to see that the product of invariant connections on principal 
\(G_\lambda\)-bundles is an invariant connection on the  principal \(\prod 
G_\lambda\)-bundle, which is the product.

\subsection{Gauge Transformations} \label{sec:gaugetra}
Let \(PG\) be a principal \(G\)-bundle over a base space \(X\).
A {\em gauge transformation\/}
\index{gauge transformation}%
is a bundle isomorphism \(\phi:PG \to PG\) that commutes with right
action. Because of the free right action of \(G\) on  \(PG\) we
can write \(\phi(p) = p\cdot\gamma(p)\) for a unique function
\(\gamma:PG \to G\). For this to commute with right action it is
necessary and sufficient that
\((p\cdot g)\cdot\gamma(p\cdot g) = (p \cdot \gamma(p))\cdot g\) and so
\begin{equation}\label{eq:gtra}
\gamma(p\cdot g) = g^{-1}\gamma(p)g
\end{equation}%
Thus the set of all gauge transformations is the set of all maps
\(\gamma:PG \to G\) satisfying (\ref{eq:gtra}).
The set of all gauge transformations of \(PG\) is called the {\em gauge
group\/}
\index{group!gauge}%
 of \(PG\) and we denote it by \(\cG(PG)\).
\index{\(G(PG)\)@\(\cG(PG)\)}%

When \(X\) is a manifold and \(G\) a Lie group, one can consider
\(\cG(PG)\) as (generally) an infinite dimensional Lie group.
One can discover what
one should take to be the Lie algebra of the gauge group by considering
a one-parameter family
\(\gamma_t(p)\) of gauge transformations
of the form \(\gamma_t(p)= \exp(t\theta(p))\)  where
\(\theta(p)\in \gg\). Property (\ref{eq:gtra}) for all \(t\) translates to
\begin{equation}\label{eq:igtra}
\theta(p\cdot g) = \Ad_{g^{-1}}\theta(p)
\end{equation}%
Maps \(\theta:PG\to\gg\) satisfying (\ref{eq:igtra}) are taken to
constitute the Lie algebra of \(\cG(PG)\) and are called {\em
infinitesimal gauge transformations\/}.
\index{gauge transformation!infinitesimal}%
  The Lie bracket of two
infinitesimal gauge transformations \([\theta_1,\theta_2]\) is
calculated pointwise  \([\theta_1,\theta_2](p) =
[\theta_1(p),\theta_2(p)]\). Since the adjoint action of \(G\) in
\(\gg\) commutes with the Lie bracket one sees that \(
[\theta_1,\theta_2] \) satisfies (\ref{eq:igtra}) and so  indeed
is likewise  an infinitesimal gauge transformation.

In a  trivialization the map \(\phi\) is represented by
\[
\Id\times\phi_U:U\times G \to U\times G
\]
 acting as \((x,g) \mapsto
(x,
\phi_U(x)(g))\). Because of commutativity with right action by \(G\) one
has \(\phi_U(x)(g) = \phi_U(x)(eg) = \phi_U(x)(e)g\) so \(\phi_U\) is
expressible through a local map \(\phi_U^0:U\to G\) as \(\phi_U(x)(g) =
\phi_U^0(x)g\). The map \(\gamma\) in a trivialization is expressed
by a map \(\Id\times\gamma_U:U\times G \to G\) satisfying
\(\gamma_U(x)(gh) = h^{-1}\gamma_U(x)(g) h\) hence \(\gamma_U(x)(h) =
h^{-1}\gamma_U(x)(e) h = h^{-1}\gamma_U^0(x) h\) and \(\gamma\) also is
expressible through a function \(\gamma_U^0:U\to G\). From
\(\phi_U(x)(g) = g\gamma_U(x)(g)\) one in fact concludes that \(\gamma_U^0
=
\phi_U^0\) and so the map \(\gamma\) can be locally suppressed.
We shall from now on drop the superscript \(0\) and write simply
\(\phi_U\).

From the above discussion and (\ref{eq:gfibmor}) 
it is easy to see that the gauge
transformation changes the transition maps \(g_{UV}\) to the equivalent
maps \(\phi_Ug_{UV}\phi_V^{-1}\). Note that there are no restrictions on
the maps \(\phi_U\) which can be an arbitrary \(0\)-cocycle with values
in \(G\). In fact, from the above discussion we can state
\begin{theorem}
Let \(\pi:PG\to X\) be a principle \(G\)-bundle. Suppose
we have a set of  trivializations \(h_U:\pi^{-1}(U)\to U\times G\) 
such that the open sets \(U\) cover \(X\).  Suppose that for each
such open set we have a maps 
\(\phi_U:U\to G\), then there is a
unique gauge transformation  \(\phi:PG\to PG\) for which the  
representative with
respect to the given trivializations are the given \(\phi_U\).
\end{theorem}%

Let now \(\omega\) be an invariant connection on \(PG\). A gauge
transformation \(\phi\) acts on \(\omega\) on the left by push-forward. The
new connection \(\phi\cdot\omega\)
\index{\(\theta\cdot\omega\)@\(\phi\cdot\omega\)}%
being defined by
\begin{equation}
(\phi\cdot \omega)_{\phi(p)}(d\phi\, v) = \omega_p(v)
\end{equation}%
In section \ref{sec:connect} we have seen that in a 
trivialization, an invariant connection is represented by the local
principal gauge potential \(A\) which is a \(\gg\)-valued
\(1\)-form on \(U\). We shall now calculate how a
gauge transformation acts on these local potentials. Identifying once
again \(T_{(x,g)}U\times G\) with \(T_xU \times \gg\), one has
\[
(\phi\cdot A )_U(x)(\xi) = (\phi\cdot \omega)^U_{(x,e)}(\xi,0) =
\omega^U _{(x,\phi(x)^{-1})}(\tilde d\phi_U^{-1}(\xi,0))
\] 
\index{\(\theta\cdot A\)@\(\phi\cdot A\)}%
where by
\(\tilde d\phi_U^{-1}\) we mean the differential of the map \((x,g)
\mapsto (x,\phi_U^{-1}(x)g)\) with the tangent spaces of \(G\) identified
with \(\gg\) through right action. Taking into account these
identifications we see that \(\tilde d\phi_U^{-1}(\xi ,0) = (\xi,
d\phi_U^{-1}(x)\xi\cdot \phi_U(x))\) where in this formula
\(d\phi_U^{-1}\) means the differential of the map \(\phi_U^{-1}:U \to
G\). From this and (\ref{eq:oa}) one has
\[
(\phi\cdot A )(x)(\xi) = \Ad_{\phi_U(x)}(A (x)(\xi) +
d\phi_U^{-1}(x)(\xi)\cdot \phi_U)
\]
 Now \(\phi_U(x)\phi_U^{-1}(x)
=
e\), so
\(d\phi_U(x)(\xi)\cdot\phi_U(x)^{-1} + \phi_U(x)\cdot
d\phi_U^{-1}(x)(\xi) = 0\)
and we  finally have the formula for a gauge transformation of the local
principal gauge potential.
\begin{equation}\label{eq:lgtra}
(\phi\cdot A )= \phi_U\cdot A \cdot\phi_U^{-1} -
d\phi_U\cdot\phi_U^{-1}
\end{equation}

It is useful to call attention to formulas (\ref{eq:vcontran}),
(\ref{eq:gammatrans}) and
(\ref{eq:locg}) which show that the transition formulas for the
representatives of a connection in two different trivializations have
the same abstract form as a gauge transformation. 

It is also useful to calculate the effect of an infinitesimal gauge
transformation on the local potential. Let the gauge transformation
\(\phi_t\) be defined through \(\phi_{t,U} (x) =
\exp(t\theta_U(x))\) and let \(\delta A _U\) be the coefficient of
\(t\) in the Taylor expansion of \(\phi_{t}\cdot A\) in (\ref{eq:lgtra}). 
One easily
calculates  that
\[
\delta A = [\theta,A ] - d\theta
\]%
We shall not need this formula in these notes, but it is very often
used in the physical literature.

\subsection{Parallel Transport}
We are in the category of manifolds. Let \(E\) be a fiber bundle with
base space \(M\), and fiber \(F\). Consider a
connection on \(E\) with vertical tangent space projectors \(\pi^v_p\).
A smooth curve in \(E\) is said to be {\em horizontal\/}
\index{horizontal!curve}%
 if at each
point its tangent lies in the horizontal tangent space of that point.
To be horizontal is to be an integral curve of a differential equation.
In fact consider a parameterized smooth curve in \(E\) and
consider its image \(e(t)=(x(t),f(t))\) in a  trivialization \(U\times
F\).
One has \(\pi^v(e'(t)) = (0,f'(t) + \Gamma(x(t),f(t))x'(t))\) and so the
condition for being horizontal is
\begin{equation}\label{eq:hc}
f'(t) + \Gamma(x(t),f(t))x'(t) = 0
\end{equation}%
This in local coordinates is an ordinary differential equation for
\(e(t)\). Note that only the components \(f(t)\) are required to obey a
differential equation and that \(x(t)\) can be freely given with
arbitrary parameterization. This allows us to determine \(f(t)\) from
\(x(t)\). Given a smooth curve \(C\) in \(M\) a {\em horizontal lifting\/}
\index{horizontal!lifting}%
of \(C\) is a horizontal curve \(\tilde C\) in \(E\) such that
\(\pi(\tilde C) = C\).
In a  trivialization this means that once
\(C\) is parameterized, then \(\tilde C\) satisfies (\ref{eq:hc})
inheriting
a parameterization from that of   \(C\). By the
existence, uniqueness, and regularity theorems for solutions of ordinary
differential equations, any smooth curve in \(M\) has at least a local
unique
horizontal lifting passing through any point \(f \in F_x\) for any
\(x\in C\).

Let \(C\) be a smooth curve in \(M\) with initial point \(x_0\) and
end point \(x_1\). Let \(f_0 \in F_{x_0}\), and assume that there is a
global horizontal lifting of \(C\) with initial point \(f_0\). The
endpoint \(f_1\in F_{x_1}\) of \(\tilde C\) is called the {\em parallel
transport
\index{parallel transport}%
 of \(f_0\) along \(C\)\/}. It is obviously unique if it exists.
If the parallel transport exists for all \(f_0\in F_{x_0}\) then the map
\(f_0 \mapsto f_1\) defines a diffeomorphism \(F_{x_0}\to F_{x_1}\).

For a vector bundle with a linear connection, equation (\ref{eq:hc}) has
the form   \(f'(t) + \Gamma(x(t))(x'(t))f(t) = 0\) which is a {\em
linear\/} equation. Thus parallel transport is always globally defined
and the parallel transport map \(F_{x_0}\to F_{x_1}\) is a linear
isomorphism.

It is useful to have explicit forms for the parallel transport map. This
is given by a construction known as time-ordered exponential integrals.
Let \(F\) be a finite dimensional vector space and consider the
following non-autono\-mous differential equation in \(F\):
\begin{equation}\label{eq:dfeatf}
\frac{df}{dt} = A(t)f
\end{equation}%
where \(A(t)\) is a  linear operator which is a \(\cC^\infty\)
function of \(t\).    By the existence, uniqueness, and regularity
theorem for the solution of ordinary differential equations, for any
\(f\in F\), there is a unique solution \(f(t)\) with \(f(a)\) given.
In differential equation theory, one generally introduces what is known
as the {\em fundamental solution\/}
\index{fundamental solution}%
 of (\ref{eq:dfeatf}), that is, an
\(\End(F)\)-valued function \(W(t,a)\) which satisfies
\begin{eqnarray}\label{eq:fsde}
\frac{\partial}{\partial t}W(t,a) &=& A(t)W(t,a) \\ \label{eq:fsic}
W(a,a) &=& I
\end{eqnarray}%
One now has 
\[
f(t) = W(t,a)f(a)
\]
In physical literature one however often sees the solution as 
given by the {\em time-ordered exponential
integral\/}, to be explained below:
\index{integral!time-ordered}
\begin{equation}
f(t) = T \exp\left(\int_a^tA(s)\,ds\right)f(a)
\end{equation}%
The operator in front of \(f(a)\) is not literally the exponential of an
integral but a symbolic way of expressing a limit of a Riemann product,
analogous to a Riemann sum. Let \([a,b]\) be an interval and partition
it as \(a=t_0 < t_1 <
\cdots < t_{N-1} < t_N =b\), let \(\Delta_it = t_{i} -t_{i-1}\), chose
\(t_i'\in[t_{i-1},t_i]\), and consider the product
\begin{eqnarray*}
\lefteqn{T\prod_{i=1}^N \exp(A(t_i')\Delta_it)} \\ 
&= &
\exp(A(t_{N}')\Delta_{N}t)
\exp(A(t_{N-1}')\Delta_{N-1}t)
\cdots
\exp(A(t_{1}')\Delta_{1}t)
\end{eqnarray*}%
Note that the order of the factors in this product is important  as the
various linear operators \(A(t)\) do not necessarily commute with each
other. The chosen order is that of decreasing values of the \(t_i\) as one
goes through the product from left to right. This is called {\em
time order\/} and the symbol \(T\) symbolizes this choice.
 One has
by definition
\begin{equation}\label{eq:trprod}
 T \exp\left(\int_a^bA(t)\,dt\right) =
 \lim_{N\to\infty} T\prod_{i=1}^N \exp\left(A(t_i')\Delta_it\right)
 \end{equation}%
 where the limit is taken in the same sense as the one that defines a
 Riemann
 integral. The limit of course is nothing more than the fundamental
solution \(W(b,a)\), thus one has 
the obvious composition property for time-ordered
exponential integrals: if \(a<b<c\), then
\begin{equation} \label{eq:prodp}
  T \exp\left(\int_a^cA(t)\,dt\right) =T\exp\left(\int_b^cA(t)\,dt\right)
 T\exp\left(\int_a^bA(t)\,dt\right)
  \end{equation}%

The time-ordered exponential integral is the non-commutative
analog of the continuous product of a function. Let \(f(t)\) be a {\em
positive\/} function on an interval \([a,b]\) and define the {\em
continuous product\/}
\index{continuous product}%
\[
\prod_a^bf(t)^{dt} =
\lim_{N\to\infty}\prod_{i=1}^Nf(t_i')^{\Delta_it}
\]
with the limit understood as in the previous paragraph. One easily shows
that if \(\ln f\) is Riemann integrable in \([a,b]\), then
\[
\prod_a^bf(t)^{dt} = \exp\left(\int_a^b \ln f(t)\, dt\right)
\]
and so the continuous product reduces to an ordinary integral, a
continuous sum, and for this reason there is no separately developed
theory for continuous products. In the non-commutative case however one
cannot reduce continuous products to integrals and a separate treatment
is necessary. To complete the identification note that in particular if
\(f(t) = \exp(a(t))\) then
\[
\prod_a^bf(t)^{dt} = \exp\left(\int_a^b
a(t) \,dt\right)
\]
Thus    \( T \exp\left(\int_a^bA(t)\,dt\right)\)  is
rightly thought of as the continuous product of \(\exp (A(t))\) ordered
with time decreasing left to right.

Let now \(A \) be an \(\End(F)\)-valued \(1\)-form  on a manifold
\(M\) and let \(C\) be a smooth oriented curve in \(M\). One can also
define the {\em path-ordered exponential integral\/}
\index{integral!path-ordered}%
\begin{equation}\label{eq:poex}
 P \exp\left(\int_CA \right) =
 \lim_{N\to\infty} P\prod_{i=1}^N \exp\left(\int_{C_i}A \right)
 \end{equation}%
where the curve \(C\) has been partitioned into {\em successive\/} arcs
\(C_1,\dots C_N\) each one inheriting its orientation from \(C\). The
limit is to be understood in relation to a fixed parameterization of
\(C\)  with the maximum parameter length of the \(C_i\) tending to zero.
 Such
a path-ordered exponential integral is an element of \(\End(F)\).

It is useful to note that in products (\ref{eq:trprod}) and
(\ref{eq:poex}) one can replace the factor
\(\exp\left(A(t_i')\Delta_it\right)\) by \(I+A(t_i')\Delta_it\) and
respectively \(\exp\left(\int_{C_i}A \right)\) by \(I +\int_{C_i}A\),
and obtain the same limits. This has the advantage of not relying on the
existence of the exponential. 

Though the time-ordered exponential
integral is a purely formal expression, it suggests the following
expansion 
\begin{equation}\label{eq:dyson}
W(b,a) = \sum_{n=0}^\infty\frac{1}{n!} \int_a^b\int_a^b\cdots\int_a^b
T(A(t_1)A(t_2)\cdots A(t_n))\,dt_1\,dt_2\cdots dt_n
\end{equation}%
where, by convention, the \(n=0\) term is \(I\), and  
\(T(A(t_1)A(t_2)\cdots A(t_n))\) means the product of the
\(A(t_i)\) in order of decreasing times. Thus 
\[
T(A(t_1)A(t_2)) = \left\{\begin{array}{cc}
A(t_2)A(t_1) & \quad\hbox{if}\quad t_1 < t_2 \\
A(t_1)A(t_2) & \quad\hbox{if}\quad t_1 > t_2
\end{array}\right.
\]
The right-hand side of (\ref{eq:dyson}) indeed does converges to \(W(b,a)\)
and is known as the {\em Dyson series\/}. 
\index{Dyson series}%
Because of the time-ordering
instruction \(T\), each term in fact is a sum of \(n!\) equal
contributions, each one an integral over a simplex
\[
\{(t_1,t_2,\dots,t_n)\,|
\,t_{\pi(1)}\ge t_{\pi(2)}\ge\cdots\ge t_{\pi(n)}\}
\] 
where \(\pi\) is a permutation of \(\{1,2,\dots,n\}\). We finally get
\begin{equation}\label{eq:picard}
W(b,a) = \sum_{n=0}^\infty
\int_a^b\int_a^{t_1}\int_a^{t_2}\cdots\int_a^{t_{n-1}}
A(t_1)A(t_2)\cdots A(t_n)\, dt_n\cdots dt_2\,dt_1
\end{equation}%
where, again by convention, the \(n=0\) term is \(I\).
This expansion can be obtained directly by applying Picard's method to
the fundamental solution equations (\ref{eq:fsde}-\ref{eq:fsic}). Note
that the existence of the exponential is not needed to define each term
of (\ref{eq:picard}) which makes it useful as a formal series
in contexts in which A(t) belongs to an algebra for which the
exponential is not defined.

We can now use path-ordered exponential integrals to
express the effect of
parallel transport. In a  trivialization of a vector bundle
the parallel transport equation reads
\[
\frac{df}{dt} = -\Gamma(x(t))(x'(t))f
\]
where \(x(t)\) is a parameterized path \(C\) in \(U\) with initial point
\(x(0)\) and final point \(x(1)\). Thus we have
\[
f(1) = T
\exp\left(-\int_0^1\Gamma(x(t))(x'(t))\,dt\right)f(0)=P\exp\left(-\int_C
\Gamma\right)f(0)
\]
and so \(P\exp\left(-\int_C \Gamma\right) \) is the parallel transport
operator for the oriented curve \(C\).

Finally, let \(G\) be a Lie group with Lie algebra \(\gg\) and \(A \)
a \(\gg\)-valued \(1\)-form on an open set \(U\) in a
 manifold \(M\). One can, for a curve contained in \(U\), likewise define
\begin{equation}%
 P \exp\left(\int_CA \right) =
 \lim_{N\to\infty} P\prod_{i=1}^N \exp\left(\int_{C_i}A \right)
 \end{equation}%
which results in an element of \(G\).

Such a path  exponential
integral solves the parallel transport equation for a principal
\(G\)-bundle with an invariant connection. In fact, consider a 
trivialization \(U\times G\)
of such a bundle
and a parameterized curve \(p(t)=(x(t), g(t))\) in  it. The tangent
vector at \(p(t)\) is
\((x'(t),g'(t))\in T_{x(t)}U\times T_{g(t)}G\).  With \(T_{g(t)}G\)
identified with \(\gg\) by right action we represent this tangent vector
now by \((x'(t), g'(t)\cdot g(t)^{-1})\in T_xU\times\gg\).  By
(\ref{eq:oa}) such a vector is horizontal if and only if
\[
g(t)^{-1}\cdot(A (x'(t)) + g'(t)\cdot g(t)^{-1})\cdot g(t)
=0
\]
 that is,  if and only if \(A (x'(t)) + g'(t)\cdot g(t)^{-1}
= 0\). Thus the parallel transport equation in a principal \(G\)-bundle
with an invariant connection is
\begin{equation}\label{eq:ptpb}%
\frac{dg}{dt} = -A(x(t))(x'(t))\cdot g
\end{equation}%
It is now easy to see that
\[
g(t) = T \exp \left(-\int_0^t A (x(s))(x'(s))\,dt\right)g(0)
\]
In fact, one has by
(\ref{eq:prodp}) that
\[
g(t+r) =  T\exp\left( -\int_t^{t+r}A (x(t))(x'(t))\,ds\right)g(t)
\]
and differentiating this with respect to \(r\) at \(r=0\) gives
(\ref{eq:ptpb}).
Thus the effect of parallel transporting a group element \(g\) over an
oriented curve \(C\) in \(U\) is to multiply it on the left by
\[
P\exp\left(-\int_C A \right)
\]
This result also shows that parallel transport is globally defined
as the path-ordered exponential integral exists for any compact
oriented curve with end points. The proof of this is analogous to the
proof of the existence of Riemann integrals of continuous functions on
compact intervals.

We now investigate how the local representatives of parallel transport
depend on the trivialization. Let \(C\) be an oriented curve in \(M\)
with initial point \(x_0\) and final point \(x_1\). Suppose parallel
transport along \(C\) is defined for all \(f_0\in F_{x_0}\). Let
\(T:F_{x_0}\to F_{x_1}\) be the parallel transport diffeomorphism \(f_0
\mapsto f_1\). If \(C\) lies in an open set \(U\) in which the bundle is
trivialized, then \(T\) is represented by a diffeomorphism \(T_U:F\to
F\). If now \(V\) is another such open set and if \(\tilde C_U\) is the
horizontal lifting of \(C\) in the bundle trivialized over \(U\), then
\(\tilde C_V =(\Id\times h_{VU})(\tilde C_U)\) is the lifting in the
bundle trivialized over \(V\). We thus have 
\(T_V(h_{VU}(x_0)(f_0)) = h_{VU}(x_1)(f_1)=h_{VU}(x_1)(T_U(f_0))\) from
which 
\begin{equation}\label{eq:pti}
T_V = h_{VU}(x_1)\cdot T_U\cdot h_{VU}(x_0)^{-1}
\end{equation} 
Note that this transformation is point-wise, that is, it does not involve
the differential of \(h_{VU}\).

\subsection{Curvature}
We are in the category of manifolds. Consider a bundle \(\pi:E \to M\)
with fiber \(F\) and a connection \(\Gamma\). Let \(x\in M \), and
\(p\in F_x\). Remember that parallel transport of \(p\) is always uniquely
defined locally along curves passing through \(x\).  We can choose local
charts
\(U\subset M\), \(W \subset V \subset F\) with
local coordinates \(x^1,\dots,x^n\)  in \(U\) and \(f^1,\dots,f^m\) in
\(V\)
 such that the bundle
trivializes over \(U\), and
given a smooth curve \(C\) in \(U\) with end points
\(x_0\) and \(x_1\) and any \(f\in W\), then parallel transport of \(f\)
exists over \(C\) and lies in \(V\).  Now parallel transport in general
depends on the path \(C\) joining the two points, and in particular if
\(x_1=x_0\) it may not be the identity. The {\em curvature\/}
\index{curvature}%
of the
connection is a measure of by how much the transport depends on \(C\),
or equivalently by how much it differs from the identity along closed
curves. A more precise statement of this, which we shall not prove, 
is the theorem that states that
the curvature is zero if and only if parallel transport is the same
along homotopy equivalent paths with the same end points. Parallel
transport may still depend on the path even with curvature zero if there
is more than one homotopy class of paths. Connections with zero
curvature are called 
\index{connection!flat}
{\em flat\/}.

In local coordinates the equation of parallel transport is:
\[
\frac{df^a}{dt} = -\sum_i\Gamma^a{}_i(x(t),f(t))\frac{dx^i}{dt}
\]%
Let us choose numbers \(\xi^1,\dots,\xi^n\) and \(\eta^1,\dots,\eta^n\)
such that the coordinates  \(x^i + t\xi^i + s\eta^i\)  define points in
\(U\) for \(t,s\in [0,1]\). Consider the piece-wise smooth curve \(C\)
defined by the border of this rectangle passing successively through the
vertices  \(x_0^i=x^i\), \(x_1^i=x^i + \xi^i\),
\(x_2^i=x^i+\xi^i+\eta^i\), \(x_3^i=x^i+\eta^i\),
and back to \(x_0\).
We shall now calculate the effect of parallel transport
along \(C\) to second order in \(\xi\) and \(\eta\).

Now to second order in \(\xi\) along the first leg, from \(x_0\) to
\(x_1\),
parallel transport can be represented by the map
\[
T_\xi^{(0)} :(x^i,f^a) \mapsto (x^i+\xi^i,
f^a-\sum_i\Gamma^a{}_i(x,f)\xi^i
+\sum_{ijb}\frac{\partial \Gamma^a{}_i}{\partial f^b}\Gamma^b{}_j
\xi^i\xi^j) 
\]
We must now compute, retaining only term up to second order in \(\xi\)
and \(\eta\), the quantity \(T_{-\eta}^{(3)}\circ T_{-\xi}^{(2)}\circ
T_\eta^{(1)}\circ T_\xi^{(0)}\) where the superscript  in \(T^{(k)}\)
means that all functions must be evaluates at the vertex point \(x_k\).
This is a straightforward though tedious calculation and we find that
the result of this parallel transport is
\[%
(x^i,f^a) \mapsto (x^i, f^a + \sum_{ij}R^a{}_{ij}(x,f)\eta^i\xi^j)
\]%
where
\begin{equation}\label{eq:gc}%
R^a{}_{ij} = \frac{\partial \Gamma^a{}_j}{\partial x^i}
-\frac{\partial \Gamma^a{}_i}{\partial x^j} +
\sum_b\frac{\partial\Gamma^a{}_i}{\partial f^b}\Gamma^b{}_j -
\sum_b\frac{\partial \Gamma^a{}_j}{\partial
f^b}\Gamma^b{}_i
\end{equation}%
In terms of the non-trivialized bundle, the quantity \(R\) at a point
\(p\in E\) is seen to be a bilinear, anti-symmetric map
\(T_{\pi(p)}M \times T_{\pi(p)}M \to V_pE\). For this reason,
\(R\) is referred to as the {\em curvature \(2\)-form\/}.
\index{\(2\)-form!curvature}%

One can use equation (\ref{eq:gc}) to calculate the curvature of a
linear connection on a vector bundle or of an invariant connection on a
principal bundle. In these cases however it is instructive to use the
path-ordered exponential integrals. For this we must recall the
Baker-Campbell-Hausdorff formula. Let  \(a\) and \(b\) be elements of a
an associative algebra.
One has, as formal power series in
\(a\) and \(b\):
\begin{equation}\label{eq:bch}%
\exp (a) \exp( b) = \exp\left(a + b + \frac{1}{2}[a,b] + []^2 + []^3 +
\cdots
\right)
\end{equation}%
where by \([]^k\) we mean a sum of terms each one of which is a
\(k\)-fold nested bracket of \(a\) and \(b\), for example \([]^2 =
\frac{1}{12}([a,[a,b]]+[b,[b,a]])\). Although exact expressions for the
\([]^k\) are known, we shall not need them explicitly beyond the terms
already shown. The same formula holds for \(a\) and \(b\) being elements
of the Lie algebra \(\gg\) of some Lie group \(G\) where the exponentials
are now actual group elements.

For a vector bundle  we can take \(V=W=F\), and denote now by
\(\Gamma=\sum_i\Gamma_i\,dx^i\) the
\(\End(F)\)-valued connection \(1\)-form.
Consider once again parallel transport along the rectangle \(C\) used
above. For the first leg of this we have the following parallel
transport
operator
\begin{equation}\label{eq:rplpt}%
P\exp\left(-\int_{C_1}\Gamma\right)
\end{equation}%
where \(C_1\) is the first segment of the path, from \(x_0\) to \(x_1\).
This to second order in \(\xi\) is
\[
T^{(0)}_\xi = \exp\left(-\Gamma(x_0)(\xi) +
(D\Gamma(x_0)\cdot\xi)(\xi)\right)
\]
where by \(D\Gamma\cdot\xi\) we mean the \(1\)-form
\[
\sum_{ij}\frac{\partial
\Gamma_i}{\partial x^j}\xi^j\,dx^i
\]
 As before, up to second order in
\(\xi\) and \(\eta\), we must calculate
\(T_{-\eta}^{(3)} T_{-\xi}^{(2)}
T_\eta^{(1)} T_\xi^{(0)}\),  where again the superscript  in \(T^{(k)}\)
means that all functions must be evaluates at the vertex point \(x_k\).
This can be done in a straightforward though tedious manner using the
Baker-Campbell-Hausdorff formula and the result, to second order
is \(\exp(R(\eta,\xi))\) where
\begin{equation}\label{eq:curvvb}%
R(\eta,\xi) = d\Gamma(\eta,\xi) + [\Gamma(\eta), \Gamma(\xi)]
\end{equation}%
 We see
this defines an \(\End(F)\)-valued \(2\)-form called the {\em curvature
\(2\)-form\/}.
\index{\(2\)-form!curvature}%
\index{\(2\)-form!\(\End(F)\)-valued}%

From (\ref{eq:pti}) one has \(\exp(R_V(\eta,\xi)=
h_{VU}(x_0)\exp(R_U(\eta,\xi))h_{VU}(x_0)^{-1} =
\exp(h_{VU}(x_0)R_U(\eta,\xi) h_{VU}(x_0)^{-1})\)
and so the curvature \(2\)-form has the
following transition formula 
\begin{equation}\label{eq:curvtran}
R_V =h_{VU}R_Uh_{VU}^{-1}
\end{equation}

We see on the right hand side the transition map of the bundle
\(\End(E)=\Hom(E,E)\) (see Section \ref{sec:bundleops}). 
This means that \(R\) can be viewed as an antisymmetric bilinear bundle map 
\(TM\times TM \to \End(E)\), that is an \(\End(E)\)-valued \(2\)-form.

For the case of a principal \(G\)-bundle one
has the group element
\[
P\exp\left(-\int_{C_1}A \right)
\]
 instead of
(\ref{eq:rplpt}). Proceeding in an entirely analogous manner, using again
the Baker-Campbell-Hausdorff formula, we conclude that up to second order
in \(\xi\) and \(\eta\), parallel transport around \(C\) corresponds to
multiplication by the group element \(\exp(F(\eta,\xi))\) where
\begin{equation}\label{eq:curvpb}%
F(\eta,\xi) = dA (\eta,\xi) + [A (\eta), A (\xi)]
\end{equation}%
defining thus a \(\gg\)-valued \(2\)-form in \(U\) called the {\em
curvature \(2\)-form\/}.
\index{\(2\)-form!curvature}%
\index{\(2\)-form!\(\gg\)-valued}%
 It is customary to write (\ref{eq:curvpb}) as
\[
F = dA  + [A ,A ]
\]

Similar to (\ref{eq:curvtran}), the transition formula for \(F\) is
\begin{equation}\label{eq:pcurvtran}
F_V= g_{VU}\cdot F_U \cdot g_{VU}^{-1}
\end{equation}

Given that \(G\) acts on \(\gg\) by adjoint action \(L\mapsto Ad_gL\)
there is an associated bundle \(P \gg = PG \times_{Ad} \gg\). We see on
the right-hand side of (\ref{eq:pcurvtran}) the transition map of \(P
\gg\) and so \(F\) can be considered as an antisymmetric bilinear
bundle map
\(TM \times TM \to P \gg\), that is a \(P \gg\)-valued \(2\)-form.

In the literature one often sees formulas slightly different from
(\ref{eq:curvvb}) and (\ref{eq:curvpb}). Sometimes there is a factor of
\(\frac{1}{2}\) in front of the second term. This is due to adopting a
different convention concerning alternating forms and exterior products
as explained in Section \ref{sec:bacon}. Sometimes there is a difference
in relative sign. This is due to conventionally using, in 
certain definitions, the negative of
what is our connection \(1\)-form. The reader beware.

If now \(\phi:PG \to PG\) is a gauge transformation, then \(\phi\) takes
horizontal curves with respect to an invariant connection \(\omega\) to
horizontal curves with respect to the connection \(\phi\cdot  \omega\). Now
in a trivialization, \(\phi\) is represented by left multiplication by
the map \(\phi_U(x)\) (see Section \ref{sec:gaugetra}),  exactly as
in a change of trivialization. This observation, along with
(\ref{eq:pcurvtran}) 
means that we have the following
gauge transformation law for the curvature:
\begin{equation} \label{eq:cug}%
 F \mapsto \phi_U\cdot F\cdot \phi_U^{-1} = \Ad_{\phi_U}F
 \end{equation}%
 Similarly we easily seen that the effect
of an infinitesimal gauge transformation (see last paragraph of Section
\ref{sec:gaugetra}) is
\[%
 \delta F = [\theta, F]
 \]%
This last equation is very frequently used in the physical literature, but
we shall not need it in these notes.

\subsection{Covariant Derivatives}\label{sec:covdev}%
Consider a fiber bundle  \(\pi:E \to M\) with fiber \(F\) and
structure group \(G\) in which all spaces are manifolds.  Suppose we are
given a connection on \(E\) with projections \(\pi^v_p\) on the vertical
tangent subspaces. Recall that a local section of \(E\) on an open set
\(U\subset M\) is a map \(\sigma: U \to E\) such that
\(\pi\circ\sigma(x) = x\). Let \(\xi\) be a tangent vector at \(x\in
U\). By the   {\em covariant derivative
\index{covariant derivative}%
 of \(\sigma\) at \(x\) in
direction \(\xi\)\/} we mean
\[%
\nabla_\xi \sigma(x) = \pi^v_{\sigma(x)} d\sigma_x \,\xi
\]%
One sees that \(\nabla_\xi\sigma(x) \in V_{\sigma(x)}E\).
Recall that for a vector bundle one has a canonical identification
\(V_pE \simeq F_{\pi(p)}\). We shall assume this identification and
consider exclusively vector bundles with \(F\) finite dimensional.
One should consider \(\nabla_\xi \sigma(x)\) as the analog of the notion
of a directional derivative of a bundle section.

One may wonder why one could not simply calculate the ordinary 
directional derivative of a section \(\sigma\). This however is not well
defined. Let \(x(t)\) be a curve
passing through \(x\) at \(t=0\) with \(x'(0)=\xi\). The directional
derivative should intuitively correspond to 
\[
\lim_{t\to 0} \frac{\sigma(x(t))-\sigma(x(0))}{t}
\]
The trouble with this is that \(\sigma(x(t))\) and \(\sigma(x(0))\) lie
in different fibers and there is no well defined way of calculating
their difference. If we have a connection though, we can parallel
transport \(\sigma(x(t))\) back along the given curve to the fiber over
\(x(0)\), and then compute the difference and  the limit. 
This procedure defines the
covariant derivative as an easy exercise shows. 

To get further insight into this situation, trivialize the bundle in an
open set \(U\) with local coordinates \(x^1,\dots, x^n\) and introduce a
basis \(f_1,\dots,f_n\) of \(F\). A section in \(U\) now is represented
by a function \(s(x)=s^i(x)f_i\) of \(U\) to \(F\). Consider the
partial derivative \(\frac{\partial s}{\partial x^j}=
\sum_i\frac{\partial s^i}{\partial x^j}f_i\), 
which would be a naive
``directional derivative". In another trivialization over the same open
set one has another representative \(\tilde s(x) =\sum_i\tilde s^i(x)f_i\)
 where 
\(\tilde s^i(x) = \sum h^i{}_k(x)s^k(x)\), and where the \(h^i{}_k\)
represent the transition map. Write this in matrix form 
as \(\tilde s = h s\). One has 
\[
\frac{\partial \tilde s}{\partial x^j}= \frac{\partial h}{\partial x^j}s
+h\frac{\partial s}{\partial x^j}
\]
If the first term were  absent, one would have that 
\(\frac{\partial s}{\partial x^j}\)
would likewise be a section of the same bundle and so partial
derivatives (and thus directional derivatives) would be well defined 
as operators on sections.  This is not so, but from the fact that the
first term is linear in \(s\) one can try to construct a new section
by an expression of the form 
\[
\frac{\partial s}{\partial x^j} + L_j s
\] 
where \(L_j(x) \in \End(F)\). Imposing a transformation law on
\(L\) so that 
\[
\frac{\partial \tilde s}{\partial x^j} + \tilde L_j \tilde s=
h(\frac{\partial s}{\partial x^j} + L_j s)
\] 
one finds 
\[
\tilde L_j = h L_j h^{-1} - \frac{\partial h}{\partial x^j}h^{-1}
\]
Comparing this to (\ref{eq:vcontran}) one sees that \(L\) must
define a connection in the bundle. Connections thus arise naturally once
one tries to introduce differential calculus for sections of bundles.

If now \(\cX\) is a vector field in \(U\) one can calculate
\(\nabla_{\cX(x)}\sigma(x)\) for each \(x\in U\),
this gives us a new local section \(\nabla_{\cX}\sigma\)
\index{\(\zeta_X\)@\(\nabla_{\cX}\)}%
of \(E\).
Trivializing over \(U\), a section
\(\sigma\) is represented by a map \(\Id\times s: U \to U\times F\)
and a linear connection by an \(\End(F)\)-valued  \(1\)-form
\(\Gamma\).  At point \(x\in U\), the differential of \(\Id\times s\)
acts as \(\xi \mapsto (\xi, ds_x\,\xi)\) so \(\nabla_\xi\sigma(x)\) is
represented by:
\begin{equation}%
ds_x\,\xi + \Gamma(x)(\xi)s(x)
\end{equation}%
where \(ds\) is the differential of \(s\). If \(f_1,\dots,f_m\) is a
basis for \(F\), then \(s(x)=\sum_{j=1}^ms^j(x)f_i\) and
\(ds_x\,\xi=\sum_{j=1}^m\xi(s^j)(x)f_i\) so for convenience we shall
write \(ds\,\xi\) as \(\xi(s)\). Note that \(\xi(s) \in F\). For a
vector field \(\cX\) one has:
\begin{equation}\label{eq:lincd}%
\nabla_\cX s = \cX(s) + \Gamma(\cX)s
\end{equation}%

Let \(W\) be a vector space and consider maps from open sets \(U\subset M\)
to \(W\). One can view these as local sections of the cartesian product 
\(M\times W\) considered as a trivial bundle with the identity map as
the one defining trivialization. One now has the canonical
identification  \(T_{(x,w)}(M \times W) \simeq T_xM \times W\) and can
define the horizontal subspace as \(H_{(x,w)}=\{0\}\times W\). In this
trivialization, \(\Gamma(x)=0\) and so
\[
\nabla_{\cX} s=\cX(s)
\]
In particular for real or complex valued partial 
functions \(f\) on \(M\) we
shall always take \(\nabla_{\cX} f=\cX(f)\).

The map \(\cX \mapsto \nabla_\cX\) fails to be a Lie algebra
homomorphism and this fact is related to the existence of curvature.
It is instructive to calculate 
\[
\nabla_\cX\nabla_\cY  -
\nabla_\cY\nabla_\cX - \nabla_{[\cX,\cY]}
\]
One has
\[
 \nabla_\cX\nabla_\cY s = \cX(\cY(s)) +
\cX(\Gamma(\cY))s +
\Gamma(\cY)\cX(s)+\Gamma(\cX)\cY(s)+\Gamma(\cX)\Gamma(\cY)s
\]
Taking into account that \(\cX(\cY(s)) - \cY(\cX(s)) =
[\cX,\cY](s)\) and, by (\ref{eq:donef}), 
that \(\cX(\Gamma(\cY)) -\cY(\Gamma(\cX) -
\Gamma([\cX,\cY]) = d\Gamma(\cX,\cY)\) one concludes that
\begin{equation}\label{eq:comcv}%
(\nabla_\cX\nabla_\cY  -
\nabla_\cY\nabla_\cX - \nabla_{[\cX,\cY]})s = R(\cX,\cY)s
\end{equation}%
where \(R\) is the curvature \(2\)-form. Note that \(R(\cX,\cY)\) at
point \(x\) depends only on the values \(\cX(x)\) and \(\cY(x)\) and not
on how these are extended to vector fields \(\cX\) and \(\cY\).

Recall that an invariant connection on a principal bundle
induces a linear connection on all
the associated vector bundles, and consequently a covariant derivative
on each such bundle. Recall also by the discussion preceding 
Theorem \ref{th:eispgl} that
any vector bundle can be considered an associated bundle of \(P\End(F)\)
and any linear connection on the bundle as one induced from an invariant
connection on the principal bundle. Furthermore given any finite family
\(E_i\), \(i=1,\dots,n\) of vector bundles with fibers \(F_i\) one can
consider them all as being associated to  a single principal bundle \(PG\)
where \(G=\End_1(F_1)\times\cdots\times \End(F_n)\) and whose action on
\(F_i\) is projection to the \(i\)-th factor followed by the natural
action of \(\End(F_i)\). With this in mind one can interpret the
discussion that follows as also pertaining to vector bundles and linear
connections even when no explicit structure group is indicated.

The covariant derivatives induced from a fixed principal bundle onto
its associated vector bundles have natural properties
with respect to various bundle constructions.
 We shall consider in
particular tensor products and bundles of linear homomorphisms, and 
deduce what may be called the Leibniz rule for the associated connections.
 Let therefore \(PG\) be a fixed principal \(G\)-bundle and \(E\) an
associated vector bundle with corresponding representation \(R\) 
of \(G\). By (\ref{eq:inprov}) and (\ref{eq:lincd}) 
one sees that in a trivialization
\begin{equation}\label{eq:asscon}
\nabla_\cX s= \cX(s) + \gr(A (\cX))s
\end{equation}

If now \(E_1\) and
\(E_2\) are two associated vector bundles defined by representations \(R_1\)
and \(R_2\) of \(G\), one easily sees that \(E_1\otimes E_2\) is the
associated bundle defined by the tensor product representation
\(R=R_1\otimes R_2\). The associated Lie algebra representation, 
as can be easily deduced from (\ref{eq:derrep}), is given
by \(\gr(L)=\gr_1(L)\otimes I + I\otimes\gr_2(L)\). Let
\(\nabla^{(1)}\) and \(\nabla^{(2)}\) be the covariant derivatives in
\(E_1\) and \(E_2\) respectively and \(\nabla\) the covariant derivative
in \(E\). 
One deduces 
\[
\nabla_\cX s_1\otimes s_2= \cX(s_1\otimes s_2) +
\gr_1(A (\cX))s_1\otimes s_2 + s_1\otimes\gr_2(A (\cX))s_2
\]
which, as 
\(\cX(s_1\otimes s_2)=\cX(s_1)\otimes s_2+s_1\otimes \cX(s_2)\),
 is to say 
\[
\nabla = \nabla^{(1)}\otimes I +
I\otimes\nabla^{(2)}
\] 
This is obviously an analog of the Leibniz
rule. Whenever dealing with bundles that are all associated to a given
principal bundle with a fixed invariant connection, we shall in general
neglect to label the symbol \(\nabla\) to indicate in which bundle the
covariant derivative is acting, as this should be clear from the
context. Thus we shall simply write \(\nabla_\cX(s_1\otimes s_2) =
(\nabla_\cX s_1)\otimes s_2 + s_1\otimes(\nabla_\cX s_2)\) which makes
this result resemble even more the Leibniz rule. This result obviously
extends to the tensor product of any finite number of associated
vector bundles.

Let \(E\) be a vector bundle with fiber \(F\) and \(E'\) the dual bundle
with fiber \(F'\), the dual of \(F\). One sees from (\ref{eq:inprov}),
 (\ref{eq:trandual}) and (\ref{eq:lincd}) that in a  trivialization,
if \(t\) represents a local section of \(E'\) that \(\nabla_\cX t=\cX(t) -
\gr(A (\cX))'t\).

Consider now the bundle \(\Hom(E_1,E_2)\). This is associated to \(PG\)
via a representation that acts on \(\phi\in\Hom(F_1,F_2)\) via
\(R(g)\phi = R_2(g)\phi R_1(g^{-1})\). The corresponding Lie algebra
representation is given by \(\gr(L)\phi = \gr_2(L)\phi - \phi\gr_1(L)\).
If now \(t\) represents a local section of \(\Hom(E_1,E_2)\) and \(s\)
represents a local section of \(E_1\) over the same open set, then
\(ts\) is a local section of \(E_2\) depending linearly on \(s\).
One has in a  trivialization,
\begin{equation}\label{eq:nabhom}%
\nabla_\cX t=\cX(t)+ \gr_2(A (\cX))t-t\gr_1(A (\cX))
\end{equation}%
Now
\(ts\) is linear in \(s\) so \(\cX(ts)=\cX(t)s + t\cX(s)\).
Using this and (\ref{eq:asscon}), one concludes that
\begin{equation}\label{eq:cdts}
\nabla_\cX ts =(\nabla_\cX t)s+t\nabla_\cX s
\end{equation}
In particular if
\(\nabla_\cX t=0\)
then \(\nabla_\cX ts = t\nabla_\cX s\).  As an example of this consider
the map \(E'\otimes E \to M\times \lF\) which pointwise is given by the
natural duality between \(F_x'\) and \(F_x\). This map corresponds to a
section of \(\Hom(E'\otimes E, M\times \lF)\). In a 
trivialization this is represented by the constant map \(t\)
with value being the natural duality \(F'\otimes F\to \lF\). Hence
\(\cX(t)=0\). Furthermore, specializing in (\ref{eq:nabhom}) to
\(\gr_2=0\) as the representation in the trivial bundle is trivial,
and to \(\gr_1(L)=-\gr(L)'\otimes I + I\otimes\gr(L)\), 
 one sees that the
last two term are also zero and so
 \(\nabla_\cX t=0\). From this we get
 \(\cX\!<f,s>=\nabla_\cX (t(f\otimes s)) =
t(\nabla_\cX(f\otimes s))\), and so \[
\cX\!<f,s> =
<\nabla_\cX f,s> + <f,\nabla_\cX s>
\]

It is often useful to consider the covariant derivative
\(\nabla_\cX\sigma\) as being a function of both \(\cX\) and \(\sigma\).
Seeing that it is linear in \(\cX\)  we can define it as a
map \(\nabla:\Gamma(E)\to\Gamma(T^*M\otimes E)\). 
\index{\(\zeta\)@\(\nabla\)}%
To see this, suppose that in some open set \(U\) one has  \(n\)
 vector
fields \(e_1,\dots,e_n\) which at each point provide a basis for the
tangent space. Let \(e^1,\dots,e^n\) be the corresponding dual set
of \(1\)-forms. We now define 
 \begin{equation}\label{eq:cdten}%
\nabla\sigma=\sum_ie^i\otimes\nabla_{e_i}\sigma
\end{equation}%
To see that this is independent of the choice of the \(e_i\), let
\(\tilde e_i\) be another choice, then one has 
\(\tilde e_i=\sum_jM_i{}^je_j\) for some field of invertible matrices
\(M_i{}^j\). One then has \(\tilde e^i = \sum_jM^i{}_jE^j\), where
\(M^i{}_j\) is the inverse of the transpose of \(M_i{}^j\). 
Is is now easy to see
that the right-hand side of (\ref{eq:cdten}) is the same using 
\(\tilde e_i\) instead of \(e_i\).
The characteristic property of
\(\nabla\) is a form of the Leibniz rule
\[
\nabla(f\sigma)=df\otimes\sigma+f\nabla\sigma
\]

Let \(E\) be a vector bundle.  
An antisymmetric
\(p\)-linear bundle map \linebreak \(\alpha: TM \times \cdots \times TM\to
E\) is  called a {\em \(E\)-valued \(p\)-forms\/},
\index{pform@\(p\)-form!\(E\)-valued}%
 or generically a
 {\em bundle-valued \(p\)-forms}.
\index{pform@\(p\)-form!bundle-valued}%
Let \(\cX_1, \dots , \cX_p\) be vector fields, then 
\(\alpha(\cX_1, \dots , \cX_p)\) is a section of \(E\). 
If \(E\) has a linear connection \(\gamma\), we can define,
analogously to (\ref{eq:dpf}) the {\em covariant exterior derivative\/}
\index{covariant derivative!exterior}%
\(d_\gamma \alpha\) of \(\alpha\) by
\begin{eqnarray}\nonumber
\lefteqn{d_\gamma \alpha(\cX_1,\dots,\cX_{p+1}) = \sum_i(-1)^{i+1} 
\nabla_{\cX_i}\alpha(\cX_1,\dots,\hat \cX_i,\dots \cX_{p+1}) +} 
\\ \label{eq:covexder} 
& & \sum_{1\le i < j \le p+1}(-1)^{i+j+1}
\alpha([\cX_i,\cX_j],X_1,\dots,\hat \cX_i, \dots,
\hat\cX_j,\dots,\cX_{p+1})
\end{eqnarray}
which as we shall see shortly is an \(E\)-valued \((p+1)\)-form. 
The ordinary exterior derivative is not well defined in this context for
the same reasons that one cannot define the ordinary directional
derivative for bundle sections. Note however that, just as for the
ordinary exterior derivative, no connection is needed on \(TM\).
We now 
determine the representative of \(d_\gamma\alpha\) in a trivialization
of \(E\). Let \(\Gamma\) be the connection \(1\)-form in the
trivialization, and \(a\) the local representative of \(\alpha\), 
then \(\nabla_{\cY}(a(\cX_1, \dots , \cX_p)) =
\cY(a(\cX_1, \dots , \cX_p)) + \
\Gamma(\cY)a(\cX_1, \dots , \cX_p)\). Using this and (\ref{eq:dpf})
one arrives at the local representative
\begin{eqnarray*}
\lefteqn{d_\gamma a(\cX_1,\dots,\cX_{p+1}) = 
da(\cX_1,\dots,\cX_{p+1}) +} 
\\ 
& & \sum_i(-1)^{i+1} 
\Gamma(\cX_i)a(\cX_1,\dots,\hat \cX_i,\dots \cX_{p+1})
\end{eqnarray*}
From this we see that locally, \(d_\gamma\alpha\) is a vector-valued 
(\(F\)-valued) \((p+1)\)-form. To show that globally 
it defines a bundle-valued 
(\(E\)-valued) \((p+1)\)-form, one needs to argue that it has the right
transition formula, but this is automatic given it's intrinsic
definition (\ref{eq:covexder}). 

We have seen that the curvature of a linear connection on a vector
bundle \(E\) can be viewed as a \(\End(E)\)-valued \(2\)-form. We now
have the {\em Bianchi Identities\/}
\begin{theorem}[Bianchi Identities]
\index{Bianchi Identities}%
Let \(E\) be a vector bundle with a linear connection \(\gamma\) and let
\(R\) be the curvature \(2\)-form of \(\gamma\), then
\begin{equation}\label{eq:bianchi}
d_\gamma R = 0
\end{equation}
\end{theorem}
\begin{proof}
Apply the Jacobi Identity for covariant derivatives 
\[
 [\nabla_{\cX},[\nabla_{\cY},\nabla_{\cZ}]]+
[\nabla_{\cY},[\nabla_{\cZ},\nabla_{\cX}]]+
[\nabla_{\cZ},[\nabla_{\cX},\nabla_{\cY}]]=0
\]
to any section of \(E\). Using (\ref{eq:cdts}, \ref{eq:comcv})
and the definition of \(d_\gamma\), one arrives at the conclusion.
\end{proof}

In contrast to the ordinary exterior derivative, we do not have
\(d_\gamma^2 = 0\). 
In fact, a simple calculation shows
\begin{eqnarray*}
\lefteqn{d_\gamma^2\alpha(\cX_1,\dots,\cX_{p+2})=}\\
& & \sum_{1\le i < j \le p+2}(-1)^{i+j+1}
R(\cX_i,\cX_j)\alpha(X_1,\dots,\hat \cX_i, \dots,
\hat\cX_j,\dots,\cX_{p+2})
\end{eqnarray*}
It is worth noting that \(d_\gamma^2\) is a \(0\)-th order differential
operator, that is, it involves no differentiation.

A linear connection on the tangent bundle \(TM\) allows the following
useful construction  
for two vector fields \(\cX\) and \(\cY\)
\[%
T(\cX,\cY) =\nabla_\cX\cY-\nabla_\cY\cX-[\cX,\cY]
\]%

Computing this in a  trivialization one has, calling the
representative of \(T\) again by the same letter, that \(T(\cX,\cY)
=\Gamma(\cX)\cY -\Gamma(\cY)\cX\). From this it is clear that at a point
\(x\in  M\), the vector \(T(\cX,\cY)\)  depends only on the vectors
\(\cX(x)\) and \(\cY(x)\) at \(x\) and not on how these are extended to the
actual vector fields \(\cX\) and \(\cY\). Thus \(T\) can be identified with
a  tensor in \(TM\otimes T^*M\otimes T^*M\) and \(T(\cX,\cY)\) is
anti-symmetric in its two arguments. This tensor is known as the {\em
torsion tensor\/}
\index{tensor!torsion}%
or simply the {\em
torsion\/}
\index{torsion}%
of the connection. A connection for which \(T=0\) is
known as {\em torsion-free\/}, or {\em torsionless\/}.
\index{connection!torsion-free}%
\index{connection!torsionless}%

\section{Manifolds}%
\subsection{Pseudo-Riemannian Manifolds}\label{sec:pseudoriemannian}%
A {\em pseudo-Riemannian\/}
\index{manifold!pseudo-Riemannian}%
manifold is a manifold \(M\) with a
non-degenerate symmetric bilinear form \(g(x)\), called the {\em
pseudo-metric\/}
\index{pseudo-metric}%
 defined in each tangent
space \(T_xM\). We shall often write \(<v,w>_x\)
\index{\(V,W_X\)@\(<v,w>_x\)}%
 in place of
\(g(x)(v,w)\) and often suppress mention of the point \(x\). We shall
assume that \(g(x)\) is a \(\cC^\infty\) function of \(x\) by which we
mean that for two \(\cC^\infty\) vector fields \(\cX\) and \(\cY\),
\(g(\cX,\cY)\) is a \(\cC^\infty\) function on \(M\).

For any symmetric bilinear non-degenerate form \(\beta\) on a real
\(n\)-dimensional vector space \(W\), there is a basis
\(e_1,\dots,e_n\) such that \(\beta(e_i,e_j) = \eta_{ij}\)
\index{\(\gamma_{ij}\)@\(\eta_{ij}\)}%
where the
matrix \(\eta = \hbox{diag}\,(1,\dots,1,-1,\dots,-1)\) with \(r\) entries
of \(1\) and \(s\) entries of \(-1\), where \(r+s=n\). The pair
\((r,s)\) is called the {\em signature\/}
\index{signature}%
 of \(\beta\). Such a basis is
called {\em orthonormal\/}.
\index{orthonormal}%
 The group of linear transformations
\(L:W\to W\) such that \(\beta(Lx,Ly) = \beta(x,y)\) is denoted by
\(\O(\beta)\)
\index{\(O(\beta)\)@\(\O(\beta)\)}%
and is know as the {\em orthogonal group\/}
\index{group!orthogonal}%
of \(\beta\).
The subgroup of \(\GL(n)\) of matrices  \(\Lambda\) such that
\(\Lambda\eta\Lambda^t = \eta\) is denoted by \(\O(r,s)\)
\index{\(O(r,s)\)@\(\O(r,s)\)}%
and is know as
the {\em pseudo-orthogonal group of signature \((r,s)\)\/}. 
\index{group!pseudo-orthogonal}%
Obviously
\(\O(\beta)= \O(-\beta)\), \(\O(r,s) \simeq \O(s,r)\), 
and \(\O(\beta)\simeq
\O(r,s)\). When \(s=0\) we write simply 
\index{\(O(n)\)@\(\O(n)\)}%
\(\O(n)\). From \(\Lambda\eta\Lambda^t = \eta\) one concludes \(\det
(\Lambda)^2 = 1\), that is \(\det(\Lambda)=\pm 1\) for any element of the
orthogonal groups. The elements of determinant \(1\) form a subgroup
called the {\em special (pseudo)-orthogonal\/} group and we denote these 
correspondingly by
\index{\(SO(\beta)\)@\(\SO(\beta)\)}%
\index{\(SO(r,s)\)@\(\SO(r,s)\)}%
\index{\(SO(n)\)@\(\SO(n)\)}%
\(\SO(\beta)\), \(\SO(r,s)\)  and \(\SO(n)\). 

In a pseudo-Riemannian manifold each \(g(x)\) has some signature. This
signature is constant on the connected components of \(M\) since to pass
continuously to
a different signature, the form would have to become degenerate at some
point. We shall assume the signature is constant on \(M\). A manifold of
signature \((n,0)\) is called a {\em Riemannian\/}
\index{manifold!Riemannian}%
 manifold. A
{\em space-time\/}
\index{space-time}%
 is a manifold of signature \((1, n-1)\), or, according
to some authors, one of signature \((n-1,1)\). Since replacing 
\(g\) by \(-g\) does not change most geometric facts, for many effects  
manifolds of signature \((r,s)\) are equivalent to
those of \((s,r)\).

If \(M\) is pseudo-Riemannian, then in a neighborhood of any point \(x_0\)
one
can, by appropriate linear combinations of any set of local coordinates, 
introduce  coordinates \(x^1,\dots,x^n\) such that
\begin{equation}\label{eq:pteta}%
<\ppv{x}{i}, \ppv{x}{j}>_{x_0}=\eta_{ij}
\end{equation}%
It should be emphasized that in general it is impossible to introduce
local coordinates so that (\ref{eq:pteta}) hold at all points in a
neighborhood of \(x_0\). The possibility of doing so depends on the
vanishing of the curvature of an appropriate connection.
In any case, consider the vectors fields \(v_i = \ppv{x}{i}\).
 In a small enough neighborhood of \(x_0\) one can apply the
usual Gram-Schmidt orthonormalization procedure to \(v_1,\dots,v_n\)
to obtain vector
fields \(e_1,\dots,e_n\) such that \(<e_i,e_j> = \eta_{ij}\). Such a
set of vector fields is called a (local) 
\index{nbein@\(n\)-bein}%
{\em \(n\)-bein\/}.

Let now \(\cF_{\O}(M)\)
\index{\(F_O(M)\)@\(\cF_{\O}(M)\)}%
be the {\em orthonormal frame-bundle\/}
\index{bundle!orthonormal frame}%
 of \(M\),
that is, the fiber at any point \(x\) is the set of all ordered
orthonormal bases
in \(T_xM\) with respect to the bilinear form \(g(x)\). This bundle is
trivialized over open sets \(U\) in \(M\) in which one has an
\(n\)-bein \(e^U_1,\dots,e^U_n\). Given an orthonormal basis
\(b_1,\dots,b_n\) at \(x\in U\) one has \(b_i = \sum_j e_j R^j{}_i\) for
some matrix \(R\in \O(r,s)\). The trivializing homeomorphism assigns
to the basis \(b\) the point \((x,R) \in U \times \O(r,s)\). If \(V\)
with \(n\)-bein \(e^V_1,\dots,e^V_n\) define now another trivialization,
then one has \(e^U_i = \sum_je^V_j G^j{}_i\) for some matrix \(G \in
\O(r,s)\)
and one can easily see that the transition map \(g_{VU}\) is given by
\(R \mapsto GR\), so \(G\) defines a transition map and the structure group
of the bundle is \(\O(r,s)\). In fact it is a {\em principal\/}
\(\O(r,s)\)-bundle as the global right action is easily defined by \(v_i
\mapsto \sum_j v_j L^j{}_i\) for \(L\in \O(r,s)\).

Using the pseudo-Riemannian structure we can reduce the structure group
of the tangent bundle from \(\GL(n)\) to \(\O(r,s)\). In fact, given an
open set \(U\) with \(n\)-bein \(e^U_1,\dots,e^U_n\) and a tangent
vector \(v\in T_xM\) for \(x\in U\) one has \(v = \sum_jv^j_Ue_j\). The
tangent bundle can now be trivialized by assigning to \(v\) the point
\((x, (v^1,\dots,v^n))\in U\times\lR^n\). As  \(v^j_V = L^j{}_iv^i_U\)
for some matrix \(L\in \O(r,s)\) the transition maps now belong to
\(\O(r,s)\) and we have expressed \(TM\) as a bundle with structure group
\(\O(r,s)\). This is an explicit example of the important process of
considering a given fiber bundle as having several structure groups.
We shall see  in subsequent sections how the reduction to \(\O(r,s)\)
allows for constructions, spin bundles in our case,
 that would be impossible with \(\GL(n)\) as the
structure group.

As before, introducing the two natural representations
\(\rho(L,z) = Lz\) and \(\rho^*(L,z) = (L^t)^{-1}z\)  of
of \(\O(r,s)\) on \(\lR^n\)
one  has the identifications:
\begin{eqnarray}\label{eq:fotm}
\cF_{\O}(M)\times_\rho {\lR}^n  &\simeq &  TM \\ \label{eq:focotm}
\cF_{\O}(M)\times_{\rho^*} {\lR}^n &\simeq &  T^*M
\end{eqnarray}%

Note that in the Riemannian case, the two
representations are identical which shows that the tangent and cotangent
bundles are then isomorphic \(\O(n)\)-bundles. Since one can introduce a 
Riemannian metric in any paracompact Hausdorff manifold, 
the tangent and cotangent bundles in these are always isomorphic 
vector bundles.

By a deliberate abuse of notation we shall also denote by
\(<\cdot,\cdot>\) the duality between \(T^*_xM \) and \(T_xM\), that is
if \(\alpha\in T^*_xM\) and \(v\in T_xM\) we shall write \(<\alpha,v>\)
for \(\alpha(v)\). We shall also write \(<v,\alpha>\) for the same
thing.

Given a pseudo-Riemannian structure \(g\) one can define a map
\(\ell:T_xM \to T^*_xM\)
\index{\(L\)@\(\ell\)}%
by the relation \(\ell(v)(w) = <v,w>\) for all
\(w\in T_xM\). By our abuse of notation this can be written as
\(<\ell(v), w> = <v,w>\). One also has the inverse map \(r:T^*_xM \to
T_xM\)
\index{\(R\)@\(r\)}%
defined by \(<r(\alpha),v> = \alpha(v)\) 
which by our abuse of
notation can also be expressed as \(<r(\alpha),v> = <\alpha,v>\).
Obviously \(r\) and \(\ell\) are inverses of each other. These maps can
be used to introduce a non-degenerate symmetric bilinear form in
each \(T^*_xM\) by the formula \(<\alpha,\beta> = <r(\alpha),r(\beta)>\).

It is useful to determine local expressions for these maps and bilinear
forms in any  trivialization of \(TM\) and \(T^*M\).  Such a 
trivialization is given by a set of vector fields \(e_1,\dots,e_n\) in an 
open set such that at any point \(x\) they form a basis for \(T_xM\). Let 
\(e^1,\dots,e^n\) be the corresponding dual set of \(1\)-forms. 

Define
\[
g_{ij} = <e_i,e_j>
\]
\index{\(G_{ij}\)@\(g_{ij}\)}%
If now \(v=\sum_i v^ie_i\) and \(w = \sum_iw^ie_i\) then
one has 
\[
<v,w> = \sum_{ij}g_{ij}v^iw^j
\]
 One will also have 
\[
r(e^i)
= \sum_j g^{ij}e_j
\]
 for some matrix \(g^{ij}\).
\index{\(G^{ij}\)@\(g^{ij}\)}%
From \(<r(e^i),e_j> = <e^i,e_j> = \delta^i{}_j\)
 one deduces \(\sum_k g^{ik}g_{kj} = \delta^i{}_j\)
so \(g^{ij}\) is the matrix inverse
of \(g_{ij}\). One has for \(\alpha = \sum_i\alpha_ie^i\) and \(\beta =
\sum_i\beta_ie^i\) that 
\[
<\alpha,\beta> =
\sum_{ij}g^{ij}\alpha_i\beta_j
\]
 Of course, by definition, 
\[
<\alpha,v> =
\sum_i\alpha_iv^i
\]

Interpreting the metric \(g\) as a section of \(T^*M\otimes T^*M\), one has
\(g=\sum_{ij}g_{ij}\,e^i\otimes e^j\). This is known as the 
{\em (pseudo)-metric tensor\/}. 
\index{tensor!metric}%
For the particular case of a coordinate basis \(e_i = \ppv{x}{i}\) 
of a set 
of local coordinates \(x^1,\dots,x^n\),
  one has \(e^i=dx^i\) and the \(g_{ij}\) are then the well-known usual
components of the  metric tensor
 \(g=\sum_{ij}g_{ij}\,dx^i\otimes dx^j\). For 
an \(n\)-bein \(g^{ij}=g_{ij}=\eta_{ij}\) 
and so \(g=\sum_{ij}\eta_{ij}\,e^i\otimes e^j\).

In the physical literature, \(r\)  is known as {\em
raising indices\/}.
\index{raising indices}%
 This is because if
\(\alpha=\sum_i\alpha_i\,dx^i\), then it is customary to write
\(r(\alpha) = \sum_i\alpha^i\ppv{x}{i}\) and so the index of
\(\alpha_i\) was ``raised". Similarly \(\ell\) is known as 
\index{lowering indices}%
{\em lowering
indices\/}.

Let \(S\in \GL(n)\), then for any integer \(n\) and any real number
\(s\) the maps \(\nu_n : S \mapsto \det (S)^{-n}\) and \(\mu_s:S
\mapsto |\det (S)|^{-s}\) define one-dimensional representations of
\(\GL(n)\). 
Remember (\ref{eq:jacocy}), the transition map given by the
Jacobian matrix of change of local coordinates.
A line bundle with fiber \(\lR\) using the transition map
\(J\) of local coordinate changes and representation \(\nu_n\) is called
the {\em bundle of signed
densities of weight \(n\)\/},
\index{bundle!signed density}%
\index{density!signed}%
 and representation \(\mu_s\), the {\em
bundle of (absolute) densities of weight \(s\)\/}.
\index{bundle!absolute density}%
\index{density!absolute}%

For a bundle
trivialized in local chart \(U\), a density \(D\) is represented by
a function \(D_U(x)\). One has in the intersection of charts \(U\)
and \(V\), for a signed density \(D_V =\det (J)^{-n}D_U\),
and for an absolute  density \(D_V =|\det (J)|^{-s}D_U\),
where  \(n\) and \(s\) are the respective weights. 

Consider now an absolute density \(D\) of weight \(1\) and its local
representatives \(D_U\) and \(D_V\) where \(U\) carries local
coordinates \(x^1,\dots,x^n\) and \(V\) coordinates \(y^1,\dots,y^n\).
For an open set \(W\subset U\cap V\) one has
\begin{eqnarray*}\lefteqn{\int\cdots\int_{y(W)} D_V(y)\, dy^1\cdots dy^n =} & & \\
\lefteqn{\int\cdots \int_{x(W)}
|\det(J)|^{-1}D_U(x)|\det (J)|\,dx^1\cdots dx^n =}& & \\
 & & \int\cdots\int_{x(W)}D_U(x)\,dx^1\cdots dx^n
\end{eqnarray*}
In the second term the factor \(|\det(J)|^{-1}\) arises due to change of
trivialization and the factor \(|\det (J)|\) due to the formula
for change of variables in integration; the two factors cancel.
This means that absolute densities of weight \(1\) can be integrated
over the manifold using local coordinates and a partition of unity:
\[
\int_M D = \sum_{U\in \cU}\int_{x_U(U)} \xi_U  D_U(x_U)\,dx_U
\]
where  \(x_U\) are local coordinates in \(U\) and the \(\xi_U\) form a
partition of unity subordinate to an atlas \(\cU\). We see thus that such a 
density defines a signed measure on \(M\) and given any function \(f\) on 
\(M\) its integral with respect to this measure is given by \(\int_M fD\).

An \(n\)-form \(\Omega\in \bigwedge^n(T^*M)\) has a representative in local
coordinates as \(D_U(x)\, dx^1\wedge\cdots\wedge dx^n\) and one sees that
the set of local coefficient \(D_U\) define a {\em signed\/}
density of weight \(1\).
This can be used to define
 an absolute density of weight \(1\) if the atlas of local
charts can be so chosen as to have \(\det (J) > 0\) for all pairs
of intersecting charts. When this is possible we say the manifold is
{\em orientable\/}, 
\index{manifold!orientable}%
and a choice of an atlas in which \(\det (J)
> 0\) is called a choice of {\em orientation\/}.
\index{orientation}%
 Once such an atlas is
chosen we say the manifold is {\em oriented\/}.
\index{manifold!oriented}%
In this case we can integrate \(n\)-forms on \(M\).

On an orientable manifold the frame-bundle \(\cF(M)\) is the disjoint
union of two bundles that correspond to frames of a fixed orientation.
Likewise the orthogonal frame bundle \(\cF_{\O}(M)\) separates into a
disjoint union corresponding to orthonormal frames of a fixed
orientation.  We shall denote any one  of these sub-bundles 
by \(\cF_{\SO}(M)\)
\index{\(F_{SO}(M)\)@\(\cF_{\SO}(M)\)}%
which is obviously a principal \(\SO(r,s)\)-bundle. The tangent and
cotangent bundles can now be considered as associated to \(\cF_{\SO}(M)\)
\begin{eqnarray*}%
\cF_{\SO}(M)\times_\rho {\lR}^n  &\simeq &  TM \\
\cF_{\SO}(M)\times_{\rho^*} {\lR}^n &\simeq &  T^*M
\end{eqnarray*}%
by restricting the  the previously used 
representations in (\ref{eq:fotm}, \ref{eq:focotm}) to \(\SO(r,s)\).

Let now  \(M\) be a pseudo-Riemannian manifold. In a
local
chart \(U\) one has the local 
representative \(\sum_{ij}g_{ij}^U\,dx^i\wedge dx^j\) of 
the metric tensor as 
defined above. From
its
definition, one easily deduces
\(\sum_{k\ell}g^V_{k\ell} J^k{}iJ^\ell{}_j = g^U_{ij}\) which means that
\(\sqrt{|\det(g^U)|} = \sqrt{|\det (g^U_{ij})|}\) is an absolute density of
weight \(1\) called the {\em volume element\/}.
\index{volume element}%
  If \(M\) is oriented, then 
\[%
\Omega = \sqrt{|\det (g^U)|}\,  dx^1\wedge\cdots\wedge dx^n
\]%
defines an \(n\)-form called the {\em volume \(n\)-form\/}.
\index{volume \(n\)-form}%

The pseudo-metric \(g\) defines a symmetric bilinear non-degenerate form
in the space of exterior 
\(p\)-covectors \(\bigwedge^p(T_x^*M)\) at a point \(x\in M\) by
the formula
\[%
<\alpha_1\wedge\cdots\wedge\alpha_p , \beta_1\wedge\cdots\wedge\beta_p> =
\det _{ij}<\alpha_i,\beta_j>
\]%
where the right-hand side is the determinant of the matrix
\(<\alpha_i,\beta_j>\).
This in turn defines the {\em Hodge star\/}
\index{Hodge star}%
 operator
\[
*:{\textstyle \bigwedge^p}(T_x^*M) \to {\textstyle
\bigwedge^{n-p}}(T_x^*M)
\]
by the relation
\begin{equation}\label{eq:hodgestar}%
\phi \wedge *\psi = <\phi,\psi>\Omega
\end{equation}%
for any two  exterior \(p\)-forms \(\phi\) and \(\psi\). One obviously has
\(*1=\Omega\).
One can get a concrete idea of the Hodge operator by considering an 
orthonormal basis 
 \(e_1,\dots,e_n\) of the tangent space at a point. Let \(e^a = r(e_a)\), 
be the corresponding covectors.
One has \(<e^a,e^b> = \eta^{ab}\) where \(\eta^{ab}\) is the inverse
matrix of \(\eta_{ab}\), obviously equal to it, introduced for
notational convenience. One has \(dx^i = \sum_ah^i{}_ae^a\) for some
matrix of coefficients \(h^i{}_a\). Thus \(\Omega =
\sqrt{|\det(g)|}\det(h)e^1\wedge\cdots\wedge e^n\). On the other hand,
\(g^{ij} = \sum_{ab} h^i{}_ah^j{}_b\eta^{ab}\) from which one deduces
\(\det(g)^{-1}=\det(h)^2 \det(\eta)\). If we denote by \(\sigma_h\) the
sign of \(\det(h)\) one concludes \(\Omega = \sigma_he^1\wedge
\cdots\wedge e^n\). One can now easily find that
\begin{equation}%
*(e^{i_1}\wedge\cdots\wedge e^{i_p})   =
\sigma_{IJ}\sigma_h({\textstyle \prod_{a=1}^p}\eta^{i_ai_a}) \,
e^{j_1}\wedge\cdots\wedge e^{j_{n-p}}
\end{equation}%
where \(\{j_1,\dots,j_{n-p}\}\) is the complementary subset,
in ascending order, to
\(\{i_1,\dots,i_{p}\}\)  in \(\{1,\dots,n\}\) and \(\sigma_{IJ}\) is
the
parity of the permutation
\[
(1,\dots,n) \mapsto
(i_1,\dots,i_p,j_1,\dots,j_{n-p})
\]
A double application of this formula gives 
\begin{equation}\label{eq:twostar}%
**\psi =
(-1)^{s+p(n-p)}\psi
\end{equation}%

If \(n\) is even, \(n=2m\), and \(\psi\) is an \(m\)-form, then \(*\psi\)
is again an \(m\)-form. An \(m\)-form \(\psi\) such that \(*\psi=\psi\)
is said to be {\em self-dual\/}, 
\index{form!self-dual}%
and one such that  \(*\psi=-\psi\)
is said to be {\em anti-self-dual\/}. 
\index{form!anti-self-dual}%
If \((-1)^{s+m}=1\), then
from (\ref{eq:twostar}) one has \(**\psi=\psi\) for any \(m\)-form. In
this case we can decompose any \(m\)-form into a sum of a self-dual and
an anti-self-dual form. Define
\[%
\psi^\pm = \frac{\psi\pm *\psi}{2}
\]%
then one sees that \(\psi=\psi^++\psi^-\) and \(*\psi^\pm=\pm\psi^\pm\).
\index{\(\xi^+\)@\(\psi^+\)}%
\index{\(\xi^-\)@\(\psi^-\)}%

Using the Hodge star one can define the differential operator \(\delta =
* d *\)
\index{\(\delta\)}%
 where \(d\) is the exterior derivative. One sees that \(\delta\) 
transforms a
\(p\)-form into a \(p-1\) form. From (\ref{eq:twostar}) one sees
that \(\delta^2=0\). The operator \(d\delta + \delta d\)
is a second order operator on the space of \(p\)-forms and in the
Riemannian case is knows as the {\em Hodge Laplacian\/}.

\subsection{The Levi-Civita Connection}
\begin{theorem}%
On any pseudo-Riemannian manifold \(M\) there is a unique invariant
connection on the orthogonal frame bundle \(\cF_{\O}(M)\) called the
Levi-Civita connection which is  characterized by the following 
properties of the
associated connection \(\nabla\) on the tangent bundle \(T(M)\):
\begin{eqnarray}\label{eq:notorsion}%
\nabla_\cX\cY-\nabla_\cY\cX &=& [\cX,\cY]\\ \label{eq:delgzero}
\cX (g(\cY,\cZ)) &=& g(\nabla_\cX\cY,\cZ) + g(\cY,\nabla_\cX\cZ)
\end{eqnarray}%
\end{theorem}%
Of these, condition (\ref{eq:notorsion}) says that  the Levi-Civita
connection is torsion-free, and condition (\ref{eq:delgzero}) says that the
covariant derivative of the pseudo-metric tensor vanishes. 
Viewing \(g\) as a section of \(T^*(M)\otimes T^*(M)\),  condition
(\ref{eq:delgzero}) is equivalent to \(\nabla g=0\) (see Section
\ref{sec:covdev}).

\begin{proof}Let \(e_1,\dots,e_n\) be vector fields in an 
open set such that at any point \(x\) they form a basis for \(T_xM\). Use 
this basis to locally trivialize \(TM\).  One has \(\nabla_{e_j}e_k = 
\sum_\ell \Gamma^\ell{}_{kj}e_\ell\) and \([e_j,e_k]=\sum_\ell 
C^\ell{}_{jk}e_\ell\) for some functions \(C^\ell{}_{jk}\).  Condition 
(\ref{eq:notorsion}) now means that
\[ 
\Gamma^\ell{}_{kj}-\Gamma^\ell{}_{jk}=C^\ell{}_{jk}
\]%
 Let now \(\Gamma_{ijk}=\sum_\ell g_{i\ell}\Gamma^\ell{}_{jk}\) and  
\(C_{ijk} = \sum_\ell g_{i\ell}C^\ell{}_{jk}\) then condition 
(\ref{eq:delgzero}) becomes
\begin{equation}\label{eq:protogam}%
e_k(g_{ij})=\Gamma_{jki} + \Gamma_{ikj} =  \Gamma_{jki} + 
\Gamma_{ijk} + C_{ijk}
\end{equation}%

Refer to equation (\ref{eq:protogam}) as \(E(ijk)\) and 
now add \(E(ijk)\) to 
\(E(jki)\) and subtract \(E(kij)\). One arrives at
\[
2\Gamma_{jki} = e_k(g_{ij}) + e_i(g_{jk})-e_j(g_{ik})+C_{kij}-C_{ijk}-
C_{jki}
\]
and finally
\begin{equation}\label{eq:levicivitagamma}%
\Gamma^i{}_{jk} = \frac{1}{2} \sum_\ell 
g^{i\ell}(e_j(g_{k\ell})+e_k(g_{j\ell})-e_\ell (g_{jk})+C_{jk\ell}-C_{k\ell 
j}-C_{\ell jk})
\end{equation}%
In  a coordinate basis, \(e_i=\ppv{x}{i}\),  one has 
\(C^\ell{}_{ij}=0\) and we arrive at the classic expression:
\begin{equation}\label{eq:levicivitacoord}%
\Gamma^i{}_{jk} = \frac{1}{2} \sum_\ell g^{i\ell}\left(\frac{\partial 
g_{k\ell}}{\partial x^j}+ \frac{\partial g_{j\ell}}{\partial x^k}-
\frac{\partial g_{jk}}{\partial x^\ell}\right)
\end{equation}%
For an \(n\)-bein \(e_i\) one has \(g_{ij}=\eta_{ij}\) 
thus \(e_k(g_{ij})=0\) 
and so 
\begin{equation}\label{eq:levicivitanbein}%
\Gamma^i{}_{jk} = \frac{1}{2} \sum_\ell \eta^{i\ell}
(C_{jk\ell}-C_{k\ell j}-
C_{\ell jk})
\end{equation}%

Thus, the connection, if it exists, is uniquely determined. 
To show 
existence it is necessary to show that (\ref{eq:levicivitagamma}), or 
alternatively the specialized form (\ref{eq:levicivitacoord}) or 
(\ref{eq:levicivitanbein}), does indeed define a connection. One way to do 
this is to show that \(\Gamma\) has the proper transformation law 
(\ref{eq:gammatrans}) under change of trivialization, which is 
straightforward if not tedious. One could also argue that as conditions 
(\ref{eq:notorsion})and  (\ref{eq:delgzero}) are 
trivialization-independent, 
and as they can be met in a unique way in any trivialization,  \(\Gamma\) 
cannot help but transform in the right way under a 
change of trivialization.
\end{proof}%

The curvature \(2\)-form associated to the Levi-Civita connection is thus 
an anti-symmetric map \(T^*M\otimes T^*M \to \End(TM)\) and so can be 
interpreted as a tensor in \(TM\otimes T^*M\otimes T^*M \otimes T^*M \). As 
such it is known as  
the {\em Riemann curvature tensor\/}.
\index{curvature!Riemann}%
 \index{tensor!Riemann curvature}%

\section{Lagrangian Theories}%
\subsection{Lagrangians}\label{sec:lagrange}%
Physical field theory is preponderately
Lagrangian theory. In its simplest terms one deals with a set of
fields \(\psi^1,\dots,\psi^m\) which for now we simply consider as
functions on a Euclidean space \(\lR^n\). For 
\(\alpha=(\alpha_1,\dots,\alpha_n)\) with \(\alpha_i\ge 0\) being
integers, denote by \(|\alpha|\) the sum 
\(\alpha_1+\cdots+\alpha_n\) and by \(\partial_\alpha\) 
the partial derivative
\[
\frac{\partial^{|\alpha|}}
{\partial x_1{}^{\alpha_1}\cdots \partial x_n{}^{\alpha_n}}
\]
 A {\em \(k\)-order Lagrangian density\/}
\index{density!Lagrangian}%
\index{Lagrangian}%
(or simply {\em Lagrangian\/}) in this
context is
just a function
\(\cL(x,\partial_\alpha\psi^a(x))\) of the point
\(x\in \lR^n\) and of the values of the fields and their partial
derivatives at \(x\) up to order \(k\) (\(\alpha=0\) corresponds to
just the field values). 
Associated to a Lagrangian  density and the fields
\(\psi^a\) is the {\em action\/}
\index{action}%
\begin{equation}\label{eq:action}%
S(\psi) = \int_{\lR^n} \cL(x,\partial_\alpha\psi^a(x))\,dx
\end{equation}%
which defines a functional on the set of fields.
The word ``action" here is borrowed from  physics where it means a quantity
which has the physical dimension of energy times time. It should not be
confused with the use we make elsewhere in these notes to mean group
action or its analogs.
The Lagrangian defines a set of differential equations that constitute
the physical laws obeyed  by the fields. These equations express the
property of the fields being a singular point of the action.  Let
\(\gamma^a\) be a set of functions of compact support and consider the
action \(S(\psi+r\gamma)\). The requirement that \(\psi\) be a singular
point of \(S\) then translates to the requirement that for all such 
\(\gamma\)
\[%
\left.\frac{d}{dr}S(\psi+r\gamma)\right|_{r=0} = 0
\]

One has
\[
\left.\frac{d}{dr}S(\psi+r\gamma)\right|_{r=0} =
\int_{\lR^n} 
\sum_a\sum_\alpha(-1)^{|\alpha|}\partial_\alpha
\frac{\partial\cL}{\partial y^a_\alpha} \partial_\alpha
\gamma^a\,dx
\]
where by  \(\frac{\partial\cL}{\partial y^a_\alpha}\) 
 we mean the partial derivative of
\(\cL\) in relation to the variable for which one substitutes 
\(\partial_\alpha\psi^a\)
to form the integrand of the action.
After integration by parts, the right-hand side becomes
\[
\int_{\lR^n}\sum_a
\sum_\alpha(-1)^{|\alpha|}\partial_\alpha
\left(\frac{\partial\cL}{\partial y^a_\alpha}\right)
\gamma^a\,dx
\]
As this must be true for all \(\gamma\) with compact support one deduces
that \(\psi\) must satisfy the {\em Euler-Lagrange equations\/} 
(one for each \(a\)):
\index{equation!Euler-Lagrange}%
\[%
\sum_\alpha(-1)^{|\alpha|}\partial_\alpha\left(\frac{\partial\cL}
{\partial y^a_\alpha}
(x,\partial_\beta\psi^b(x))\right)
=0
\]%
This is often abbreviated to
the strange-looking expression
\[%
\sum_\alpha(-1)^{|\alpha|}\partial_\alpha
\frac{\partial\cL}{\partial\left(\partial_\alpha \psi^a\right)} =0
\]%
When one of the variables \(x^i\) corresponds to time, these are the
dynamical equation of a physical theory described by \(\cL\). That all
fundamental physical theories are described by Lagrangians is a
remarkable fact, which is not truly understood. Although quantum theory
sheds some light on this, as the action itself and not just its critical
points has physical meaning, we shall simply accept this fact in these
notes.

Once it is realized that physical fields should really be thought of
as sections of fiber bundles,
 (\ref{eq:action}) has to be modified. In
first place, \(\lR^n\) should be replaced by a manifold \(M\) (such as 
physical
space-time), and in
second place, the \(\psi^a\) ought to be considered local representatives
of fiber-bundle sections in appropriate trivializations. One should then
replace (\ref{eq:action}) by
\begin{equation}\label{eq:newaction}%
S(\psi) =\sum_{U\in\cU}\int_U\xi_U
\cL_U(x_U,\psi_U{}^a(x_U),\psi_U{}^a_i(x_U))\,dx_U
\end{equation}%
where \(\cU\) is an atlas for \(M\) of  coordinate charts over
which the bundles in question trivialize, \(x_U\) are local coordinates
in \(U\),  \(\psi_U\) are representatives of the sections, and \(\xi_U\)
is a partition of unity subordinate to \(\cU\). The \(\cL_U\) are local
representatives of what would be the Lagrangian understood in global
form. For this to make sense, the resulting action \(S\) must not depend
on the details of the construction in (\ref{eq:newaction}), that is, on
the choice of local coordinates and trivializations. While it is easy to
restate the definition of a Lagrangian theory so that this is automatic,
it is by no means entirely straightforward to provide examples, and even
less so of describing all possibilities.

 Most physical theories are defined by first order Lagrangians, though
higher order ones do occur.

Let \(\pi:E\to M\) be any bundle with fiber \(F\)
(we are in the category of manifolds) and
consider at any point \(x\in M\) the following relation on
sections of \(E\). Let \(f:E \to \lR\) be any \(\cC^\infty\) function
and let \(\cX_1,\dots,\cX_k\) be any \(k\) vector fields on \(M\). We
say two sections \(\sigma_1\) and \(\sigma_2\) have a {\em contact of
order \(k\)\/}
\index{contact of order \(k\)}%
 at \(x\) if for all such \(f\) and all such \(\cX_i\) one
has for all \(0\leq \ell \leq k\)
\[%
(\cX_1\cX_2\cdots\cX_\ell f\circ\sigma_1)(x) = (\cX_1\cX_2\cdots\cX_\ell
f\circ\sigma_2)(x)
\]%
Having contact of order \(k\) is obviously an equivalence relation.
It is easy to see that in a trivialization \(U\times F\) with local
coordinates \(x^1,\dots,x^n\) in \(U\) and \(y^1,\dots,y^m\) in a local
chart \(W\) in \(F\) that if a section \(\sigma\) in a neighborhood
of \(x\) is given by \(y^a = s^a(x)\), then \(\sigma_1\) and
\(\sigma_2\) have contact of order \(k\) at \(x\) if and only if \(s_1^a\) 
and
\(s_2^a\) coincide at \(x\) along with all partial derivatives up to and
including order \(k\). By the {\em \(k\)-jet\/}
\index{kjet@\(k\)-jet}%
 of \(\sigma\) at \(x\)
we mean the equivalence class of \(\sigma\) under the relation of having
contact of order \(k\). The {\em \(k\)-jet bundle\/}
\index{bundle!kjet@\(k\)-jet}%
  \(J^k(E)\) is
the bundle of \(k\)-jets of sections of \(E\) at all point \(x\in M\).
This is a manifold for
which local coordinates are defined in terms of the coordinates \(x^i\)
and \(y^a\) introduced above, where to the equivalence class
\([\sigma]_x\) we associate the coordinates
\[
\left(x^1,\dots,x^n,s^1(x),\dots,s^m(x),\frac{\partial s^1(x)}{\partial
x^1},\dots,\frac{\partial^k s^m(x)}{\partial (x^n)^k}\right)
\]
where one has to include all the partial derivatives up to order \(k\)
of all the components \(s^a\). The rather complex 
exact expression for the transition
maps of this bundle will not be important for us, it can be calculated
from the chain rule.
Given a section \(\sigma\) of \(E\) we define its {\em \(k\)-th jet
extension\/}
\index{jet extension}%
 by \(j^k\sigma(x) = [\sigma]_x\) which is a section of
\(J^k(E)\).

Lagrangians as we have defined them are functions of the fields and
their derivatives, so they are naturally defined 
on \(J^k(E)\)
which we see incorporates in a global manner sections and their 
derivatives up to order \(k\). One must, 
to be able to calculate the action, integrate the
Lagrangian density. This will be possible if we define the Lagrangian
density as a bundle map
\[%
\cL: J^k(E) \to \cV(M)
\]%
where \(\cV(M)\)  is the bundle of volume densities, that is, absolute
densities of weight
\(1\). If \(M\) is oriented, this bundle can be replaced by
\(\bigwedge^n(T^*M)\), the bundle of \(n\)-forms on \(M\).  The
action of a section \(\sigma\) is now defined as
\begin{equation}%
S(\sigma) = \int_M \cL\circ j^k\sigma
\end{equation}%
The action is a functional on \(\Gamma(M)\), the set of smooth sections
of \(E\). In most cases of interest, this set is formally an infinite
dimensional differential
manifold.  Singular point of \(S\) on this manifold correspond to
solutions of the Euler-Lagrange differential equations
whose local form can be calculated as before.

Gauge theory is a Lagrangian theory of connections on principal fiber
bundles along with sections of vector bundles, invariant under action of
the gauge group. To be able to talk about a Lagrangian defined for a
connection on a principal bundle we must see how a connection can be
seen as a section of a bundle. To this end remember formula
(\ref{eq:locg}) giving the relation between the local principal gauge
potentials in two trivializations. We interpret this now as expressing the
transition map of a fiber bundle. If we trivialize in a local chart one
has  \(A  = \sum_iA_i^U\,dx^i\) where \(A_i^U \in \gg\). If \(G\) is a 
matrix group, then the \(A_i\) are matrices, known in the physical
literature  as {\em gauge potentials\/}. 
\index{gauge potential}%
If we
take the \(A_i^U\) as local representatives of \(\omega\) we see that
our bundle must have fiber \(\gg^n\). One has from (\ref{eq:locg}) that
\(A_i^V = \sum_j J^*_i{}^j (g_{VU}\cdot A_j^U\cdot
g_{VU}^{-1} + \gamma_j)\) 
where \(\gamma_i\in\gg\) are defined by \(-dg_{VU}\cdot
g_{VU}^{-1} = \sum_i \gamma_i\, dx^i\), and \(J^*_i{}^j\) is the matrix of
the adjoint action of 
\(\GL(n)\) corresponding to the 
cocycle of change of local coordinates (\ref{eq:jacocy}), that is,
the inverse of the transpose of \(J\).
This is a transition map of an affine 
bundle with the following data: The group \(\tilde G\) of the bundle as
a set is
\(\GL(n) \times G \times \gg^n\). The group multiplication law is
\[
(J,g,L_1,\dots,L_n)(J',g',L_1',\dots,L_n') = (JJ', gg', \Ad_g
L_1'+L_1,\dots,\Ad_gL_n'+L_n)
\]
The action of \(\tilde G\) on the fiber
is given by 
\[
(J,g,L_1,\dots,L_n) \cdot (K_1,\dots,K_n) =
(K_1',\dots,K_n')
\]
 where 
\[
K_i' = \sum_j J^*_i{}^j(\Ad_gK_j + L_j)
\]
 and
the transition map of the bundle is \(\tilde g_{VU} = (J_{VU}, g_{VU},
\gamma_1,\dots,\gamma_n)\). The bundle so constructed is called the {\em
connection bundle\/} 
\index{bundle!connection}%
of \(PG\). With this construction, invariant
connections on \(PG\) correspond to sections of the connection bundle.
One can now introduce Lagrangian densities of invariant connections.

A gauge transformation \(\phi:PG \to PG\) induces a transformation on an
invariant connection \(\omega \to \phi\cdot \omega\) and consequently a
transformation of a section \(A\) of the connection bundle given
essentially
by (\ref{eq:lgtra}).
We say a Lagrangian theory for a gauge potential is {\em gauge
invariant\/}
\index{action!gauge invariant}%
 if the action satisfies \(S(\phi\cdot A) = S(A)\) for all
gauge transformations \(\phi\).  Often the Lagrangian density itself is
gauge invariant which guarantees the invariance of the action. In this
case, as the transition map relating two trivializations
(\ref{eq:locg})
is also in the
form of a gauge transformation, the Lagrangian density automatically
takes care of the requirement that the action not depend on the choice
of trivialization in the construct (\ref{eq:newaction}) above.

As a first example, let \(M\) be any oriented pseudo-Riemannian 
manifold with a principal
\(G\)-bundle. Let \(A\) be a section of the connection bundle and \(F
\) the local representative of
the corresponding curvature two-form. The {\em Yang-Mills\/}
\index{Yang-Mills}%
 Lagrangian
density
for a principal \(G\)-bundle of a matrix group \(G\) is given by
\begin{equation}\label{eq:YM}%
\cL_{YM} = - k\Tr(F\wedge *F)
\end{equation}%
where \(k\) is a conventional constant, and 
\(*\) is the Hodge star (see (\ref{eq:hodgestar})). 
This equation has to be well interpret as \(F\) has values in \(\gg\)
which is a linear space of matrices. Given a \(\gg\)-valued \(p\)-form
\(\alpha\) and \(q\)-form \(\beta\), each one is a matrix
\(\alpha^a{}_b\) and \(\beta^a{}_b\) of ordinary \(p\)- and \(q\)-forms
respectively.  What is meant by \(\alpha \wedge
\beta\) is the matrix \(\gamma^a{}_b \) of \(p+q\)-forms where
\(\gamma^a{}_b= \sum_c \alpha^a{}_c\wedge\beta^c{}_b \).
Note that because of the trace and formulas (\ref{eq:locg}) and
(\ref{eq:cug}), the resulting \(n\)-form is well defined over the
manifold and is gauge invariant.
We thus have a gauge-invariant theory, called the Yang-Mills
theory.
Recall that the curvature \(2\)-form can be interpreted as a 
\(P \gg\)-valued \(2\)-form. The Euler-Lagrange equations for the 
Yang-Mills theory are \(d_A*F=0\) where \(d_A\) is the covariant
exterior derivative. Along with the Bianchi Identities (\ref{eq:bianchi}), 
the Yang-Mills
\(2\)-form \(F\) therefore satisfies

\begin{eqnarray}\label{eq:dgf}
d_A F & = & 0 \\ \label{eq:deltagf}
d_A*F & = & 0
\end{eqnarray}
Note that if \(A\) leads to a self-dual or anti-self-dual curvature,
 \(F = \pm *F\),  then by the
Bianchi Identities, which always hold,
the Yang-Mills equations are automatically satisfied.

Another much studied example is the Chern-Simons theory. Let \(M\) be an 
oriented
\(3\)-manifold and consider a principal bundle of a matrix group \(G\).
The Chern-Simons Lagrangian is
\begin{equation}\label{eq:chernsimons}%
\cL_{CS} = k\,\Tr\,(A \wedge dA + \frac{2}{3}A\wedge A
\wedge A)
\end{equation}%
where \(k\) is a constant. 
The Euler-Lagrange equations are \(F=0\), that
is, the connection must be flat.  
The transformation properties of this Lagrangian under gauge transformations 
of \(A\) are a lot more sophisticated than for the  Yang-Mills theory since 
\(A\)
transforms in a more complicated fashion than \(F\). In fact it is not even 
immediately clear that the action is well defined, as 
(\ref{eq:chernsimons}) 
is not invariant under gauge transformations, so one cannot join
overlapping 
trivializations as is done in the Yang-Mills case using the 
construction of (\ref{eq:newaction}).  However for some 
groups, such as 
\(\SU(n)\), the principal bundles are trivial and one can use 
one global trivialization. The gauge invariance continues to be subtle.
Under a gauge transformation defined  by the function \(\phi:M\to G\), 
(\ref{eq:chernsimons})  by
\[%
\cL_{WZ}(\phi)=\frac{k}{3}\,\Tr\,(d\phi\,\phi^{-1}\wedge 
d\phi\,\phi^{-1}\wedge d\phi\,\phi^{-1})
\]%
which is known as the {\em Wess-Zumino\/}
\index{Wess-Zumino}%
term. 
If \(\phi_s\) is a smooth curve in the gauge
group (in the sense that \((x,s)\mapsto \phi_s(x)\) is \(\cC^\infty\)), 
then an easy calculation shows that the derivative of \(\cL_{WZ}(\phi_s)\) 
with respect to \(s\) is an exact differential. 
Thus on a manifold without boundary, 
the Chern-Simons action 
is invariant under 
those transformations that can be joined to the identity by a 
smooth curve in 
the gauge group (in fact for those in the component of the identity), 
but is  not generally  invariant. This subtle 
property is part 
of the reason for the great interest in this theory.

\subsection{Minimal Coupling}\label{sec:mincoup}%
For simplicity consider initially a system of first-order partial
differential equations on \(\lR^n\) for a set of \(m\) real functions
\(\psi^a\), \(a=1,\dots,m\) given by
\begin{equation}\label{eq:bspin}%
\sum_{i=1}^n L_i\frac{\partial\psi}{\partial x^i} +M\psi = 0
\end{equation}%
where the \(L_i\) and \(M\) are \(m\times m\) constant matrices.
There is no loss of generality in supposing real functions, for if the
\(\psi\) were complex, one could consider the \(2m\) real functions
comprised of their real and imaginary parts, and these would satisfy
a system of equations of the same form as (\ref{eq:bspin}). Suppose
furthermore that there is a matrix group \(G\) of \(m\times m\) matrices
that commute with the \(L_i\) and \(M\). If now \(\psi\) is a solution
of (\ref{eq:bspin}) and \(\Lambda \in G\) then \(\Lambda\psi\) is also a
solution and we say that \(G\) is a {\em global\/}
\index{group!global symmetry}%
 symmetry group of the
system. By ``global" one means in this case that the same matrix
\(\Lambda\) is applied to \(\psi(x)\) for all points \(x\).  If however
now we apply a \(G\)-valued function \(\Lambda(x)\) to \(\psi(x)\) and
consider the new function \(\tilde\psi(x)=\Lambda(x)\psi(x)\), then
\(\tilde\psi\) is not a solution of the new system. One however
easily shows  that \(\tilde\psi\) satisfies the equation
\[%
\sum_{i=1}^n L_i(\ppv{x}{i}
-\frac{\partial\Lambda}{\partial x^i}\Lambda^{-1})\tilde\psi
+M\tilde\psi = 0
\]%
We see therefore that the group of {\em local\/} 
\index{group!local symmetry}%
transformations
\(\psi(x)\mapsto\Lambda(x)\psi(x)\) is not a symmetry group of the system
as the system itself changes under the transformation. One could however
enlarge the system to make the local transformations symmetries. One
interprets the term \(-\frac{\partial\Lambda}{\partial x^j}\Lambda^{-1}\) as
representing the change under the action of the local  group
of an additional
set of functions represented by an  \(n\)-tuple  of \(m\times m\)
matrices \(A_i(x)\). Rewrite now the original equation as
\begin{equation}\label{eq:agspin}%
\sum_{i=1}^n L_i(\ppv{x}{i}+A_i)\psi +M\psi = 0
\end{equation}%
and extend the local group  to now act on both \(\psi\) and \(A\) as
\begin{eqnarray}%
\psi(x) &\mapsto & \Lambda(x)\psi(x)\\ \label{eq:urgauge}
A_i(x) &\mapsto &\Lambda(x) A_i(x) \Lambda(x)^{-1} -
\frac{\partial\Lambda(x)}{\partial x^i}\Lambda(x)^{-1}
\end{eqnarray}%

Comparing (\ref{eq:urgauge}) to (\ref{eq:lgtra})
one sees that it is of the same form as the change in a gauge potential
due to the action of a gauge group. Gauge transformations thus arise
naturally if one tries to make a global symmetry group local.
Furthermore, comparing the operator \(\ppv{x}{i}+
A_i\) to (\ref{eq:lincd}) one sees that it is of the form of a
covariant derivative with respect to the connection whose gauge
potential is \(A\).
The change in going from (\ref{eq:bspin}) to
(\ref{eq:agspin}) is precisely that of replacing ordinary partial
derivatives with the corresponding covariant derivatives.
The new set of functions comprised of \(\psi\) and \(A\)
is however not satisfactory
from various aspects. Whereas the \(\psi\) are governed by a system of
differential equations, the \(A\) are not. From a physical point of view
this not natural as all physical fields are to be thought of as
fundamentally dynamical objects.  What is lacking is thus a system of
differential equations that would govern \(A\). Since on \(A\) the local
group acts in the same way as a gauge transformation, it is natural to
posit that \(A\) is in fact a local gauge potential and that that it is
governed by a gauge-invariant system of differential equations.
One saw in Section \ref{sec:lagrange}
that Lagrangian gauge theories provide such equations.  Particular
cases of equation
(\ref{eq:bspin}) can likewise be obtained from a Lagrangian. Suppose
there is an invertible \(m\times m\) matrix \(R\) such that \(RL_i=-
L_i^tR^t\) and \(RM=M^tR^t\), and define \(\bar\psi = \psi^tR\). The
Lagrangian density
\begin{equation}\label{eq:psilag}
\cL_\psi=\bar\psi(\sum_iL_i\ppv{x}{i}+M)\psi
\end{equation}%
is easily shown to have (\ref{eq:bspin}) as the
Euler-Lagrange equations. Many important physical equations of free 
non-interacting  particles exemplify this particular case.
Let now
\(\cL_A\) be a gauge-invariant Lagrangian density for a gauge-invariant
theory for the local potential \(A\) of a principal \(G\)-bundle for the
\(m\times m\) matrix group \(G\). The sum of the two, \(\cL = \cL_\psi +
\cL_A\) gives rise to the system of Euler-Lagrange equations consisting
of the independent systems (\ref{eq:bspin}) and the Euler-Lagrange
equations of \(\cL_A\). Let now \(\cL_\psi'\) be the result of replacing
in \(\cL_\psi\) the partial derivative \(\ppv{x}{i}\)
by the covariant derivative \(\nabla_i=\ppv{x}{i}+
A_i\), then \(\cL_{(\psi,A)}=\cL_\psi' +
\cL_A\) has as its Euler-Lagrange system one that
now couples the fields \(\psi\) and \(A\). 
For instance, if \(\cL_A\) is the Yang-Mills Lagrangian, then 
(\ref{eq:deltagf}) changes to 
\begin{equation}\label{eq:delgfj}
*d_A*F = j
\end{equation}
where \(j\) is a \(1\)-form called the {\em current \(1\)-form\/}
constructed from the \(\psi\) field (\(j\) is proportional to 
\(\sum_i\bar\psi L_i\psi\,dx^i\) for our example (\ref{eq:psilag})). 
This process is called {\em
minimal coupling\/}
\index{minimal coupling}%
 and is the method used in physical theories to pass
from a set of equations such as (\ref{eq:bspin}) representing 
non-interacting free fields to coupled interacting fields whose interaction
is mediated by local gauge potentials as dynamical objects. In this
sense physical theories
such as the standard model of elementary
particles
 are constructed from practically nothing, 
at least as far as their general form is concerned. Confining ourselves
to second order partial differential equations, the choices for the
Lagrangian density of noninteracting systems, such as \(\cL_\psi\), and
the choices for gauge invariant \(\cL_A\) are severely limited. 
If interacting theories are to be defined by minimal coupling, 
then there is only an
extremely reduced number of possibilities. It is a most remarkable
fact that such theories are so successful in describing the vast majority
of physical interactions. 

\section{Electromagnetism}
\subsection{Maxwell's Equations}\label{sec:maxwell}
The very first physical gauge theory is classical Maxwell
electrodynamics, though it was not originally presented as such. 
In appropriate physical units, Maxwell's equations
\index{equation!Maxwell's}%
for the electric field
\index{field!electric}%
 \(\bf E\) and magnetic  field
\index{field!magnetic}%
 \(\bf B\) are:
\begin{eqnarray}\label{eq:divb}%
{\bf\nabla} \cdot {\bf B} &=& 0 \\ \label{eq:curle}
{\bf\nabla} \times {\bf E} +\frac{\partial {\bf B} }{\partial t} &=& 0 \\
\label{eq:dive}{\bf\nabla} \cdot {\bf E} &=& \rho \\ \label{eq:curlb}
{\bf\nabla} \times {\bf B} -\frac{\partial {\bf E} }{\partial t} &=& 
{\bf J} 
\end{eqnarray}%
where \(\rho\) is the charge density
 and \({\bf J}\) the 
electric current density.
Equations (\ref{eq:divb}) and (\ref{eq:curle}) are know as the {\em
homogeneous\/} Maxwell's equations and the other two as the 
{\em non-homogeneous\/} ones. One of the immediate consequences of 
Maxwell's
equations is the conservation law for electric charge:
\[%
{\bf \nabla}\cdot {\bf J} + \frac{\partial\rho}{\partial t} = 0
\]%

From the homogeneous equations we deduce that there is a function
\(V\), called the {\em electric, or scalar, potential\/},
\index{potential!electric}%
and a vector field \({\bf A}\), called the {\em magnetic, or vector,
 potential\/}
\index{potential!magnetic}%
\index{potential!vector}%
such that 
\begin{eqnarray*} 
{\bf B} &=& {\bf \nabla}\times {\bf A} \\ 
{\bf E} &=& -{\bf \nabla}V - \frac{\partial {\bf A}}{\partial t}
\end{eqnarray*}%

Our next step is to interpret these equations in four-dimensional 
space-time, that is, \(\lR^4\) with a constant pseudo-Riemannian metric 
with signature \((1,3)\). We introduce  linear coordinates 
\((x^0,x^1,x^2,x^3)=(t, x, y,z)\) such that the corresponding coordinate 
vector fields \(\ppv{x}{\mu}\) are orthonormal. 
We denote by  \((D_x,D_y,D_z)\) the components of
 a vector \({\bf D}\)
in \(\lR^3\).

Consider now the \(1\)-form 
\[%
A = \sum_\mu A_\mu\, dx^\mu=-V\, dt + A_x\, dx + A_y\, dy + A_z\, dz 
\]%
A direct calculation of \(F=dA\)
 now provides
\begin{eqnarray}\label{eq:fda}\nonumber
\lefteqn{F= \frac{1}{2}\sum_{\mu\nu}F_{\mu\nu}\,dx^\mu\wedge dx^\nu =}
 \\ \nonumber
\lefteqn{-E_x\,dt\wedge dx -E_y\, dt\wedge dy -E_z\, dt\wedge dz +}\\ 
& & B_x\, dy\wedge dz +B_y\, dz\wedge dx + B_z\, dx\wedge dy
\end{eqnarray}%
The tensor \(\sum_{\mu\nu}F_{\mu\nu}\,dx^\mu\wedge dx^\nu\) is called 
the 
{\em electromagnetic tensor\/}
\index{tensor!electromagnetic}%
which combines the electric and the magnetic fields into a single
object.

The homogeneous Maxwell's equations are now seen to just be \(dF=0\) which 
follows immediately from \(dF = d^2A =
0\). To formulate the non-homogeneous equation introduce the \(1\)-form 
\begin{equation}\label{eq:jform}%
j = \sum_\mu j_\mu\, dx^\mu= -\rho\, dt + J_x\, dx + J_y\, dy + J_z\, dz 
\end{equation}%
A simple calculation, using the Hodge star operator, now shows that 
\(\delta F = *d*F = j\) so that
Maxwell's equation are now reduced to 
\begin{eqnarray}\label{eq:df}%
dF &=& 0 \\ \label{eq:delf}
\delta F &=& j
\end{eqnarray}%
Note that the conservation of charge is now expressed as \(\delta j =0\) 
which follows immediately from (\ref{eq:delf}) and \(\delta^2=0\).

The last step in interpreting electromagnetism as a gauge theory is to
consider the \(1\)-form \(A\) as the local principal gauge potential of
an invariant connection on a principal \(\U(1)\)-bundle and the
electromagnetic \(2\)-form \(F\) as its curvature. 
The Lie algebra of \(\U(1)\) is just the real line \(\lR\), so the local
principal gauge potential and the curvature are just normal real-valued
differential forms.
Note that \(\U(1)\) is 
abelian and so equation (\ref{eq:curvpb}) for the curvature 
gives just \(F=dA\). The form of  a gauge transformation is also simplified.
 A function
\(\phi:U \to \U(1)\) can be locally expressed as \(\phi =
e^{i\Lambda}\). An easy calculation now shows that the corresponding
gauge transformation (\ref{eq:lgtra}) is \(A \mapsto A -d\Lambda\).
Source-free (\(j=0\)) electromagnetism is a Lagrangian theory
with the conventional Lagrangian density being
\[
\cL_{EM}= -\frac{1}{2}F\wedge *F
\]
Comparing this with (\ref{eq:YM}) we see that electromagnetism is just
a Yang-Mills theory for the group \(\U(1)\). To get a theory with
sources, it is customary to use minimal coupling to couple the
electromagnetic field to other fields. Compare
(\ref{eq:df},\ref{eq:delf}) to (\ref{eq:dgf},\ref{eq:delgfj}). 

\subsection{Dirac's Magnetic Monopole}%

Interpreting electromagnetism as a gauge theory on \(P\U(1)\) does not
offer any special advantages except when one considers quantum theory or
extensions to situations that transcend classical Maxwellian theory. One
such advantage is seen in trying to define magnetic monopoles. The
homogeneous Maxwell's equations state that there are no magnetic
sources, that is, classical electromagnetism has no magnetic monopoles.
\index{magnetic monopole}%
 In analogy with electrostatics, a magnetostatic 
magnetic monopole situated at the origin in \(\lR^3\) would be a source of
a magnetic field away from the origin given by 
\begin{equation}\label{eq:mpole}%
{\bf B} = g\frac{{\bf r}}{r^3}
\end{equation}%
Here \({\bf r}=(x,y,z)\) is the position vector,  \(r\) it's norm, and
\(g\) is a physical constant called the  {\em magnetic charge\/}.
\index{charge!magnetic}%
There is no magnetic potential \({\bf A}\) defined away from the origin
such that \({\bf B}={\bf \nabla}\times {\bf A}\) since then by Stokes's
theorem the flux of \({\bf B}\) through a spherical surface centered at
the origin would be zero, contradicting it's explicit form 
(\ref{eq:mpole}). Dirac's solution to this problem was to introduce a
magnetic potential that is singular along a curve (the ``Dirac string")
joining the origin to infinity. Such a singularity is non-physical
since the curve can be chosen arbitrarily. Reformulating the monopole in
terms of principal bundles does away with this singularity. 

Let \(M\) be \(\lR^3\) minus the origin. Cover \(M\) with two open sets
 which are \(M\) minus one of the half \(z\)-axes:
\begin{eqnarray*}
U_s &=& M \setminus \{(0,0,z)\,|\,z > 0\} \\
U_n &=& M \setminus \{(0,0,z)\,|\,z < 0\} 
\end{eqnarray*}%
In \(U_s\) introduce the \(1\)-form 
\[
A_s = g\frac{y}{r(r-z)}\,dx -g\frac{x}{r(r-z)}\, dy
\]
 Likewise in 
\(U_n\) introduce the \(1\)-form
\[
A_n = -g\frac{y}{r(r+z)}\, dx +g\frac{x}{r(r+z)}\, dy 
\]
 It is easy to
show that in each open set the exterior derivative of each \(1\)-form 
is precisely \(B_x\, dy\wedge dz +B_y\, dz\wedge dx+B_z\, dx\wedge dy\)
where \({\bf B}\) is the monopole field 
(\ref{eq:mpole}). One has
\begin{equation}\label{eq:gamon}%
A_n-A_s=2g\frac{y}{x^2+y^2}\, dx-2g\frac{x}{x^2+y^2}\,dy
\end{equation}%
Suppose now that \(2g\) is an integer, then one easily calculates that
(\ref{eq:gamon}) is
 equal to \(-d\phi\cdot \phi^{-1}\) where \(\phi:U_s\cap U_n \to
\U(1)\) is given by 
\[
\phi(x,y, z) = \frac{(x+iy)^{2g}}{|x+iy|^{2g}}
\] 
that is, \(\phi=e^{2gi\theta}\), where \(\theta\) is the 
azimuthal spherical
coordinate.
But this, by Theorem
\ref{th:gluea}, is just the condition that the two  \(1\)-forms 
define an invariant connection on a
principal \(\U(1)\)-bundle over \(M\). The curvature of this connection
is precisely the monopole field (\ref{eq:mpole}) interpreted as a 
\(2\)-form. Note that no singular Dirac string is needed to define the
monopole now. The individual \(1\)-forms when interpreted as vector
potentials have Dirac stings; for \(A_s\) it is the positive \(z\)-axis,
and for \(A_n\) the negative one.
The Dirac monopole is not a realistic candidate for a possible physical
particle,  but
more complex gauge theories allow for the existence of realistic
magnetic monopoles. Experimental searches for magnetic monopoles have up
to now failed to find any. If they do exist, they are very rare in the
universe. Note that in the above description the magnetic charge \(g\)
is {\em quantized\/}, that is, it assumes discrete values since \(2g\)
must be an integer. In a more detailed quantum mechanical analysis, when
both electric and magnetic charges are allowed one finds that it is the
product of the two charges that must be quantized. Physicists
found this result quite intriguing for, as the argument goes, 
if there is just one magnetic monopole in the universe, quantization of
electric charge, observed empirically, follows.

\section{Spin}

\subsection{Clifford Algebras}\label{sec:cliffa}%
Let \(V\) be a vector space over a base-field \(\lF\) of characteristic
different from \(2\), \(\beta\) a symmetric bilinear form on \(V\), and
\(q(v)=\beta(v,v)\) the corresponding quadratic form. Let
\(T(V)=\sum_{p=0}^\infty T^p(V)\) be the full tensor algebra of \(V\)
where \(T^0(V)=\lF\) and for \(p>0\), \(T^p(V)=V^{p\otimes}=
V\otimes\cdots\otimes V\)
is the \(p\)-fold tensor product of \(V\) with itself. Let \(I\subset
T(V)\) be the two-sided ideal generated by elements of the form
\(v\otimes v + q(v)\) for all \(v\in V\). The {\em Clifford Algebra\/}
\index{algebra!Clifford}%
\(\Cl(V,q)\) is defined as the quotient \(T(V)/I\). 
Since \(V\cap I=\{0\}\),
the inclusion \(V\subset T(V)\) descends to an
 inclusion of \(V\) into \(\Cl(V,q)\) and we shall thus 
identify \(V\) with
a subspace of \(\Cl(V,q)\). We have for elements of \(V\) the
 fundamental Clifford relations
\begin{equation}\label{eq:qliff}%
v^2+q(v)=0
\end{equation}%
Substituting \(v+u\) for \(u\) in (\ref{eq:qliff}), one easily 
deduces the equivalent 
polarized form of (\ref{eq:qliff}):
\begin{equation}\label{eq:cliff}%
uv+vu+2\beta(u,v)=0
\end{equation}%
Note that when \(q=0\) then \(\Cl(V,q)=\bigwedge(V)\), the exterior
algebra of \(V\).

As an arbitrary element of \(T(V)\) is a sum
of tensor products of elements of \(V\) one sees that \(V\) generates
\(\Cl(V,q)\) as an algebra (by convention an empty product of elements
of \(V\) is equal to \(1\))

Let now \(v_1,v_2,\dots,v_m\in V\) and consider the element
\(v_1v_2\cdots v_m\in\Cl(V,q)\). Let \(\pi\) be any permutation of
\(\{1,\dots,m\}\). By repeated use of (\ref{eq:cliff}) one sees that
\begin{equation}\label{eq:clperm}%
v_1v_2\cdots v_m = \sigma_\pi v_{\pi(1)}v_{\pi(2)}\cdots v_{\pi(m)} + w
\end{equation}%
where \(w\) consists of terms which are products of {\em at most
\(m-2\)\/} factors \(v_i\) and \(\sigma_\pi\) is \(\pm 1\) according to
whether the permutation is even or odd. Among the vectors
\(v_1,\dots,v_m\) there might be some that are equal. One can now use
(\ref{eq:clperm}) to bring any two such next to each other, and then use
(\ref{eq:qliff}) to eliminate them in favor of a numerical coefficients.
Proceeding in this manner we see that any product \(v_1v_2\cdots v_m\)
can now be rewritten as a linear combination of products of at most
\(m\) factors chosen from the same set of vector
and in which all factors in each product are different.

To save on notation we shall often suppress some or all of the data
\((V,q)\) from \(\Cl(V,q)\) when the context makes clear what is meant.

\begin{theorem}\label{th:ucliff}%
\(\Cl(V,q)\) satisfies the following {\em universal property\/}. 
\index{universal property}%
Let
\(\cA\) be any associative algebra with unit \(e\),
and \(\phi:V\to \cA\) a map that
satisfies \(\phi(v)^2+q(v)e=0\), then \(\phi\) extends to a unique algebra
homomorphism \linebreak \(\phi^\sharp:\Cl(V,q)\to \cA\).
\end{theorem}%
\begin{proof}%
By the universal property of the tensor algebra, \(\phi\) extends to a
unique algebra homomorphism \(\hat\phi:T(V)\to \cA\).
Now \(\hat\phi\) vanishes on the ideal I, and so descends to an extension
\(\phi^\sharp\) to \(\Cl(V,q)\) showing existence. Uniqueness follows
immediately from the fact that \(V\) is a set of generators for \(\Cl\).
\end{proof}%

Denote by \(\O(\beta)\)
\index{\(O(\beta)\)@\(\O(\beta)\)}%
 the set of linear transformations \(\gamma:V\to V\)
such that \(\beta(\gamma v,\gamma w)=\beta(v,w)\), or equivalently
\(q(\gamma v)=q(v)\). 
If \(q\) is non-degenerate, then
then  \(\O(\beta)\) is the indicated orthogonal group, otherwise
\(\O(\beta)\) contains non-invertible elements. In any case it is a
semigroup. 
Let now \(\gamma\in \O(\beta)\) and consider the map \(V\to\Cl(V,q)\) given by
\(v\mapsto \gamma\cdot v\in\Cl\). Since \((\gamma\cdot v)^2 = q(\gamma\cdot v)=q(v)\),
we have by Theorem \ref{th:ucliff} an extension of this map to all of
\(\Cl\) thus associating to each \(\gamma\in \O(\beta)\) an element of
\(\End(\Cl(V,q))\) defining thus  an action
of \(\O(\beta)\) on \(\Cl\) by algebra endomorphisms (isomorphisms
in case \(q\) is non-degenerate). We call this the {\em
canonical action\/}
\index{action!canonical}%
 of \(\O(\beta)\) on \(\Cl\).

An {\em ordered semigroup\/} 
\index{ordered semigroup}%
\(S\) is a semigroup with a partial order
such that if \(a,b\in S\) with \(a< b\) then  \(ra< rb\) and
\(ar< br\) for all \(r\in S\).

Let \(S\) be an ordered semigroup. An associative algebra \(\cA\) is an
{\em \(S\)-filtered\/}
\index{algebra!filtered}%
 algebra if there are {\em subspaces\/}
\(\cA^{(s)}\) for \(s\in S\) such that \(\cA=\bigcup_{s\in S}\cA^{(s)}\),
\(\cA^{(s)}\cA^{(r)}\subset\cA^{(sr)}\), and if \(r\le s\) then
\(\cA^{(r)}\subset \cA^{(s)}\).

A Clifford algebra \(\Cl\) is canonically an \(\lN\)-filtered algebra
where the set of natural numbers forms an ordered semigroup under 
addition and 
the usual order.
 Define \(\Cl^{(r)}\) as the
image of \(\oplus_{p=0}^r T^p(V)\) under the quotient map.

Let \(S\) be a semigroup. An associative algebra \(\cA\) is an
{\em \(S\)-graded\/}
\index{algebra!sgraded@\(S\)-graded}%
 algebra if there are  vector {\em subspaces\/}  \(\cA_s\)
for \(s\in S\) such that
\(\cA=\oplus_{s\in S}\cA_s\) and
\(\cA_s\cA_r\subset\cA_{sr}\). Note that if \(e\in S\) is the identity 
then \(\cA_e\) is a subalgebra.
An element \(a\in\cA\) is said to be {\em homogeneous\/}
\index{homogeneous element}%
 if
\(a\in\cA_s\) for some \(s\), in which case we shall call \(s\)
the {\em degree\/}
\index{degree}%
 of \(a\) and denote it by \(|a|\).

Associated to any \(S\)-filtered algebra there is a canonical
\(S\)-graded algebra  \(Gr(\cA)=\oplus_{s\in S}Gr_s(\cA)\)
where \(Gr_s(\cA) = \cA^{(s)}/(\bigcup_{r<s}\cA^{(r)})\).

\begin{theorem}\(Gr(\Cl(V,q))\simeq\bigwedge(V)\)
\end{theorem}%
\begin{proof} The  map \(V^p\to Gr_p(\Cl)\) given by
\((v_1,v_2,\dots,v_p)\mapsto [v_1v_2\cdots v_p]\) is
\(p\)-linear  and, by (\ref{eq:clperm}), is anti-symmetric. 
Thus it descends
to a map \(\gamma_p:\bigwedge^p(V)\to Gr_p(\Cl)\).
The direct sum of these  maps is an
algebra homomorphism \(\gamma:\bigwedge(V) \to Gr(\Cl(V,q))\),
which is obviously surjective as \(V\) generates \(\Cl\). 
An element \(\phi\) is in the kernel of
\(\gamma\) if and only if each \(p\)-homogeneous part is in the kernel
of \(\gamma_p\), so suppose \(\phi\) is \(p\)-homogeneous. It is thus a sum
of terms of the form \(\alpha_1\wedge\cdots\wedge\alpha_p\). Consider the
corresponding sum of Clifford products \(\alpha_1\cdots\alpha_p\) in
\(\Cl^{(p)}\). If this sum is zero
in \(Gr_p\) this means that upon a finite number of applications
of (\ref{eq:clperm}) it can be reduced to an element of \(\Cl^{(p-1)}\).
This fact is independent of the actual quadratic form \(q\) that is used
and is equally valid for \(q=0\) but this means that \(\phi\) was already
zero in \(\bigwedge^p(V)\) and so \(\gamma\) is injective.
\end{proof}%

As a corollary we have
\begin{theorem} \(\Cl(V,q)\) and \(\bigwedge(V)\) have the same
dimension, in particular if \(V\) is finite dimensional of dimension
\(n\), both algebras have dimension \(2^n\).
\end{theorem}

It is in fact easy to establish a bijection between
linear bases of \(\Cl(V,q)\) and \(\bigwedge(V)\) which is done in the
proof of the following theorem
\begin{theorem}\label{th:cliffbasis}
Let \((v_\alpha)_{\alpha\in A}\) be a Hammel basis for \(V\). Assume
\(A\) is totally ordered. In \(\Cl(V,q)\), the products 
\(v_{\alpha_1}v_{\alpha_2}\cdots v_{\alpha_p}\)  with
\(\alpha_1<\alpha_2<\cdots<\alpha_p\) and any \(p\ge 0\)
 form a basis for \(\Cl(V,q)\) as a
vector space. By convention the product for \(p=0\) is \(1\).
\end{theorem}
\begin{proof}
Since 
finite tensor products of the \(v_\alpha\) form a basis for \(T(V)\), the 
finite Clifford products of the \(v_\alpha\) generate  \(\Cl(V,q)\). 
By what was said above, 
any finite 
product of the \(v_\alpha\) is a linear combination of terms of the form  
\(v_{\alpha_1}v_{\alpha_2}\cdots v_{\alpha_p}\) with 
\(\alpha_1<\alpha_2<\cdots<\alpha_p\).   The image of 
\(v_{\alpha_1}v_{\alpha_2}\cdots v_{\alpha_p}\) in 
\(\bigwedge(V)\) is \(v_{\alpha_1}\wedge 
v_{\alpha_2}\wedge\cdots\wedge v_{\alpha_p}\). As these images are linearly 
independent in \(\bigwedge(V)\),  the original terms must be linearly 
independent in \(\Cl(V,q)\) and so form a basis.
\end{proof}

The most commonly seen gradings are \(\lZ\) and \(\lN\) gradings.
A Clifford algebra has however a canonical \(\lZ_2\)-grading.
Let \(\Cl_0\) be the image under the quotient map
of the {\em even\/} tensor powers \(\oplus_{p=0}^\infty T^{2p}(V)\)
and \(\Cl_1\) the image
of the {\em odd\/} tensor powers \(\oplus_{p=0}^\infty T^{2p+1}(V)\).
 A
\(\lZ_2\)-graded algebra \(\cA\) is also called a {\em superalgebra\/}
\index{superalgebra}%
where \(\cA_0\) is called the {\em bosonic subalgebra\/}
\index{bosonic subalgebra}%
 and \(\cA_1\) the {\em
fermionic subspace\/}
\index{subspace!fermionic}%
 (which is not a subalgebra).

Given two \(\lZ_2\)-graded algebras \(\cA\) and \(\cB\), the {\em
twisted, or graded, tensor product\/} 
\index{twisted tensor product}%
\index{graded tensor product}%
\(\cA\hat\otimes\cB\) of the two is a
\(\lZ_2\)-graded algebra which, as a vector space, coincides with the
ordinary tensor product \(\cA\otimes\cB\) but for which the multiplication,
given homogeneous
elements \(a_1,a_2\in \cA\) and \(b_1,b_2\in\cB\),
 is defined by
\((a_1\otimes b_1)(a_2\otimes b_2)=
(-1)^{|a_2||b_1|}(a_1a_2)\otimes (b_1b_2)\). The sign on the right hand
side
is negative exactly when in going from the left to the right-hand side,
two fermionic elements exchange position. The \(\lZ_2\)-grading of
\(\cC=\cA\hat\otimes\cB\) is given by
\(\cC_0=(A_0\otimes B_0)\oplus(A_1\otimes B_1)\) and
\(\cC_1=(A_0\otimes B_1)\oplus(A_1\otimes B_0)\).

Suppose now that \(V=V_1\oplus V_2\) is a direct sum decomposition such
that if \(v=v_1\oplus v_2\) then \(q(v)=q(v_1)+q(v_2)\). We call this an
\index{orthogonal decomposition}%
{\em orthogonal\/} decomposition. Let \(q_i\) for \(i=1,2\)
be the restriction of \(q\) to \(V_i\). One has
\begin{theorem}\label{th:twipcl}%
 \(\Cl(V,q)\simeq
\Cl(V_1,q_1)\hat\otimes\Cl(V_2,q_2)\).
\end{theorem}%
\begin{proof} Consider the map \(f:V \to
\Cl(V_1,q_1)\hat\otimes\Cl(V_2,q_2)\) given by
\(f(v) = v_1\otimes 1+1\otimes v_2\). Since the \(v_i\) are
fermionic, one calculates that \(f(v)^2=v_1^2\otimes 1+1\otimes
v_2^2=q(v_1)+q(v_2)=q(v)\). By Theorem \ref{th:ucliff} this map extends
to an algebra homomorphism
\(f^\sharp:\Cl(V,q)\to \Cl(V_1,q_1)\hat\otimes\Cl(V_2,q_2)\). Let 
\((v_\alpha)_{\alpha\in A}\) be a Hammel basis for \(V_1\) and 
\((w_\beta)_{\beta\in B}\) a Hammel basis for \(V_2\). 
Assume \(A\) and \(B\) 
disjoint and consider a total order on \(A\cup B\) in which any element of 
\(A\) is less than any element of \(B\). By Theorem \ref{th:cliffbasis} 
terms of the form 
\(v_{\alpha_1}v_{\alpha_2}\cdots v_{\alpha_p}
w_{\beta_1}w_{\beta_2}\cdots w_{\beta_q}\) with 
\(\alpha_1<\alpha_2<\cdots<\alpha_p<\beta_1<\beta_2<\cdots<\beta_q\) form a 
basis for \(\Cl(V,q)\) and the corresponding terms 
\(v_{\alpha_1}v_{\alpha_2}\cdots v_{\alpha_p}\otimes
w_{\beta_1}w_{\beta_2}\cdots w_{\beta_q}\) form a basis for 
\(\Cl(V_1,q_1)\hat\otimes\Cl(V_2,q_2)\). But one has 
\(f^\sharp(v_{\alpha_1}v_{\alpha_2}\cdots v_{\alpha_p}
w_{\beta_1}w_{\beta_2}\cdots w_{\beta_q})=v_{\alpha_1}v_{\alpha_2}\cdots 
v_{\alpha_p}\otimes
w_{\beta_1}w_{\beta_2}\cdots w_{\beta_q}\) and so \(f^\sharp \) is an algebra 
isomorphism. 
\end{proof}%

If \(V\) is real and finite dimensional of dimension \(n\), 
there is a basis
\(e_1,\dots,e_n\) such that \(\beta(e_i,e_j)=\diag(1,\dots,1,-1,\dots,-
1,0,\dots,0)\) where there are \(r\) entries of \(1\), \(s\) entries of
\(-1\) and \(t=n-r-s\) entries of \(0\).  If \(V\) is complex, then
one can choose a basis for which \(\beta(e_i,e_j)=
\diag(1,\dots,1,0,\dots,0)\) with \(r\) entries of \(1\),
and \(t=n-r\) entries of \(0\). The bilinear form \(\beta\) is
non-degenerate if and only if \(t=0\). Non-degenerate forms are thus in
the real case classified by their signature \((r,s)\) and in the complex
case they are all equivalent.
Bases such as the ones here introduced are called {\em orthonormal\/}.
\index{basis!orthonormal}%
 Choose one such. By our 
previous discussion the products
\(e_{i_1}e_{i_2}\cdots e_{i_p}\) with strictly increasing indices
form a basis for \(\Cl\).

We can now use Theorem \ref{th:twipcl} to determine the Clifford
algebras of real and complex finite-dimensional vector spaces. Assume
the quadratic form is non-degenerate. In the real case denote by
\((r,s)\) the signature of the form and the corresponding Clifford
algebra  by \(\Cl_{\lRe}(r,s)\).
\index{\(Cl_{R}(r,s)\)@\(\Cl_{\lRe}(r,s)\)}%
 In the complex case we denote the 
Clifford
algebra by 
\index{\(Cl_{C}(n)\)@\(\Cl_{\lCe}(n)\)}%
\(\Cl_{\lCe}(n)\).

Consider the real case for \(n=1\) and let \(e\) be an orthonormal basis.
Then \(q(e)\)
is either \(1\) or \(-1\). In the first case \(e^2=1\)
and using the basis
\(f_1=\frac{1}{2}(1+e)\) and \(f_2=\frac{1}{2}(1-e)\) one finds that
\((\alpha_1f_1+\alpha_2f_2)(\beta_1f_1+\beta_2f_2)=
(\alpha_1\beta_1f_1+\alpha_2\beta_2f_2)\). Hence
\(\Cl_{\lRe}(1,0)=\lR\oplus\lR\). In the second case \(e^2=-1\)
and identifying
\(e\) with the imaginary unit \(i\)  of the complex numbers,
one easily sees that
 \(\Cl_{\lRe}(0,1)=\lC\).
In the complex case for \(n=1\) the orthonormal basis \(e\) satisfies
\(e^2=1\) and proceeding just as in the real case we conclude
\(\Cl_{\lCe}(1)=\lC\oplus\lC\). By Theorem \ref{th:twipcl} one concludes
\(\Cl_{\lRe}(r,s) = (\lR\oplus\lR)^{r\hat\otimes}
\hat\otimes\lC^{s\hat\otimes}\) and \(\Cl_{\lCe}(n) =
(\lC\oplus\lC)^{n\hat\otimes}\) which allows us to explicitly calculate
each Clifford algebra. We shall not do this calculation here but present
the rather intriguing result. Let \(\lR\), \(\lC\), and \(\lH\) denote
respectively the real, complex and quaternionic field. For any one of
these fields \(\lF\), let  \(\lF(n)\) denote the algebra of \(n\times n\)
matrices with elements in \(\lF\). For the real case consider now
the following correspondence:
\[
\begin{array}{c|c|c|c|c|c|c|c}%
 0  & 1  & 2 &  3 &  4 &  5 &  6  & 7 \\ \hline
\lR & \lC & \lH & \lH\oplus\lH & \lH & \lC & \lR & \lR\oplus\lR
\end{array}
\]
To calculate \(\Cl_{\lRe}(r,s)\) find \(r-s \) modulo \(8\) on the first
line of the table. The Clifford algebra will be either \(\lF(m)\) or
\(\lF(m)\oplus\lF(m)\) where the scheme of the expression is read from
the second line of the table and \(m\) is chosen so that the resulting
dimension of the
algebra is exactly \(2^{r+s}\).
For example \(\Cl_{\lRe}(1,3) \simeq \lR(4)\) and
\(\Cl_{\lRe}(3,1) \simeq \lH(2)\).
For the complex algebra \(\Cl_{\lCe}(n)\),
the correspondence
is much simpler:
\[
\begin{array}{c|c}%
 0  & 1   \\ \hline
\lC & \lC\oplus\lC
\end{array}
\]
where the first line is \(n\) modulo \(2\). Thus
\(\Cl_{\lCe}(4)=\lC(4)\).

Let \(\cA\) be an associative algebra with unit
\(e\). An {\em \(\cA\)-module\/}
\index{module!\(\cA\)-}%
  is a vector space \(M\), with the same
base
field \(\lF\) as \(\cA\), along with a
map \(\cdot:\cA\times M \to M\)  such that
\begin{enumerate}%
\item \((a\cdot(b\cdot m))=(ab)\cdot m\)
\item \((a+b)\cdot m=a\cdot m + b\cdot m\)
\item \((\alpha a)\cdot m = \alpha(a\cdot m)\)
\item \(e\cdot m = m\)
\end{enumerate}
where \(a,b\in \cA\), \(m\in M\), and \(\alpha\in\lF\).

This definition bears a strong resemblance to the definition a group
action and  can be thought of as defining the action of an
algebra on a vector space by means of linear maps. We shall refer to
the map \(\cdot\) as an {\em algebra action\/}. 
\index{action!algebra}%
Much of the following
material on \(\cA\)-modules will also bear a strong resemblance to
analogous
notions concerning group actions. In particular if by \(\cA^*\) we denote
the  group of
invertible elements of \(\cA\), then the restriction of the algebra action
of an \(\cA\)-module \(M\) to \(\cA^*\times M\) results in a
representation of the group \(\cA^*\).

A vector spaces \(V\) is an \(\End(V)\)-module where for \(B\in
\End(V)\) and \(v\in V\) one defines 
\(B\cdot v= Bv\). In the same way, if \(\lF\) is any field, then
\(\lF^n\) is a \(\cM(n,\lF)\)-module.
The set of local sections \(\Gamma(U)\) of a vector
bundle is a \(\cF(U)\)-module of the algebra of maps \(U\to \lF\) under
pointwise multiplication.

Given an \(\cA\)-module \(M\), the map \(m\mapsto a\cdot m\) defines an
element \(L(a)\in\End(M)\) and the map \(a\mapsto L(a)\) is an algebra
homomorphism \(\cA\to \End(M)\). Reciprocally given any vector space
\(V\) and an algebra homomorphism \linebreak \(L:\cA\to \End(V)\) 
one turns \(V\)
into an \(\cA\)-module by defining \(a\cdot v = L(a)v\). \(\cA\)-modules
are also know as {\em representations\/}
\index{representation}%
 of \(\cA\).

Any algebra \(\cA\) is itself automatically an \(\cA\)-module where the
map \linebreak \(\cdot:\cA\times\cA\to\cA\) is algebra multiplication. 
This action is
known as the {\em regular representation\/}. 
\index{representation!regular}%

Given an algebra \(\cA\) with base field \(\lF\) and an extension of the
field \(\hat{\lF} \supset \lF\), by an \(\hat{\lF}\)-module of \(\cA\) we
mean an action of \(\cA\) on a vector space \(W\) over the extended
field \(\hat{\lF}\) by \(\hat{\lF}\)-linear endomorphisms, that is,
a homomorphism of  \(\lF\)-algebras \(\cA\to \End_{\hat{\lFe}}(W)\). The
case of interest for us is that of a {\em complex\/} module
\index{module!complex}%
 of a {\em
real\/} algebra.

We say an \(\cA\)-module \(M\) is {\em irreducible\/}
\index{module!irreducible}%
 if there is no
proper non-zero subspace \(W\subset M\) such that \(\cA \cdot W \subset W\).
Given two \(\cA\)-modules \(M_1\) and \(M_2\), an {\em intertwiner\/}
\index{intertwiner}%
 is
a linear map \(T:M_1\to M_2\) such that \(a\cdot Tm = T(a\cdot m)\).
Intertwiners are morphisms in the category whose objects are 
\(\cA\)-modules. Two \(\cA\)-modules are said to be {\em equivalent\/}
\index{module!equivalent}%
 if they
are isomorphic in this category. This means there is an invertible
intertwiner.

We state without proof the
following theorem.
\begin{theorem}%
Let \(\lF\) be any of the fields \(\lR\), \(\lC\), or \(\lH\) then, up to
isomorphism:
\begin{enumerate}%
\item \(\lF(n)\) has a unique irreducible representation, namely the
natural action of \(\lF(n)\) on \(\lF^n\)
\item \(\lF(n)\oplus\lF(n)\) has two irreducible representations given
by projection onto one of the summands followed by the natural
representation of the summand.
\end{enumerate}%
\end{theorem}%

As a consequence of this theorem we can assert that a Clifford algebra
has either one or two irreducible modules depending on
whether it is of the form
\(\lF(m)\) or \(\lF(m)\oplus\lF(m)\).

Let
\(q\) be a quadratic form on a real vector space \(V\). On the
complexified space \(V_{\lCe} = V\otimes_{\lRe}\lC\) one can introduce the
complexified
quadratic form \(q_{\lCe}\) defined uniquely by requiring 
\(q_{\lCe}(v\otimes
z)=z^2q(v)\).  Define \(\Cl_{\lCe}(V,q)\) as
\(\Cl(V,q)\otimes_{\lRe}\lC\). One has
\begin{theorem}
\(\Cl_{\lCe}(V,q)\simeq \Cl(V_{\lCe},q_{\lCe})\).
\end{theorem}%
\begin{proof}%
Consider the map \(V_{\lCe}\to \Cl(V,q)\otimes_{\lRe}\lC\)
defined by \(v\otimes z \mapsto v\otimes z\in \Cl(V,q)\otimes_{\lRe}\lC\).
One has in \(\Cl(V,q)\otimes_{\lRe}\lC\) that \((v\otimes z)^2 =
v^2\otimes z^2=-q(v)z^2 =-q_{\lCe}(v\otimes z)\) which by the universal
property of Clifford algebras means that the map extends to an algebra
homomorphism \(\Cl(V_{\lCe},q_{\lCe})\to\Cl(V,q)\otimes_{\lRe}\lC\).
Theorem \ref{th:cliffbasis} shows now that this homomorphism establishes a
bijection between complex linear bases of the two algebras. 
\end{proof}%

Consider now a {\em complex\/} \(\Cl(V,q)\)-module \(W\).  
One sees that there is then a natural unique 
extension of Clifford multiplication to turn \(W\) into a 
\(\Cl_{\lCe}(V,q)\)-module, just set \((a\otimes z)\cdot w=z(a\cdot w)\).

Let now \(e_1,\dots,e_n\) be an orthonormal basis for \(V\) and 
consider the 
element \(\eta=e_1e_2\cdots e_n\).  One easily finds \(\eta^2=(-
1)^{r+\frac{n(n-1)}{2}}\) where \((r,s)\) is the signature of \(q\). This 
means that there is an integer \(m\) such that the 
\(\Cl_{\lCe}(V,q)\) element  
\(\omega=i^m\eta\), called the {\em volume element\/}, 
\index{volume element}%
satisfies 
\(\omega^2=1\). If \(r+\frac{n(n-1)}{2}\) is even, then  one can take
\(\omega=\eta\) and the volume element belongs to \(\Cl(V,q)\) itself.
The volume element is, up to an overall sign, 
independent of 
the choice of orthonormal basis. Indeed if \(e_1',\dots,e_n'\) is another 
basis, then \(e_i'=Se_i\) for some \(S\in \O(V,q)\). As the \(e_i\) 
anticommute with each other, one has \(\eta'=\det(S)\eta=\pm\eta\). This 
ambiguity results in a sign ambiguity in the definition of \(\omega\).
Note that if we change to another basis preserving orientation, then this 
ambiguity disappears, and so if we deal with a fixed orientation,
there is  no ambiguity.

 Let now  
\begin{equation}\label{eq:cliffids}
p_{\pm}=\frac{1\pm \omega}{2}
\end{equation}
It is easy to verify that 
\(p_{\pm}\) are idempotents \(p_{\pm}^2=p_{\pm}\) that \(p_+p_-=p_-p_+=0\) 
and that \(p_++p_-=1\). The sign ambiguity in \(\omega\) lead to an 
ambiguity 
in the distinction between \(p_+\) and \(p_-\).
When \(n\) is odd, \(\eta\) (and consequently also \(\omega\) and 
\(p_{\pm}\)) commutes with elements of \(V\) and therefore 
is in the center of 
the algebra, when \(n\) is even, \(\eta\) (and consequently 
also \(\omega\)) 
anticommutes with elements of \(V\). For even \(n\), any element \(v\in V\) 
intertwines \(p_+\) and \(p_-\), that is, \(p_+v=vp_-\) and \(p_-v=vp_+\). 

If \(W\) is any complex module, then one has a canonical 
decomposition \(W=W^+\oplus W^-\) where \(W^{\pm}=p_{\pm} W\). 
For \(n\) odd 
one has \(\Cl(V,q)W^{\pm}\subset W^{\pm}\) and so for an irreducible module 
one of the submodules \(W^{\pm}\) must be zero. For even \(n\), one has 
\(\Cl(V,q)W^{\pm}\subset W^{\mp}\), and since for \(v\in V\) 
with \(q(v)\neq 
0\) one has \(v^2=-q(v)\), Clifford multiplication by  \(v\) establishes a 
linear isomorphism between \(W^+\) and \(W^-\). In this case an irreducible 
module decomposes into two {\em subspaces\/} of equal dimension which are 
invariant under the {\em even\/} part (which is a subalgebra) of 
the Clifford 
algebra, and which are mapped into each other under the odd part. 
Entirely analogous considerations apply whenever \(W\) is a real
\(\Cl(V,q)\)-module and the volume element belongs to \(\Cl(V,q)\)
itself.

\subsection{Spin Groups}\label{sec:spingroups}%

Consider a Clifford algebra \(\Cl(V,q)\) for an \(n\)-dimensional vector
space \(V\) with a non-degenerate quadratic form \(q\). 
Let \(\alpha\) be the algebra isomorphism \(\alpha:\Cl \to
\Cl\)
\index{\(\alpha\)}%
which (see Theorem \ref{th:ucliff})
is induced by the map \(V\to \Cl\) given by
\(v\mapsto -v\).
By the same theorem, \(\alpha^2=I\)
as this holds on \(V\). 
Denote by \(\Cl^*\) the
group of invertible elements of the
algebra. This contains a subgroup \(\tilde\cP\) of elements \(\phi\) such
that
\(\alpha(\phi)V\phi^{-1}\subset V\). Denote by \(\tilde{\Ad}_\phi\), the
{\em twisted\/} adjoint action
\index{action!twisted adjoint}%
\(\tilde{\Ad}_\phi\psi=\alpha(\phi)\psi\phi^{-1}\). One has for \(v\in
V\) with \(q(v)\neq 0\) that \(v^{-1}= -q(v)^{-1}v\) so
\(q(v)\tilde{\Ad}_vw=
 vwv= -2\beta(w,v)v -wv^2 = -2\beta(w,v)v +q(v)w\) and we have
\begin{equation}\label{eq:tadw}%
\tilde{\Ad}_vw=  w-2\frac{\beta(w,v)}{q(v)}v
\end{equation}%
This shows in particular that \(\{v\in V\,|\,q(v)\neq 0\}\subset
\tilde\cP\).
The map \(\phi\mapsto\tilde{\Ad}_\phi\) defines a group homomorphism
\(\tilde\cP\to\GL(V)\).

\begin{theorem} \(\tilde{\Ad}(\tilde\cP)= \O(V,q)\)
\end{theorem}%
\begin{proof} Let \(w=\tilde{\Ad}_\phi v\), then
\(q(w) = -w^2 = \alpha(w)w=\alpha(\alpha(\phi)v\phi^{-1})
\alpha(\phi)v\phi^{-1}\) which, using the fact that  \(\alpha\)
is a homomorphism and that \(\alpha^2=I\),
reduces to \(\phi\alpha(v)v\phi^{-1}\) which is equal to \(q(v)\).
That the image of \(\tilde \cP\) under \(\tilde{\Ad}\) is all of
\(\O(V,q)\) can be deduced from (\ref{eq:tadw}) as this formula defines
an orthogonal reflection in a hyperplane, and by the
Cartan-Dieudonn\'e theorem any
element of \(\O(V,q)\) is a product of a finite number of such reflections.
\end{proof}%

Now in a certain sense the group \(\tilde\cP\) is too big as any scalar
multiple of an element in \(\tilde\cP\) defines the same element of
\(\O(V,q)\)
under \(\tilde{\Ad}\). Just how this redundancy is to be removed to define
a more convenient subgroup of \(\tilde\cP\) depends in part on the base
field and on the application in mind. We shall here consider essentially
two cases, the real finite-dimensional case and the complexified real
case.

Let thus \(V\) be finite dimensional and \(q\) non-degenerate. 
We define
the group \(\Pin(V,q)\subset\tilde\cP\)
\index{\(Pin(V,q)\)@\(\Pin(V,q)\)}%
as being generated by elements \(v\in V\)  with \(q(v)=\pm 1\)
and \(\Spin(V,q)\subset\Pin(V,q)\)
\index{\(Spin(V,q)\)@\(\Spin(V,q)\)}%
 as the subgroup of even elements.
We state, without proof:
\begin{theorem} In the real case 
\(\tilde{\Ad}\) defines a two-to-one
covering of \(\O(V,q)\) by \(\Pin(V,q)\) and a two-to-one covering
of \(\SO(V,q)\) by \(\Spin(V,q)\). Elements \(\phi\) and \(-\phi\) map to
the same element of the orthogonal groups.
\end{theorem}%

For the case of  Clifford algebras over \(\lC\) the spin group is not
necessarily the most useful object. One often has to deal with a {\em
complexified\/} Clifford algebra and the underlying real structure can
be used to construct a complex extension of the real spin groups.

The group \(\Pin^c(V,q)\subset\Cl_{\lCe}(V,q)\)
\index{\(Pin^c(V,q)\)@\(\Pin^c(V,q)\)}%
is defined as the subgroup
generated by \(\Pin(V,q)\otimes 1\) and \(1\otimes \U(1)\), 
and the group \(\Spin^c(V,q)\) as the subgroup generated by 
\(\Spin(V,q)\otimes 1\) and \(1\otimes \U(1)\).
These group
should not be confused with \(\Pin(V_{\lCe},q_{\lCe})\) and
 \(\Spin(V_{\lCe},q_{\lCe})\). Note that
\(u\otimes z\) and \((-u)\otimes (-z)\) define the same element in
\(\Pin^c(V,q)\). One can show that this is the only ambiguity. 
As \(u\) and \(-u\) define the same element of \(\O(q)\), 
one still has canonical maps \(\lambda^c:\Pin^c(V,q)\to \O(q)\) and its
restriction 
\(\Spin^c(V,q)\to \SO(q)\).

Given a \(\Cl(V,q)\)-module \(W\) it is often useful to have a symmetric
bilinear or hermitian sesquilinear form on \(W\) that is invariant under
the action of given subgroups (such as \(\Pin\) or \(\Spin\) or their
subgroups) of \(\Cl^*\).  We shall address this question only partially.
For any \(\Cl(V,q)\)-module \(W\) and any \(\phi\in\Cl(V,q)\) we denote
by \(\hat\phi\) 
\index{\(\theta\)@\(\hat\phi\)}%
the endomorphism in \(\End(W)\) that corresponds to the
algebra action by \(\phi\). 

 Consider first a real \(\Cl(n,0)\)-module \(W\) and introduce 
any inner product \((\cdot,\cdot)_0\) on it. Let \(e_1,\dots e_n
\in \lR^n\) be an orthonormal basis. The elements
\(\pm 1\), along with \(\pm e_{i_1}e_{i_2}\cdots e_{i_p}\) 
 with \(i_1 < i_2 \cdots i_p\)  and
\(1\le p \le n\) form, under Clifford multiplication, a finite
group of order \(2^{n+1}\) which we shall denote by \(\lG\).  
We now define a new inner product
\[%
(\phi,\psi) =\frac{1}{2^{n+1}}
\sum_{\gamma\in\lG}(\hat\gamma\phi,\hat\gamma\psi)_0 
\]%
It is clear now that one has \((\hat e_i\phi,\hat e_i\psi)=(\phi,\psi)\), 
in other
words, Clifford action by \(e_i\) is orthogonal. Furthermore one has 
\((\hat e_i\phi,\psi)=(\hat e_i^2\phi,\hat e_i\psi)=
-(\phi,\hat e_i\psi)\) and Clifford 
action by \(e_i\) is anti-symmetric. Let now 
\(v=\sum_i\lambda_ie_i\), then 
\((\hat v\phi,\hat v\psi)=\sum_i\lambda_i^2(\hat e_i\phi,\hat e_i\psi)+ 
\sum_{i \neq 
j}\lambda_i\lambda_j(\hat e_i\phi,\hat e_i\psi)\) The first sum is 
\((\sum_i\lambda_i^2)(\phi,\psi)=q(v)(\phi,\psi)\) and the second sum 
vanishes since for \(i\neq j\) one has \((\hat e_i\phi,\hat e_j\psi)=-
(\phi,\hat e_i\hat e_j\psi)=(\phi,\hat e_j\hat e_i\psi)=
-(\hat e_j\phi,\hat e_i\psi)\). For \(q(v)=1\) 
therefore one has \((\hat v\phi,\hat v\psi)=(\phi,\psi)\),  
and so \((\cdot,\cdot)\)  is 
\(\Pin(n,0)\)-invariant. 

A precisely analogous construction can be used in case of a complex 
\(\Cl(n,0)\)-module 
\(W\) to construct a  \(\Pin^c(n,0)\)-invariant hermitian 
inner product on \(W\). In this case Clifford action by \(e_i\) are
anti-hermitian. 

We now consider space-time signature and let \(W\) be a complex module of 
\(\Cl(1,n-1)\). Choose  an orthonormal basis
\(e_0,e_1,\dots,e_{n-1}\), starting the labeling for convenience with \(0\). 
In what follows Greek indices run from \(0\) to \(n-1\) and roman indices 
from \(1\) to \(n-1\). In the
complexified algebra \(\Cl_{\lCe}(1,n-1)\), the elements \(e_0,
ie_1,\dots, ie_{n-1}\) are orthonormal for the complexified quadratic
form. So there is a hermitian inner product in \(W\) for which all
these elements act as anti-self-adjoint isometries. 
We thus have \(\hat e_0^*=-\hat e_0\) and \(\hat e_i^* = \hat e_i\).
From this one has \(\hat  e_0\hat e_\mu\hat e_0 = \hat e_\mu^*\) and so
for a {\em real\/} vector \(v=\sum_\mu v^\mu e_\mu\) one has 
\(\hat  e_0\hat v\hat e_0 = \hat v^*\). For \(a\in W\) define now \(\bar
a\in W'\) by \(\bar a b = \bar a(b) = (a,\hat e_0 b)\). The map
\(a\mapsto \bar a\) is an {\em anti-linear\/} isomorphism \(W \to W'\).
The hermitian product \((a,b) \mapsto \bar a b\) is non-degenerate but 
in general not positive definite. Let \(v_1,\dots, v_p\) be real vectors
and \(\phi=v_1\cdots v_p\in \Cl\). Let \(\phi^t = v_p\cdots v_1\). 
One has \((\hat\phi a, \hat e_0\hat\phi
b) =(a, \hat\phi^* \hat e_0\hat \phi b) = 
(a, \hat e_0 \hat \phi^t\hat \phi b) =
\prod_iq(v_i) (a,b)\). This means that the new hermitian product is
invariant by action of \(\phi\) if \(\prod_iq(v_i)=1\). This in
particular is true, though we shall not prove it here,
for the component of identity of \(\Spin(1,n-1)\). 
Because of this, the hermitian product
\(\bar a b\) is extremely useful in physical theories. An entirely
analogous treatment can be made for signature \((n-1,1)\) as well.

\subsection{Spin Bundles}\label{sec:spinbundles}
Let \(M\) be a pseudo-Riemannian manifold with pseudo-metric \(g\). Let
\(q_x\) be the quadratic form defined in \(T_xM\) by the pseudo-metric.
One now has the Clifford algebra \(\Cl_x=\Cl(T_xM,q_x)\)
defined at each point \(x\) of the manifold. Since the
pseudo-metric has the same signature \((r,s)\) at every point,
each \(\Cl_x\) is isomorphic to \(\Cl(r,s)\). The Clifford algebras
\(\Cl_x\) are fibers of an algebra bundle, which we denote by \(\Cl(TM)\),
with fiber \(\Cl(r,s)\). Let \(U\) be an open set with coordinate
functions \(x^1,\dots,x^n\) and let \(v_i = \ppv{x}{i}\) be the
coordinate vector fields. Local coordinates for \(a \in \Cl_x\) can be
taken to be the \(x^i\) along with the \(2^n\) coefficients
\(a_{i_1\cdots i_p}\)
in the expansion \(a = \sum_{p=0}^n a_{i_1\cdots
i_p}v_{i_1}\cdots v_{i_p}\). This defines \(\Cl(TM)\) as a manifold. 
To define the bundle structure, pick a fixed
orthonormal basis \(f_1,\dots,f_n\) of \(\lR^{r+s}\subset \Cl(r,s)\).
Let now \(U\) be an open set with \(n\)-bein \(e_1,\dots,e_n\) and
define, using Theorem \ref{th:ucliff}, the  isomorphism \(\Cl_x\to\Cl(r,s)\)
 by extension from \(e_i\mapsto f_i\).
These can now be used to define a map \(h_U:\pi^{-1}(U)\to
U\times\Cl(r,s)\), establishing thus the defining  trivializations.

By the above construction of \(\Cl(TM)\) it is easy to see that this
bundle is associated to the principal \(\O(r,s)\)-bundle \(\cF_{\O}(M)\)
of ordered orthonormal bases, where the action of \(\O(r,s)\) on
\(\Cl(r,s)\) is the canonical one. Thus
\(\cF_{\O}(M)\times_\rho\Cl(r,s)\simeq\Cl(TM)\), where \(\rho\) is the
canonical action. For an oriented manifold we can restrict \(\rho\) to
\(\O(r,s)\) and get \(\Cl(TM)\) as a bundle associated to
\(\cF_{\SO}(M)\).

A {\em spin bundle\/}
\index{bundle!spin}%
 on an pseudo-Riemannian manifold \(M\) is
a vector bundle \(S\), with fiber \(W\), whose fiber \(W_x\)
at each point \(x\)
is a \(\Cl_x\)-module.
This means that there has to be a bundle map \(\Cl(TM)\times S \to S\)
which restricted to \(\Cl_x\times W_x\) makes \(W_x\) into a 
\(\Cl_x\)-module. One generally assumes that all 
these pointwise modules are equivalent.
We say a spin bundle \(S\) is {\em irreducible\/} if \(W\) is an
irreducible module.
\index{bundle!spin!irreducible}%

An example is \(S=\Cl(TM)\) where the algebra action is algebra
multiplication. Now given any \(\Cl(r,s)\)-module \(W\) it is not in
general possible to construct a spin bundle with fiber \(W\) so that the
pointwise modules are equivalent to \(W\).

Abstracting from the algebra bundle \(\Cl(TM)\), one can state the
following problem. Given an algebra bundle \(A\) with fiber
\(\cA\), and an \(\cA\)-module \(W\), does there exist a vector bundle
\(S\) with fiber \(W\) and a bundle map \(A\times S \to S\) such that
restricted to fibers this map makes \(W_x\) into a 
\(\cA_x\)-module equivalent to \(W\). 
In general the answer is no as the following
example shows.

\begin{example}%
Let \(\cA=\lR\oplus\lR\), and \(X=S^1\). Consider \(S^1\) as \([0,1]\)
with endpoint identified. Let the bundle \(A\) be obtained from
\([0,1]\times\cA\) by the identification \((0,(a,b))\sim (1,(b,a))\).
Let \(W=\lR\) with algebra action \((a,b)\cdot r = ar\).
\end{example}%

Note that \(\cA\) has
two one-dimensional ideals given by \(I_1=\lR\times\{0\}\) and
\(I_2=\{0\}\times \lR\). Now \(I_2\cdot \lR=0\). Let us call \(I_2\) the
{\em annihilating ideal\/}. If a bundle \(S\)
existed which answered to the posed problem, there would be defined at
each point an annihilating ideal of \(\cA_x\) varying smoothly with \(x\).
However the bundle of one-dimensional ideals of the \(\cA_x\) is the double
cover \(S^1\to S^1\)
and so it has no global section. Thus the bundle \(S\)  does not exist.

There is no easy answer to the general problem. For the case of
Clifford algebras, one knows of  topological criteria 
for orientable   manifolds \(M\)
that guarantee that {\em any\/} \(\Cl(r,s)\)-module can define a
spin bundle. Such manifolds are know as {\em spin manifolds\/}.
\index{manifold!spin}%
Of particular interest are the irreducible modules. Notice
that some modules, such as \(\Cl(r,s)\) itself, always define a spin
bundle.
Spin manifolds are special in  that any module defines one.

An {\em (r,s)-spin structure\/}
\index{structure!spin}%
 for an oriented
pseudo-Riemannian manifold \(M\) with signature \((r,s)\) is a
principal \(\Spin(r,s)\)-bundle \(P\Spin(r,s)\) over \(M\) and a bundle
map \(\xi:P\Spin(r,s)\to \cF_{\SO}(M)\) which is equivariant with respect to
the right actions, that is, \(\xi(p\cdot u)=\xi(p)\cdot \lambda(u)\),
where \(\lambda:\Spin(r,s)\to \SO(r,s)\) is the canonical double cover. 
The manifold \(M\) is
said to be an {\em (r,s)-spin manifold\/}
\index{manifold!spin}%
 if it has an
\((r,s)\)-spin structure. When no signature is mentioned,  the
Riemannian case is understood.

\begin{example}\label{ex:spins1}
There are two inequivalent spin structures on the  circle \(S^1\). 
Pick any   Riemannian metric and an orientation. 
Identify \(\pi:\cF_{\SO}(S^1)\to S^1\) with the identity
map \(\Id:S^1\to S^1\). One has \(\Spin(1) \simeq \lZ_2\) so there are
two inequivalent principal \(\Spin(1)\) bundles given by Example
\ref{ex:pz2}. The two inequivalent spin structures 
\(\xi:P\Spin(1) \to \cF_{\SO}(S^1)\), each one given by one of the principal
bundles, are given by the unique obvious maps. 
\end{example}
In the physical literature the first structure is known as the {\em
Ramond\/} 
\index{Ramond}%
structure and the second one as the {\em Neveu-Schwartz\/}
\index{Neveu-Schwartz}%
structure.

Notice that given any action of \(\SO(r,s)\), one can define a
\(\Spin(r,s)\) action by composing it with \(\lambda\).
From this it is clear
that using the spin structure, any bundle associated to \(\cF_{\SO}(M)\) can
be redefined as a bundle associated to \(P\Spin(r,s)\). In particular,
this is true of
the Clifford bundle \(\Cl(M)\). The action of \(\Spin(r,s)\) on
\(\Cl(r,s)\) via composition with \(\lambda\) and the canonical action
of \(\SO(r,s)\)  is equivalent to the adjoint action, that is, if \(u\in
\Spin(r,s)\) and \(\phi\in\Cl(r,s)\) then \(\lambda(u)\cdot\phi=\Ad_u\phi
= u\phi u^{-1}\). Using this fact we can now create spin bundles from
any \(\Cl(r,s)\)-module \(W\). Since \(\Spin(r,s)\subset\Cl(r,s)\), \(W\)
carries a representation, call it \(\rho\) of \(\Spin(r,s)\) being simply
the restriction of the algebra action. We can now form the associated
bundle \(S=P\Spin(r,s)\times_\rho W\). To define the action of
\(\Cl_x\) on \(W_x\) making \(S\) into a spin bundle, consider the map:
\begin{equation}\label{eq:spinbdef}%
\Id\times\cdot:P\Spin(r,s)\times
\Cl(r,s) \times W \to P\Spin(r,s)\times W
\end{equation}%
There is an action of \(\Spin(r,s)\) on the left-hand space given by
\((r,\phi,z)\mapsto (ru^{-1}, \Ad_u\phi, u\cdot z)\) under the
quotient of which one has  the associated bundle
\(\Cl(M)\times S\). Likewise there is an action on the right-hand space
given by \((r,z)\mapsto (ru^{-1},  u\cdot z)\) under the
quotient of which one has the associated bundle
\(S\). These two actions are compatible with the horizontal map as one
has \(\Ad_u\phi\cdot (u\cdot z) = (u\phi u^{-1}u)\cdot
z=u\cdot(\phi\cdot z)\), which thus defines a bundle map
\(\Cl(M)\times S \to S\), and so a spin-bundle.

\begin{example}Using the two spin structures of Example \ref{ex:spins1}
one constructs two inequivalent spin bundles on \(S^1\). As
\(\Cl(1,0)\simeq \lC\), the two bundles are the quotients of
\(P\Spin(1)\times \lC\) which results in the trivial bundle \(S^1\times
\lC\) for the first \(P\lZ_2\) bundle of Example \ref{ex:pz2} and 
for the second in a
Moebius-band type construction of Example \ref{ex:moebius} using \(\lC\)
as the fiber instead of \([-1,1]\). 
\end{example}%
The two bundles are inequivalent as \(\Spin(1)\)-bundles 
since the cocycle 
of Example \ref{ex:pz2} is not a \(\lZ_2\) coboundary, as is easily seen.

Whereas a spin structure solves the problem of defining a spin bundle
starting from any \(\Cl(r,s)\)-module, a weaker condition, that of a
spin\({}^c\) structure solves the problem for {\em complex\/} modules. 
Since as was seen before, 
the algebra action on a complex module \(W\) can
be extended to an algebra action of \(\Cl_{\lCe}(r,s)\), one naturally
has an
action of \(\Spin^c(r,s)\) on \(W\) extending that of \(\Spin(r,s)\).

An {\em \((r,s)\)-spin\({}^c\) structure\/} 
\index{structure!spin\({}^c\)}%
for an oriented
pseudo-Riemannian manifold \(M\) with signature \((r,s)\) is a
principal \(\Spin^c(r,s)\)-bundle \(P\Spin^c(r,s)\),  a
principal \(\U(1)\)-bundle \(P\U(1)\) over \(M\), and a bundle
map \(\xi:P\Spin^c(r,s)\to \cF_{\SO}(M)\times P\U(1)\)
which is equivariant with respect to
the right actions, that is,
\(\xi(p\cdot(u, z))=\xi(p)\cdot (\lambda(u), z^2d)\),
where \(\lambda:\Spin(r,s)\to \SO(r,s)\) is the canonical double cover. 
The manifold \(M\) is
said to be an {\em \((r,s)\)-spin\({}^c\) manifold\/} 
\index{manifold!spin\({}^c\)}%
if it has an
\((r,s)\)-spin\({}^c\) structure. 

Suppose now that \(M\) is an oriented \((r,s)\)-spin\({}^c\) manifold
and let \(\xi\) be a \((r,s)\)-spin\({}^c\) structure.
Because of the canonical map \(\lambda^c:\Spin^c(r,s)\to \SO(r,s)\) one
sees that any bundle associated to
 \(\cF_{\SO}(M)\) can
be redefined,
using the spin\({}^c\) structure, as a bundle associated to
\(P\Spin^c(r,s)\).
This of course is again true of
the Clifford bundle \(\Cl(M)\). One has also the complexified Clifford
bundle  \(\Cl(M)\otimes \lC\), viewed either as the tensor
product of \(\Cl(M)\) with the trivial bundle with fiber \(\lC\), or as
a bundle associated to \(P\Spin^c(r,s)\) through the canonical action of
\(\Spin^c(r,s)\) on \(\Cl_{\lCe}(r,s)\).

Let now \(W\) be a complex
\(\Cl(r,s)\)-module.  We can now carry out a construct entirely
analogous to the one that constructed a spin bundle from a spin
structure to now create a complex spin bundle from the spin\({}^c\)
structure. To do so, note that there is an action, call it \(\rho\), of
\(\Spin^c(r,s)\) on \(W\) and so one can form the associated bundle
\(S=P\Spin^c(r,s)\times_\rho W\).  We can proceed to define the action
of \(\Cl_x\otimes_{\lRe}\lC\) on \(W_x\) making \(S\) into a complex
spin bundle. Similar to what we did  before,
 consider now  the map
\[
\Id\times\cdot:P\Spin^c(r,s)\times
\Cl_{\lCe}(r,s) \times W \to P\Spin^c(r,s)\times W
\]%
We can now repeat almost word by word the paragraph following 
(\ref{eq:spinbdef}) to define 
a bundle map
\(\Cl_{\lCe}(M)\times S \to S\), and so a spin-bundle.

If either \(r+\frac{n(n-1)}{2}\) is even or we are
dealing with a complex module, then either \(\Cl(V,q)\), 
or respectively \(\Cl(V,q)\otimes\lC\), contains a
volume element.  In the corresponding
situations for spin-bundles on 
oriented pseudo-Riemannian manifolds,
one can choose a global section \(\omega\) of the
Clifford bundle which at each point \(x\in M\) is the volume element of
\(\Cl_x\). This is because there is no sign ambiguity in choosing
the volume element if we only consider orthonormal basis with the same
orientation. We now have  global idempotents \(p_\pm\) defined by
(\ref{eq:cliffids}) pointwise.
Now \(W=W^+\oplus W^-\)
and the corresponding bundle \(S\) splits into two sub-bundles 
\(S^\pm=p_\pm S\). If \(n\) is even, these have the same fiber dimension,
and are  then sometimes
called {\em half-spin\/}
\index{bundle!half-spin}%
bundles.

Connections on spin bundles generally arise through their being associated
to principal bundles. Let us first consider a spin structure
\linebreak
\(\xi:P\Spin(r,s)\to \cF_{\SO}\) on a pseudo-Riemannian manifold \((M,g)\). 
Let 
\(S\) be a
spin-bundle associated to \(P\Spin(r,s)\) and \(\omega\) the Levi-Civita
connection on \(\cF_{\SO}\). Because \(\xi\) is two-to-one on fibers and
equivariant, parallel transport in \(\cF_{\SO}\) lifts in a unique way to
transport in \(P\Spin(r,s)\) and this defines a unique lifting of
\(\omega\) to an invariant connection \(\tilde\omega\) on \(P\Spin(r,s)\)
which then is transferred to \(S\) in the way  explained in Section
\ref{sec:connect}.  As \(\Cl(M)\) is
associated to \(\cF_{\SO}\), and consequently also to  \(P\Spin(r,s)\) as
explained above in this section, this bundle also gains a connection
associated to \(\tilde\omega\). The connections in  \(\Cl(M)\) and \(S\)
are related through the Leibniz rule:
\begin{equation}\label{eq:leibclspin}%
\nabla_\cX a\psi=(\nabla_\cX a)\psi+a\nabla_\cX\psi
\end{equation}%
for \(a\in \Gamma(\Cl(M))\) and \(\psi\in\Gamma(S))\) and where the product
is Clifford action. To see this note that the Leibniz rule holds
for sections of  \(\Cl(M)\otimes S\) by virtue of Section \ref{sec:covdev}.
 The Clifford multiplication map \(
\Cl(M)\otimes S \to S\) is represented in a  trivialization by a
constant map \(\mu:\Cl(r,s)\otimes W \to W\). Following the argument of
Section  \ref{sec:covdev},  one sees
that \(\nabla \mu=0\) and so the Leibniz rule holds.

Now whereas a spin structure allows us to use the Levi-Civita connection to 
uniquely define an associated connection on spin bundles, the same is not 
true for a spin\({}^c\) structure \(\xi:P\Spin^c(r,s)\to \cF_{\SO}(M)\times 
P\U(1)\)
because of the presence of of the factor \(P\U(1)\).  However, given an 
invariant connection \(\alpha\) on \(P\U(1)\), one can combine it with the 
Levi-Civita connection \(\omega\) on \(\cF_{\SO}\) to get the 
product connection (see end of Section \ref{sec:connect})
\(\omega \times \alpha\) on \(\cF_{\SO}(M)\times P\U(1)\), and just 
as before lift 
it uniquely by \(\xi\) to an invariant connection on 
\(P\Spin^c(r,s)\) which 
can now be used to define the associated connection on its associated 
bundles. This reproduces the connection associated to \(\omega\) on all 
bundles that are also associated to \(P\Spin(r,s)\), in particular those 
associated to \(\cF_{\SO}(M)\), but on \(S\) the connection 
is defined by both 
\(\omega\) and \(\alpha\). The resulting covariant derivative continues to 
obey the Leibniz rule (\ref{eq:leibclspin}) for the same reason as 
in the previous paragraph.

One can use the idea behind the construction of the spin bundles 
to introduce 
further structure in them. For instance suppose that on the 
\(\Cl(r,s)\)-module \(W\) one has a symmetric bilinear  form 
\((\cdot,\cdot)\) that is invariant with respect 
to Clifford algebra action by 
elements of \(\Spin (r,s)\). Consider now the map
\[%
\Id\times(\cdot,\cdot):P\Spin(r,s)\times
W \times W \to P\Spin(r,s)\times \lR
\]%
There is an action of \(\Spin(r,s)\) on the left-hand space given by
\((r,\phi,\psi)\mapsto (ru^{-1}, u\phi, u\psi)\) under the
quotient of which one has  the associated bundle
\(S\times S\). Likewise there is an action on the right-hand
space
given by \((x,r,s)\mapsto (x, ru^{-1},s)\) under the
quotient of which one has  the trivial bundle \linebreak
\(M\times\lR\). These two actions are compatible with the horizontal map as 
one
has \( (u\phi,u\psi) = (\phi,\psi)\) by hypothesis. 
One now has the  bundle map
\(S\times S \to M\times \lR\), defining a smoothly varying bilinear 
form, which we continue to denote by \((\cdot,\cdot)\), in each fiber
\(W_x\). 
Suppose we have done such a construction, then the pointwise bilinear 
form can be identified with a section \(t\) of \(S'\otimes S\) where \(S'\) 
is the dual bundle to \(S\). Now \(S'\otimes 
S=P\Spin(r,s)\times_{\rho^*\otimes\rho}(W'\otimes W)\), where 
\(\rho^*\) is the dual representation.  The bundle 
\(S'\otimes S\) inherits a connection \(\omega\) from the 
connection on the bundle \(P\Spin(r,s)\) which in turn is inherited from 
the Levi-Civita connection on \(\cF_{\SO}(M)\). 
A trivialization of \(S'\otimes S\) is 
obtained as a quotient of \(U\times \Spin(r,s)\times (W'\otimes W)\). Let  
\(\cX\) be any vector field. One now has in this trivialization  
\(\nabla_{\cX} t =
\cX(t) +(-\rho(\omega(\cX))'\otimes I + I\otimes \rho(\omega(\cX)))t\).
The first term vanishes since \(t\) is represented by a {\em constant\/} 
section. The vanishing of the second term is just the infinitesimal 
expression of the invariance of  \(t\) under the action of \(\rho^*\otimes 
\rho\).
Thus \(\nabla t = 0\) and so for two sections \(\phi\) 
and \(\psi\) of \(S\) 
and any vector field \(\cX\) we have
\begin{equation}\label{eq:inners}%
\cX(\phi,\psi)=(\nabla_{\cX}\phi,\psi) +(\phi,\nabla_{\cX}\psi)
\end{equation}%

In entirely a similar fashion, starting with a
\(\Spin^c(r,s)\)-invariant hermitian sesquilinear form on a complex
\(\Spin(r,s)\)-module \(W\), one can construct a pointwise hermitian
sesquilinear form on the associated spin bundle \(S\), satisfying
property (\ref{eq:inners}).

\subsection{The Dirac Operator} 

In the standard model of elementary particles, matter is represented by
spinor fields and interactions by connections on principal bundles. The
Dirac operator is the basic differential operator acting on spinor
fields, and the forces between the particles described by these fields
is achieved through the minimal coupling ideas of Section
\ref{sec:mincoup}. This imparts particular importance and usefulness to
the Dirac operator. It is remarkable that the Dirac operator has also
shown to have a fundamental mathematical importance in manifold theory.

Let \(S\) be a spin bundle on a pseudo-Riemannian manifold \(M\). Assume
\(S\) has a linear connection and let
\(\nabla\) be the corresponding covariant derivative. Recall (see end of 
Section \ref{sec:covdev}) that one
can consider \(\nabla\) as a map \linebreak
\(\nabla:\Gamma(S)\to
\Gamma(T^*M\otimes S)\).
The pseudo-Riemannian metric provides an isomorphism 
\(r:T^*M\to TM\)
(see Section \ref{sec:pseudoriemannian})
which extends to an isomorphism \(r^\sharp:\Gamma(T^*M\otimes S) \to
\Gamma(TM\otimes S) \). Clifford algebra action finally gives a map
\(\Gamma(T^*M\otimes S)\to\Gamma(S)\). We define the Dirac
operator \(D:\Gamma(S)\to \Gamma(S)\) as the composition of these three
maps: 
 \begin{diagram} \Gamma(S)& \rTo^\nabla & 
\Gamma(T^*M\otimes S) &\rTo^{r^\sharp} &
 \Gamma(TM\otimes S)& \rTo^\cdot & \Gamma(S)
\end{diagram}%

It is instructive to calculate the local form of this operator in a
trivialization. Let \(e_1,\dots,e_n\) be a set of vector fields in an open 
set \(U\) which at each point form a basis for the tangent space, and let 
\(e^1,\dots,e^n\) be  the corresponding dual \(1\)-forms.
Let \(\gamma_i(x)\in \End(S_x)\) be
the endomorphism that corresponds to Clifford action by \(e_i\), and
let \(\nabla_i\) denote \(\nabla_{e_i}\).
 One has
\(\nabla \psi = \sum_i e^i\otimes \nabla_i \psi\). Under \(r^\sharp\) this
becomes \(\sum_ir(e^i)\otimes \nabla_i \psi\). One has 
\(r(e^i)=\sum_jg^{ij}e_j\). Under Clifford action one finally has \(D\psi 
=\sum_{ij}g^{ij}\gamma_i\nabla_j \psi\). If we set 
\(\gamma^i=\sum_jg^{ij}\gamma_j\) then one can write \(D\psi 
=\sum_j\gamma^j\nabla_j \psi\). Two particular useful cases is to take for 
\(e_i\)  an \(n\)-bein or the coordinate vector fields 
\(\ppv{x}{i}\) for a set of local coordinates 
\(x^1,\dots,x^n\).

\begin{example}%
A complex representation of \(\Cl(0,3)\) is provided by associating to the
canonical  
orthonormal basis \((e_1, e_2, e_3)\) of \(\lR^3\subset \Cl(0,3)\) 
the following corresponding \(2\times 2\) matrices 
known as the 
{\em Pauli spin matrices}.
\index{Pauli spin matrices}%
\[
\begin{array}{ccc}%
\sigma^1 = \left(\begin{array}{cc}0 & 1 \\ 1 & 0\end{array}\right) &
\sigma^2=\left(\begin{array}{cc}0 & -i \\ i & 0\end{array}\right) &
\sigma^3=\left(\begin{array}{cc}1 & 0 \\ 0 & -1\end{array}\right)
\end{array}
\]
The corresponding spin bundle on \(\lR^3\) is just  \(\lR^3\times
\lC^2\) and the Dirac operator acting on \(\psi:\lR^3\to \lC^2\) is then
\[
\left(\begin{array}{cc}%
\frac{\partial}{\partial z} & \frac{\partial}{\partial x}-
i\frac{\partial}{\partial y} \\
\frac{\partial}{\partial x} + i\frac{\partial}{\partial y} &
\frac{\partial}{\partial z}
\end{array}\right)
\]
\end{example}%

One easily calculates that \(D^2 = \Delta I\) where \(\Delta\) is the
three-dimensional Laplacian. It is this property of being a ``square
root" of the Laplacian that is at the root of the great usefulness of
the this operator.

We briefly describe now  the historical case of four-dimensional
space-time \(\lR^4\) with the notations and conventions of Section
\ref{sec:maxwell}, and consider an irreducible complex module of 
\(\Cl(3,1)\). The reason for choosing signature \((3,1)\) and not 
that of the space-time,  \((1,3)\), is that in the 
physical literature the Clifford
algebra that is normally associated to the quadratic form \(q\) is what
in the mathematical literature is normally associated to the form 
\(-q\). The reader must also beware that some authors consider the 
signature
of space-time to be \((3,1)\) and some even use imaginary
coordinates. 

Since \(\Cl_{\lCe}(4) \simeq \lC(4)\) the unique 
irreducible complex module of
\(\Cl(3,1)\) can be taken to be \(\lC^4\). The spin bundle \(S\)
 is trivial
and can be taken to be \(\lR^4\times \lC^4\). Section of this bundle are
called {\em Dirac spinors\/}. 
\index{Dirac spinor}%

The famous {\em Dirac equation\/}
\index{equation!Dirac}%
is 
\[
(iD - mI)\psi = 0
\]%
where \(m\) is a constant that in a physical
particle theory corresponds to the mass of the particle. 

Following the discussion at the end of Section \ref{sec:spingroups}
there is a hermitian product \(\bar\psi\phi\) on spinors 
which is invariant
under the action of the component of the identity of \(\Spin(3,1)\).
In the physical literature this is known as {\em Lorentz invariance\/}.

The Dirac equation is the
Euler-Lagrange equation of a Lagrangian theory with
\[
\cL_D = \bar\psi(iD-mI)\psi\Omega
\]
 where \(\Omega =dx^0\wedge dx^1 \wedge dx^2 \wedge dx^3\) is the volume
element of space-time. 
By the discussion in Section \ref{sec:spinbundles} the bundle
\(S\) splits into a direct sum of two sub-bundles \(S=S^+\oplus S^-\). 
Sections of these sub-bundles are called {\em Weyl spinors\/}. As the
Dirac operator maps \(S^\pm\) to \(S^\mp\), the Dirac equation only
makes sense for Weyl spinors if \(m=0\), that is Weyl spinors correspond
to massless particles.   The group \(\Spin(3,1)\) maps each
sub-bundle into itself, but the two representation are
not equivalent.
In the physical literature it is said that
they differ by  
{\em helicity\/}, 
\index{helicity}%
which has to do with the intrinsic angular momentum carried by
the particles.

The Dirac equation has a global \(U(1)\) symmetry  whose action is
to multiply \(\psi\) by a unimodular complex number. Applying the
minimal coupling idea of Section \ref{sec:mincoup} to make the \(\U(1)\)
symmetry local, modifies the Dirac equation to 
\begin{equation}\label{eq:diracem}
(iD +qA -mI)\psi = 0
\end{equation}
Where \(A\) is a vector field viewed as a section of the 
Clifford bundle (and so has Clifford algebra action on spinors) 
and  physically identified with
the electromagnetic potential of Section \ref{sec:maxwell} through the
correspondence, provided by the metric, of vector fields and \(1\)-forms. 
We have
introduced a physical constant \(q\) identified with the
\index{charge!electric}%
 {\em electric
charge\/}.

There is an antilinear map \(\psi\to \psi_c\), which we shall not detail
here, that takes a solution of (\ref{eq:diracem}) to one  of 
\((iD -qA -mI)\psi = 0\)
with the opposite sign of the charge. This map is called
{\em charge conjugation\/}, 
\index{charge conjugation}%
though this is a bit of a misnomer, as
in particle physics it relates particles and anti-particles and should
be more properly called {\em matter-anti-matter conjugation\/}, which is
defined also for neutral particles. Charge conjugation maps a Weyl
spinor of one helicity to one of the other.

As \(\Cl(1,3) \simeq \lR(4)\), there is a real irreducible spin bundle 
based on the module \(\lR^4\). Sections of this spin bundle are called
{\em Majorana spinors\/}. In the physical literature, Majorana spinors
are generally taken as sections of the complex Dirac spin bundle which
satisfy a certain real-linear {\em reality condition\/}. As Majorana
spinors do not have a global \(U(1)\)-symmetry, one cannot use the
minimal coupling scheme to couple them to the electromagnetic field, and
so they correspond to neutral particles, more specifically 
to particles that are identical to their anti-particles.

Using again the minimal coupling ideas of Section \ref{sec:mincoup}, the
Lagrangian density for a Dirac spinor coupled to electromagnetism is taken
to  be
\[
\cL = \bar\psi(iD +q A -mI)\psi
\Omega -\frac{1}{2} F\wedge *F
\]
where \(F\) is the electromagnetic \(2\)-form (\ref{eq:fda}). The 
Euler-Lagrange equations for this Lagrangian are
\begin{eqnarray*}
&(iD +qA -mI)\psi =0&\\
&\delta F = q\bar\psi\cdot\psi&
\end{eqnarray*}
where \(\bar\psi\cdot\psi\) is the \(1\)-form that takes a vector \(v\) to 
\(\bar\psi v\psi\).
This gives an example of the current \(1\)-form (\ref{eq:jform}).
This provides a rather accurate theory of electrons and positrons (the
anti-particle of the electron) interacting with the electromagnetic
field at low energies and treated quasi-classically, that is, the
particles are treated quantum-mechanically and the electromagnetic field
classically. A full quantum mechanical treatment starts with the same
Lagrangian but follows a procedure of {\em quantization\/} of both the
\(\psi\) and \(A\) fields and results in a {\em quantum field theory},
known
as {\em quantum
electrodynamics, or QED\/}, one of the most precise physical theories
ever constructed.

Consider now the product \(L=S^+\otimes \lC^N\) of \(S^+\) (similar
considerations apply to the other half-spin bundle) with the trivial
\(\lC^N\)-bundle, and the equation \begin{equation}\label{eq:dpsi}
D\otimes I \psi = 0 \end{equation} for a section \(\psi\) of
\(L\). Introducing the canonical basis \(e_1,\dots,e_N\)
of \(\lC^N\), one can write \(\psi = \sum_{i=1}^N\psi_i\otimes e_i\) and
(\ref{eq:dpsi}) corresponds to \(N\) independent identical Dirac
equations \(D\psi_i=0\). Equation (\ref{eq:dpsi}) has a global \(\U(N)\)
symmetry whose action, given \(T \in \U(N)\), is \(T\cdot \psi =
I\otimes T\psi\). We are now in the context of Section \ref{sec:mincoup}
and can make the symmetry global through minimal coupling introducing an
invariant connection on a principal \(\U(N)\)-bundle. Extend the
hermitian product on \(S^+\) to \(L\) by setting
\((\psi\otimes u,\phi\otimes v)= \bar\psi\phi(u, v)\) where
\((u,v)=\sum_{i=1}^Nu_i^*v_i\) is the usual hermitian inner product
on \(\lC^N\). Equation (\ref{eq:dpsi}) is the Euler-Lagrange equation of
a Lagrangian density given by \(\cL_\psi = (\psi,(D\otimes I)\psi)\Omega\). 
The
minimally coupled Dirac-Yang-Mills theory is governed by the Lagrangian
density 
\[ \cL_{DYM} = (\psi,(D\otimes I+g\hat A)\psi)\Omega -
\frac{1}{2}\,\Tr\,(F\wedge *F) 
\] 
where \(g\) is a physical constant
called the {\em coupling constant\/}, \(A=\sum_iA_idx^i\) is 
the gauge potential of
the connection, \(F\) the curvature \(2\)-form of the connection, and
\(\hat A\) is defined as the composition 
\begin{diagram}
\Gamma(L) & \rTo^{I\otimes A} & \Gamma(T^*M\otimes
L) &\rTo^{r^\sharp} & \Gamma(TM\otimes L)&
\rTo^{\cdot\otimes I} & \Gamma(L) 
\end{diagram}%
where, for concreteness' sake, \((I\otimes A)(\psi\otimes u)=\sum_idx^i\otimes
\psi\otimes A_iu\), \linebreak
\((\cdot \otimes I) (v\otimes\psi\otimes u) = v\psi\otimes u\), 
and \(r^\sharp\) is the obvious extension of the isomorphism
\(r:T^*M \to TM\). 

The Euler-Lagrange equations of this theory are
\begin{eqnarray*}
&(D\otimes I+g\hat A)\psi = 0& \\ 
&{* d_A*F} = g(\psi,(\cdot \otimes \cdot)\psi)&
\end{eqnarray*}
where the right-hand side of the second equation has to be interpreted
as a \(1\)-form whose value at a vector \(v\) is 
\(g(\psi,(v \otimes \cdot)\psi)\) which in turn must be interpreted as a
hermitian \(N\times N\) matrix, an element of \(\gu(N)\). Concretely the
matrix 
elements are \(g\bar\psi_i v\psi_j\). Hermiticity is assured by properties
of the hermitian product on \(S^+\). 

It is essentially this construct that gave rise to the original Yang-
Mills theory. A quantized elaboration of it is behind the Standard Model of
elementary particle interactions, the spinor fields representing
fundamental matter, and the gauge potentials the forces acting on it. An
extra field called the {\em Higgs field\/} 
\index{field!Higgs}%
must be added to provide appropriate masses to the particles, but we
won't elaborate on this here.

\subsection{The Seiberg-Witten Equations}
We are now in condition to introduce the Seiberg-Witten equations. 
Let \(M\) 
be a \(4\)-dimensional oriented Riemann manifold. Such a manifold 
always has 
a spin\({}^c\) structure, though we shall not prove this here. 
Chose one such 
structure and introduce an irreducible complex spin bundle \(S\).  
Introduce 
now a connection in \(P\Spin^c\) which is a lifting of the 
product connection 
in \(\cF_{\cO}(M)\times P\U(1)\) consisting of the Levi-Civita connection on 
\(\cF_{\cO}(M)\) and an invariant connection \(\alpha\) on \(P\U(1)\). 
This connection then induces one in the 
associated bundle \(S\). By considerations introduced in section 
\ref{sec:cliffa} one can introduce a fiberwise hermitian inner product 
\((\cdot, \cdot)\) in this bundle satisfying (\ref{eq:inners}) with respect 
to the  covariant derivative. 
 The spin bundle \(S\)
splits into the direct sum of two half-spin bundles \(S^{\pm}\). 
Denote 
by \(D_\alpha^+\) the Dirac operator restricted to \(S^+\). Let \(F_\alpha\) 
be the curvature \(2\)-from  of the connection, and \(F_\alpha^+\) 
its self-dual part.
Let now \(e_1,\dots,e_4\) be an 
orthonormal set of tangent vectors at some point, 
\(e^1,\dots,e^4\) the dual 
basis of covectors, and \(\psi \in S^+\). Define
\[
\sigma(\psi)=\sum_{ij}(e_i\psi,e_j\psi)e^i\wedge e^j
\]%
It is not hard to verify that \(\sigma(\psi)\) is independent of the choice 
of orthonormal basis and that it is a purely imaginary,  
self-dual \(2\)-form. 

 The famous Seiberg-Witten equations for a half-spinor field 
\(\psi\in\Gamma(S^+)\) and the connection \(\alpha\) are now:
\begin{eqnarray}%
F_\alpha^+ &=& \frac{i}{4}\sigma(\psi)\\
D_\alpha^+\psi &=& 0
\end{eqnarray}%

Unfortunately we shall not explore the remarkable properties of these
equations in these notes.

\section*{Appendix}%

\appendix

\section{Basic Conventions}\label{sec:bacon}%

If \(X\) is an object of any category, by \(\End(X)\) we mean the set of
morphism of \(X\) to itself, that is \(\Hom(X,X)\). By \(\Aut(X)\) we
mean the set of elements of \(\End(X)\) that are invertible.
Depending on the category, the sets \(\End(X)\) and \(\Aut(X)\) may have
additional structures which are always assumed. Thus for linear spaces,
\(\End(X)\) is an algebra. Of course, \(\Aut(X)\) is always a group. 

If \(W\) is a vector space over a 
field \(\lF\), then we denote by \(W'\)
\index{\(W'\)}%
 its dual, that is, the set of
linear maps \(\phi:W\to \lF\). We shall sometimes denote  \(\phi(w)\)
by \(<\phi,w>\). Suppose \(W\) finite dimensional and let 
\(e_1,\dots,e_n\) be a basis. We denote by \(e^1,\dots,e^n\) the
corresponding dual basis,
\index{basis!dual}%
 that is, one defined by
\[
<e^i,e_j> = \delta^i{}_j
\]
where the {\em Kroneker
symbol\/} 
\index{\(\delta^i{}_j\)}%
\(\delta^i{}_j\) is defined as 
\[
\delta^i{}_j = \left\{\begin{array}{cc} 1 & \hbox{if}\quad i=j \\
0 & \hbox{if}\quad i\neq j\end{array}\right.
\]

We use the physicist's habit of indicating components by indices,
sometimes they are subscripts and sometimes superscripts, as for \(e^i\),
\(e_j\), and  \(\delta^i{}_j\) above. The placement is not capricious,
but the reasons will not be explained here. The reader should not
confuse a superscript index with a power.

Given \(\phi_1,\phi_2,\dots,\phi_p\) elements of \(W'\), 
our convention as to
what \(\phi_1\wedge\phi_2\wedge\cdots\wedge\phi_p\) means as an anti-symmetric
\(p\)-linear form on \(W\) is 
\begin{equation}\label{eq:pformcon}%
(\phi_1\wedge\phi_2\wedge\cdots\wedge\phi_p)(w_1,w_2,\dots,w_p) =
\det_{ij}<\phi_i,w_j>
\end{equation}%
where by \(\det_{ij}a_{ij}\) we mean the determinant of the matrix
\(a_{ij}\). There is a contending convention in which the right-hand side
of (\ref{eq:pformcon}) is divided by \(p!\). There are very good
reasons for adopting either one of these, so the reader must beware.

If \(A:W \to V\) is a linear map between two vector spaces over \(\lF\),
we denote by \(A':V'\to W'\)
\index{\(A\)}%
 the {\em dual map\/}
\index{map!dual}%
defined by
\((A'\phi)(x) = \phi(Ax)\). For an \(n\times m\) matrix \(M\) over
\(\lF\), we denote by \(M^t\) the matrix transpose. 

The group \(\Aut(W)\) of invertible linear maps \(W \to W\) is called
the {\em general linear group of W\/} 
\index{group!general linear}%
and shall be denoted by \(\GL(W)\).
\index{\(GL(W)\)@\(\GL(W)\)}%
For \(W=\lF^n\) we shall write \(\GL(n, \lF)\)
\index{\(GL(n,F)\)@\(\GL(n, \lF)\)}%
and when \(\lF=\lR\) we shall simply write 
\(\GL(n)\). 
\index{\(GL(n)\)@\(\GL(n)\)}%
The identity of \(\GL(W)\) we shall usually denote by 
\index{\(I\)}
\(I\).

All manifolds are assumed to be Hausdorff and paracompact.
If \(M\) is a manifold we denote by \(T_xM\) the
space of tangent vectors at \(x\in M\) and by \(TM\) the 
{\em tangent bundle\/}.
\index{bundle!tangent}%
Likewise we denote by \(T^*_xM\) the set of covectors at  \(x\in M\) and
by \(T^*M\) the
{\em cotangent bundle\/}. 
\index{bundle!cotangent}%
We assume the reader is familiar with tensor fields and the
corresponding tensor bundles such as \(TM\otimes T^*M \otimes T^*M\), etc.
Likewise for differential forms. 

Adopting convention (\ref{eq:pformcon}) one has the following formula
for a \(1\)-form \(\alpha\) and vector fields \(\cX\) and \(\cY\):
\begin{equation}\label{eq:donef}%
d\alpha(\cX,\cY) = \cX(\alpha(\cY)) -\cY(\alpha(\cX)) -\alpha([\cX,\cY])
\end{equation}%
With the contending convention there would be an overall factor of
\(\frac{1}{2}\) on the right-hand side. 

The analog of (\ref{eq:donef}) for \(p\)-forms is
\begin{eqnarray}\nonumber
\lefteqn{d\alpha(\cX_1,\dots,\cX_p) = \sum_i(-1)^{i+1} 
\cX_i(\alpha(\cX_1,\dots,\hat \cX_i,\dots \cX_{p+1})) +}\\ \label{eq:dpf}
& & \sum_{1\le i < j \le p+1}(-1)^{i+j+1}
\alpha([\cX_i,\cX_j],X_1,\dots,\hat \cX_i, \dots,
\hat\cX_j,\dots,\cX_{p+1})
\end{eqnarray}
where a hat over an argument means that it is missing.

\section{Parameterized Maps}%
\label{sec:parmaps}%
A map \(f:X\times Y\to Z\) can be viewed alternatively as a family of
maps from \(Y\) to \(Z\), parameterized by \(x\in X\). Formally, there
is an associated map \(f^\flat: X\to \Hom(Y,Z)\) given by
\(f^\flat(x)(y)=f(x,y)\).  When \(Y\) reduces to a one-point set,
\(f\) is essentially a map from \(X\) to \(Z\), and \(f^\flat\) is then
thought of as a parameterized family of elements of \(Z\). Under this
circumstance one usually drops mention of the set \(Y\).
By abuse of notation we shall drop the indicator
\(\flat\) and use the same symbol for both maps, letting context clarify.
Again, depending on the context, the most convenient
 view of such maps could be
either as maps from cartesian products, or as parameterized maps. In
fiber-bundle theory, the parameterized version is the most useful for
many of the maps encountered there, and we shall use special notational
conventions for these.
When all such maps are parameterized by the same space \(X\) we
shall  often, for sake of brevity, not indicate the parameter \(x\)
which is tacitly understood. Thus if \(g:X\times W \to Y\) we shall
write \(f\circ g\) for the parameterized map 
corresponding to the map  \(X\times W \to Z\) given
by \((x,w) \mapsto f(x,g(x,w))\). Alternatively we can write \(f\circ
g(x)(w)=(f(x)\circ g(x))(w)\). Similar construct hold when the maps, for
each \(x\in X\) take values in algebraic objects. Thus if \(f(x)\in V\)
where \(V\) is a vector space, and \(L(x)\in \End(V)\) is a
parameterized family of endomorphism of \(V\), then by \(Lf\) we mean
the map \(x\mapsto L(x)f(x)\).

\section{Vector-valued Differential Forms}\label{sec:vvdf}%
Let \(M\) be a manifold. Recall that a covector at \(x\in M\)
is a linear functional on \(T_xM\), that is, a linear map \(T_xM \to
\lR\). Let now \(W\) be a real vector space. A linear map \(\omega:T_xM
\to W\) is called a {\em \(W\)-valued covector at \(x\)\/}, or
generically a {\em vector-valued\/}
\index{form!vector-valued}%
covector. If one has a
\(W\)-valued covector \(\omega(x)\) for each \(x\in U\subset
M\)  then one speaks of a {\em \(W\)-valued\/} differential
forms in \(U\). Sometimes we shall write \(\omega_x\) instead of
\(\omega(x)\).

Let \(\omega\) be a
\(W\)-valued form and \(\phi\in W'\) be an element of the dual space,
then \(\phi\circ \omega\)
is a usual differential form, that is, for each \(x\), 
a map \(T_xM \to \lR\) given by
\(v\mapsto \phi(\omega_x(v))\). We say that \(\omega \) is \(\cC^\infty\)
if \(\phi\circ \omega\) is \(\cC^\infty\) for all  \(\phi\in W^*\).

Higher order {\em vector-valued \(p\)-forms\/}
\index{pform@\(p\)-form!vector valued} are defined analogously. 
A \(W\)-valued \(p\)-form in
an open set \(U\) is, for each, 
\(x\in U\)  a totally anti-symmetric \(p\)-linear map \(T_xM \times
\cdots\times T_xM\to W\). 
For any \(\phi\in W'\) and any \(W\)-valued
\(p\)-form \(\alpha\), \(\phi\circ \alpha\) is an ordinary \(p\)-form,
and as before we say \(\alpha\) is is \(\cC^\infty\)
if \(\phi\circ \alpha\) is \(\cC^\infty\) for all  \(\phi\in W'\).

The exterior differential \(d\) can now be extended to \(W\)-valued
forms. Given any \(\cC^\infty\) \(W\)-valued
\(p\)-form \(\alpha\), we define the \(W\)-valued \(p+1\)-form \(d\alpha\)
as that form for which for all \(\phi\in W'\) one has
\(\phi\circ (d\alpha)=d(\phi\circ\alpha)\). The
reader can easily verify that this defines  \(d\alpha\) uniquely. As
before, \(d^2=0\).

\printindex

\end{document}